\documentclass[]{jkpaper}
\usepackage{tikz}
\usetikzlibrary{decorations.markings}
\usetikzlibrary{decorations.pathreplacing}
\usetikzlibrary{arrows.meta}
\usetikzlibrary{patterns}
\usetikzlibrary{decorations.pathmorphing}
\usepackage{tikz-cd}
\usepackage{bigints}
\usepackage{biblatex}
\usepackage{aas_macros}
\usepackage[outline]{contour}
\usepackage{hyperref}
\usepackage{url}
\usepackage{amssymb}
\usepackage{soul}

\theoremstyle{plain}

\theoremstyle{definition}

\newtheorem*{eg*}{Example}

\newcommand{\OIST}{\raisebox{-0.08em}{\includegraphics[height=0.8em]{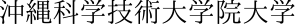}}}

\newcommand{\eps}{\varepsilon}

\newauthornote{\jjvk}{JJVK}{blue!80}
\newcommand{\stkout}[1]{\ifmmode\text{\sout{\ensuremath{#1}}}\else\sout{#1}\fi}

\newcommand{\sins}{\mathop{\lrcorner}}
\newcommand{\fins}{\cdot}
\newcommand{\sd}{\mathrm {d}}
\def\extd{\mathrm {d}}
\newcommand{\fd}{\delta}
\newcommand{\sL}{\mathcal{L}}
\newcommand{\fL}{\mathbb{L}}
\newcommand{\DD}[1]{\text{D}_A#1}

\title{Linearization \texorpdfstring{\textcolor{black!40!white}{(in)}}{(in)}stabilities and crossed products}
\author{Julian De Vuyst\texorpdfstring{\textsuperscript{1,a}}{}, Stefan Eccles\texorpdfstring{\textsuperscript{1,b}}{}, Philipp A.\ H\"ohn,\texorpdfstring{\textsuperscript{1,c}}{} and Josh Kirklin\texorpdfstring{\textsuperscript{2,d}}{}}
\institution{\texorpdfstring{\textsuperscript{1}}{}Qubits and Spacetime Unit,\texorpdfstring{\\}{ } Okinawa Institute of Science and Technology \emph{(}\OIST\emph{)},\texorpdfstring{\\}{ } 1919-1 Tancha, Onna-son, Kunigami-gun, Okinawa, Japan 904-0495
\texorpdfstring{\\\vspace*{1.5em}}{ } \texorpdfstring{\textsuperscript{2}}{}Perimeter Institute for Theoretical Physics,\\ 31 Caroline Street North, Waterloo, ON, N2L 2Y5, Canada}

\email{
\textsuperscript{a}\emaillink{julian.devuyst@oist.jp}\\
\textsuperscript{b}\emaillink{stefan.eccles@oist.jp}\\
\textsuperscript{c}\emaillink{philipp.hoehn@oist.jp}\\
\textsuperscript{d}\emaillink{jkirklin@pitp.ca}
}

\abstr{
Modular crossed product algebras have recently assumed an important role in perturbative quantum gravity as they lead to an \emph{intrinsic} regularization of entanglement entropies by introducing quantum reference frames (QRFs) in place of explicit regulators. This is achieved by imposing certain boost constraints on gravitons, QRFs and other fields. Here, we revisit the question of how these constraints should be understood through the lens of perturbation theory and particularly the study of linearization (in)stabilities, exploring when linearized solutions can be integrated to exact ones. Our aim is to provide some clarity about the status of justification, under various conditions, for imposing such constraints on the linearized theory in the $G_N\to0$ limit as they turn out to be of second-order. While for spatially closed spacetimes there is an essentially unambiguous justification, in the presence of boundaries or the absence of isometries this depends on whether one is also interested in second-order observables. 
 Linearization (in)stabilities occur in any gauge-covariant field theory with non-linear equations and to address this in a unified framework, we translate the subject from the usual canonical formulation  into a systematic covariant phase space language. This overcomes theory-specific arguments, exhibiting the universal structure behind (in)stabilities, and permits us to cover arbitrary generally covariant theories. We comment on the relation to modular flow and illustrate our findings in several gravity and gauge theory examples.
}

\bibliography{refs}
\begin{document}
\maketitleandtoc

\section{Introduction}

One of the key observations in quantum gravity in recent years is that relational descriptions of quantum field degrees of freedom, including gravitons, lead to an intrinsic regularization of entanglement entropies  without the need to introduce an explicit UV regulator \cite{Chandrasekaran:2022cip}. While independent of holography, this observation was motivated by certain transitions in the type of regional von Neumann algebras in holography \cite{leutheusser2023causal,leutheusser2023emergent,Witten:2021unn} and has set off a large wave of follow-up works \cite{Jensen:2023yxy,Witten:2023qsv,Witten:2023xze,Kudler-Flam:2023qfl,Kudler-Flam:2024psh,AliAhmad:2023etg,Aguilar-Gutierrez:2023odp,Chen:2024rpx,Gomez:2023upk,Gomez:2023wrq,Fewster:2024pur,Kaplan:2024xyk,AliAhmad:2024wja,AliAhmad:2024vdw,Kolchmeyer:2024fly}. It has brought about  novel proofs of the generalized second law \cite{Faulkner:2024gst}, technical improvements to proofs of the Bekenstein bound and the quantum null energy condition \cite{Kudler-Flam:2023hkl}, a proposal for formulating the generalized second law beyond the semiclassical regime using quantum reference frames \cite{KirklinGSL}, and the observation that gravitational entropy of subregions is observer (or more precisely quantum reference frame) dependent \cite{DeVuyst:2024pop,DVEHK}. 

The seminal observation by Chandrasekaran, Longo, Penington and Witten (CLPW) in \cite{Chandrasekaran:2022cip} pertains to perturbative quantum gravity in the $G_N\to0$ limit and is based on so-called modular crossed product algebras, which essentially augment the algebra of regional QFT observables by a certain modular flow that acts as an automorphism of the algebra. Technically, the crux is that the original regional QFT algebra is a type $\rm{III}$ von Neumann algebra, which means that it possesses no trace and hence no well-defined entropies, while the crossed algebra is of type $\rm{II}$ and so possesses traces, density operators and well-defined renormalized entropies \cite{Takesaki1979,Sorce:2023fdx,Witten:2021unn}.

Physically, the crossed algebra can be realized as the algebra of regional QFT observables dressed either to the worldline of an observer with a clock or to the boundary, while including the clock or boundary ADM Hamiltonian. This algebra is then invariant under a constraint of the form
\begin{equation}\label{2clockconstraint}
     C = H_{\rm QFT}+\sum_i H_i\,,
\end{equation}
where $H_{\rm QFT}$ denotes the modular Hamiltonian (i.e.\ the generator of modular flow) of the regional algebra and its complement, and $i$ runs over the different observers with clocks and/or disconnected components of a global boundary when the spacetime is not spatially closed \cite{DeVuyst:2024pop,DVEHK}. Modular flow depends on a choice of KMS (or vacuum) state for the global QFT and does not in general admit a geometric interpretation. It does, however, admit such an interpretation in the case of a bifurcate Killing horizon, in which case the modular Hamiltonian $H_{\rm QFT}$ coincides with the boost Hamiltonian  associated with that horizon \cite{haag2012local,Bisognano:1975ih,Sewell:1982zz,sorce2024analyticity} (see \cite{Jensen:2023yxy,sorce2024analyticity} for a conjecture going beyond this case). This construction can thus be realized in de Sitter space with its cosmological horizon, as well as in black hole spacetimes. When the spacetime has an asymptotic boundary, the sum in Eq.~\eqref{2clockconstraint} contains a boost ADM Hamiltonian for each disconnected piece of the global boundary.

As pointed out in \cite{DeVuyst:2024pop,DVEHK,Fewster:2024pur,AliAhmad:2024wja}, the observer and boundary clocks constitute quantum reference frames (QRFs) for the modular flow; in particular, the crossed product algebra is generated by the relational observables describing the boost evolution of the regional degrees of freedom relative to a clock QRF and the reorientation transformations of that QRF \cite{DeVuyst:2024pop,DVEHK}. In an effort to go beyond the idea of explicitly introducing observers with clocks, \cite{Chen:2024rpx,Kudler-Flam:2024psh} instead extracted suitable clock degrees of freedom from an inflaton field. However, also these are  dynamical reference frames in the classical theory \cite{Goeller:2022rsx,Carrozza:2022xut} and QRFs in the quantum case. Thus, the regional QFT degrees of freedom in perturbative quantum gravity have a well-defined renormalized\footnote{\label{footnote: renormalized entropy}We say \textit{renormalized} entropy because for the relevant type II Von Neuman algebras, the entropies are defined up to a state-independent additive constant.  Said another way, these algebras have well-defined entropy differences.  In the case of type $\rm{II}_1$ factors, there is a maximum entropy state that is conventionally set to have entropy of zero (all other entropies then being negative), while for type $\rm{II}_\infty$ factors there is no maximal entropy state.  Type $\rm{I}$ factors, such as those familiar from finite dimensional quantum mechanics, are further set apart by also having minimal entropy states.  For information on Von Neumann algebras and type classifications we refer the reader to \cite{WittenRevModPhys.90.045003,Sorce:2023fdx,Takesaki1979,Witten:2021unn}.} entropy when described relative to a (modular gauge flow associated) QRF, but not otherwise.

Most works on the modular crossed product in gravity demonstrate that \emph{if} one imposes boost constraints of the schematic form \eqref{2clockconstraint} then one finds a type reduction in the regional von Neumann algebras.
What, in our opinion, has been somewhat lacking in the literature is an explicit justification for why one should impose such constraints on the perturbative quantum gravity theory (with  exceptions \cite{Jensen:2023yxy,Kudler-Flam:2023qfl,Kaplan:2024xyk} that we comment on below). Of course, CLPW note, for instance, that isometries are gauge in spatially closed universes\footnote{Sometime this distinction is given to spatially \textit{compact} universes rather than spatially \textit{closed} universes with the understanding that spacetimes with physical boundaries are excluded by fiat.  The relevant criterion is then whether or not spatial slices have an asymptotic boundary, which occurs only in the noncompact case.}, so that one must impose their generators as constraints \cite{Chandrasekaran:2022cip}. But there are still subtleties to address. The reason, as we shall see, is that these constraints \eqref{2clockconstraint} can \emph{not} be realized as linearized gauge constraints in the linearized theory in which we are ultimately interested in the strict $G_N\to0$ limit: the bulk contribution of linearized constraints associated with Killing fields vanishes \emph{identically}. Instead, they turn out to be second-order constraints. Consequently, it must be clarified why
\begin{itemize}
     \item[(a)] one should impose second-order constraints when exploring the linearized theory, 
    \item[(b)] it is legitimate to consider only a single (or finitely many) second-order constraint(s) as in Eq.~\eqref{2clockconstraint} while the diffeomorphism gauge group of gravity is infinite-dimensional, and
    \item[(c)] one should impose such constraints in spacetimes with boundaries where isometries are typically not gauge.
\end{itemize}
Given the physical significance of the observation by CLPW  in \cite{Chandrasekaran:2022cip}, it appears to be of some importance to clarify how one should understand the crossed product-generating constraints from the point of view of perturbation theory. This article is our attempt to provide such an explanation.  

Exact solutions in general relativity are hard to come by, so one typically expands around exact symmetric solutions, as these are analytically the most tractable. Ironically, it turns out that perturbation theory is not well-defined without further ado for exactly these physically relevant cases. This challenge is known as linearization (in)stabilities in general relativity \cite{fischer1980structure,Arms:1982ea,Marsden_lectures,Arms:1986vk,Deser:1973zza,Moncrief1,Moncrief:1976un,Arms1,Arms:1979au,fischer1980structure,Arms:1982ea,Saraykar:1981qm,Saraykar2,Deser:1973zzb,choquet1979maximal,Khavkine:2013iei,Altas:2019qcv} and provides the bulk of the clarification.

\begin{figure}
    \centering
    \begin{tikzpicture}
        \draw[yellow!80!black,very thick,fill=yellow!30] (0.5,0.5) ellipse (4 and 2);
        \begin{scope}
            \clip (1,0) .. controls (1.5,0.2) and (2,1) .. (2,1.25) -- (2,1.3) -- (2,1.25)
            .. controls (2,1) and (2.5,0.2) .. (3,0) -- (2,-1);
            \fill[inner color=yellow!80, outer color=yellow!30] (2,1.3) circle (1.5);
        \end{scope}
        \begin{scope}
            \clip (2,1.3) circle (1.5);
            \draw[very thick,yellow!60!black] (1,0) .. controls (1.5,0.2) and (2,1) .. (2,1.25) -- (2,1.3) -- (2,1.25)
            .. controls (2,1) and (2.5,0.2) .. (3,0);
        \end{scope}
        \draw[yellow!50!black,shift={(2,1.2)},rotate=-88,-latex] (0,0) -- (0.6,0);
        \draw[red!80!white,shift={(2,1.3)},rotate=-15,-latex] (0,0) -- (0.6,0);
        \fill[yellow!30!black] (2,1.3) circle (0.07) node[black,above] {\small$\lie_\xi g = \lie_\xi \phi = 0$};
        
        \draw[yellow!50!black,shift={(-1,0.8)},rotate=10,-latex] (0,0) -- (0.6,0);
        \draw[yellow!50!black,shift={(-1,0.8)},rotate=100,-latex] (0,0) -- (0.4,0);
        \fill[yellow!30!black] (-1,0.8) circle (0.07) node[black,left] {\small$(g,\phi)$};
        \node[yellow!50!black] at (0.5,-0.7) {\Large$\delta S = 0$};
    \end{tikzpicture}
    \caption{Depicted is the space of solutions to the Einstein equations on spatially closed spacetimes. Away from configurations with Killing symmetries, this space has the structure of a smooth (but infinite-dimensional) manifold. However, it is singular near configurations with Killing symmetries due to linearization instabilities. These give extra constraints on linearized perturbations, making certain perturbations spurious/unphysical (such as the vector shown in \textcolor{red!80!white}{red}), and hence reducing the dimension of the solution tangent space around such symmetric configurations. The overall space of solutions may thus be understood as stratified with cusp-like conical singularities at the symmetric configurations as shown.}
    \label{Figure: linearization instability singularities}
\end{figure}
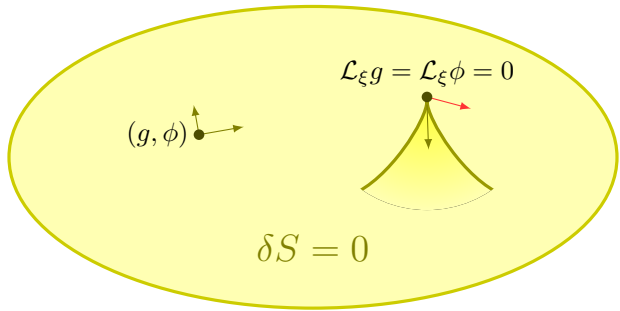

As we shall see, in spatially closed spacetimes with Killing horizons (such as the de Sitter case of \cite{Chandrasekaran:2022cip}), there exists an essentially unambiguous justification for (a) and (b). More precisely, in spatially closed spacetimes with Killing symmetries a linearized classical solution can be integrated to an exact one \emph{if and only if} all the second-order Killing symmetry gauge generators, called \emph{Taub charges} \cite{Taub1,taub2011variational,Taub2},  are imposed to vanish, such as in Eq.~\eqref{2clockconstraint}. If a linearized solution does not satisfy these constraints, it is spurious and thereby unphysical, rendering the perturbative expansion inconsistent. The results are conical singularities in the space of exact solutions to general relativity, see fig.~\ref{Figure: linearization instability singularities}: linearized solutions not obeying the Taub charge constraints correspond to vectors at a background solution that are not tangential to the space of solutions. Crucially, there is thus only a \emph{finite number} of second-order constraints that must be imposed to avoid inconsistencies, one for each independent Killing field, and one can safely ignore all other high-order diffeomorphism constraints as $G_N\to0$.

These are rigorous classical results. While we cannot make an equally strong statement in the quantum theory for lack of an exact non-perturbative theory to which we can relate the perturbative expansion, there are strong arguments for the necessity and sufficiency of imposing the corresponding quantum Killing generators also as constraints in the quantum theory to avoid inconsistent solutions \cite{Moncrief:1978te,Moncrief:1979bg,Higuchi:1991tk,Higuchi:1991tm,Losic:2006ht}. Notably, these arguments already led Moncrief to suggest in \cite{Moncrief:1978te,Moncrief:1979bg} that one should include internalized observers to address thermodynamical questions in de Sitter space. The gravitational crossed product algebras and ensuing entropy discussions can be viewed as a realization of this idea.

One might still wonder why it then suffices to consider only the boost Hamiltonian in Eq.~\eqref{2clockconstraint} to discuss intrinsically regulated regional entanglement entropies. The answer is that the presence of the clocks breaks the isometry group down to one of the form $\mathbb{R}\times G$, where $\mathbb{R}$ denotes the boost time translations and $G$ is some compact Lie group depending on the spacetime \cite{Chandrasekaran:2022cip,Chen:2024rpx}. One could include QRFs also for the extra $G$-factor \cite{delaHamette:2021oex} in order to ensure invariance under all Taub charge constraints. However, owing to the direct product structure and the compactness of $G$, this will not influence the von Neumann type conversion \cite{Fewster:2024pur,AliAhmad:2024eun}.

We will also see that the case of spacetimes with an asymptotic boundary or without Killing symmetries is different:  no instabilities arise \cite{Deser:1973zzb,choquet1979maximal,fischer1980structure,Arms:1982ea} and the justification for imposing the second-order constraint in Eq.~\eqref{2clockconstraint} on the linearized theory has to be sought elsewhere.\footnote{Here we are restricting the discussion to spatially connected spacetimes. In spacetimes with multiple connected components, there will be separate linearization instabilities in each spatially closed component with Killing symmetries.} In a partial response to (c), they are the bulk gauge constraint contribution to the Hamiltonian associated with the boost field but not needed for consistency of the linearized theory. Of course, there is nothing stopping one from imposing these constraints if one wishes to go to higher order. However, one must now be careful not to open Pandora's box (b): why Eq.~\eqref{2clockconstraint} but not any of the infinitely many other second-order constraints? As we will argue, it depends on whether or not one is interested also in a limited number of certain second-order degrees of freedom. But overall, this case  stands on arguably weaker grounds as far as the $G_N\to0$ limit is concerned.

To be sure, it was previously noted in \cite{Jensen:2023yxy,Kaplan:2024xyk} that the crossed product-generating boost constraints of CLPW \cite{Chandrasekaran:2022cip} can be understood in terms of linearization instabilities. The second-order nature was also explicitly emphasized on more general grounds in \cite{Jensen:2023yxy} and for black hole spacetimes in~\cite{Kudler-Flam:2023qfl} (though no connection with linearization instabilities was made). However, the matter as regards (a--c) has in our view not been sufficiently settled. Therefore, we revisit this discussion and provide a more explicit and comprehensive connection of gravitational crossed product algebras with the study of linearization (in)stabilities, in the hope to bring some clarity about the status of justification of the relevant constraints under various conditions. This will further lead us to revisit the case of general subregions in spacetimes without isometries \cite{Jensen:2023yxy} with perhaps a slightly different point of view.  Our discussion will also be more general, as it encompasses not only general relativity but arbitrary generally covariant theories, thus also arbitrary matter content, and backgrounds with or without matter. All of these lead to boost constraints of the form Eq.~\eqref{2clockconstraint}. 

Our work also offers novelties for the study of linearization (in)stabilities. While the bulk of the original literature is formulated in a canonical ADM language and focused on vacuum general relativity \cite{fischer1980structure,Arms:1982ea,Marsden_lectures,Deser:1973zza,Moncrief1,Moncrief:1976un,Arms:1982ea,Deser:1973zzb,choquet1979maximal}, or general relativity with specific matter content such as Maxwell \cite{Arms1}, scalar \cite{Saraykar:1981qm,Saraykar2}, or Yang-Mills fields \cite{Arms:1979au,Arms:1982ea,Altas:2021htf}, our main construction will instead be based on the modern language of the covariant phase space \cite{Kijowski:1973gi,Kijowski:1976ze} and cover arbitrary gauge-covariant theories with  generally covariant theories as a special case. Linearization instabilities arise generically in any gauge-covariant theory with non-linear equations of motion on spatially closed spacetimes. Besides illustrating our results in general relativity with various matter content, we will also exemplify them in pure Yang-Mills and Chern-Simons theories, showing moreover that no instabilities arise in their special Abelian incarnations: Maxwell and Abelian Chern-Simons theory, both of which have linear equations of motion. Covariant frameworks for exploring linearization instabilities in more generality have appeared before in \cite{Arms:1986vk} and \cite{Khavkine:2013iei}, however, neither invoking covariant phase space language. As far as we are aware, our work appears to be the first to explore the subject of linearization (in)stabilities systematically with the powerful covariant phase space tools. 

Instabilities in general gravity theories have begun to be explored in \cite{Altas:2017fcp,Altas:2018dci} and our results  provide a comprehensive framework for this. Crossed products also arise in lower-dimensional gravity models such as Jackiw-Teitelboim (JT) gravity \cite{Penington:2023dql,Kolchmeyer:2023gwa}. In line with our discussion, the existence of singularities in the space of solutions of JT gravity on circular Cauchy slices has recently been established in \cite{Alonso-Monsalve:2024oii} using the covariant phase space (but no connection with linearization instabilities was made).

One point to stress is that the price to pay for the generality of our results is a certain formality. This is the standard level of formality in general covariant phase space discussions and means that we will not delve into such details as proving (non-)existence results for solutions of certain partial differential equations arising from the field equations. This, however, would be necessary in order to rigorously establish the occurrence or absence of an instability. The prime reason for the limited scope of the original canonical literature on linearization (in)stabilities \cite{Moncrief1,Moncrief:1976un,fischer1980structure,Arms:1982ea,Arms1,Arms:1979au,Saraykar:1981qm,Saraykar2,Marsden_lectures,Deser:1973zzb,choquet1979maximal} is that these works proved such existence results, which depend on the details of the field equations at hand. These older results are thus stronger in the latter sense, while our results elucidate the universal structure behind linearization (in)stabilities, showing in particular how one can overcome theory-specific arguments in a unifying manner. 

The rest of this manuscript is organized as follows: we begin with a somewhat more traditional and explicit exposition of linearization (in)stabilities in perturbative general relativity with matter on a vacuum background in Sec.~\ref{sec:gravitonsandmatter}; this part does not invoke the covariant phase space. In Sec.~\ref{ssec_modflow}, we connect this discussion with modular crossed products. In Sec.~\ref{sec_matter}, we briefly explain what changes when matter is included in the background and which challenges arise to generalizing certain arguments to that case. In Sec.~\ref{sec: Constraints in the covariant phase space}, we  exhibit our general covariant phase space analysis of linearization (in)stabilities in arbitrary gauge-covariant theories. In this section, we also illustrate the results in general relativity  (minimally and non-minimally) coupled to various sources of matter, comment briefly on higher curvature theories and discuss connections with modular flow. The gauge theory examples are discussed in Sec.~\ref{sec: gauge theory examples}, and in Sec.~\ref{sec_generalregions} we comment on the case of general subregions in gravity, before concluding in Sec.~\ref{sec_conc}. For better legibility, we have moved some details into several appendices.

\section{Gravitons and matter on a matter-free background}
\label{sec:gravitonsandmatter}

We begin by exploring gravitons and matter on a vacuum background. To this end, we consider Einstein gravity (minimally or non-minimally) coupled to matter governed by the action 
\begin{equation}\label{fullaction}
    S[g,\phi]=\frac{2}{\kappa^2}\int_\mathcal{M}\epsilon(R-2\Lambda)+S_M[g,\phi]+S_\partial[q,\phi_\partial]\,,
\end{equation}
where $\kappa=\sqrt{32\pi G_N}$ and $G_N$ is Newton's constant. In our discussion, the Einstein-Hilbert term could, in fact, be replaced with a general gravity action (e.g., see \cite{Altas:2017fcp,Altas:2018dci}), but for definiteness we restrict to Einstein gravity and $D+1=4$ dimensions. Here, $S_M$ is the matter action (later to include the clock QRFs), where in standard covariant phase space convention $\phi$ is a collective label for all matter degrees of freedom and  $S_\partial$ is a suitable boundary term that is typically of order $O(\kappa^{-2})$ (e.g.\ the Gibbons-Hawking-York term), where $q$ is the induced boundary metric and $\phi_\partial$ the pullback of the matter fields to the boundary. For the moment, we will consider the matter non-perturbatively\footnote{By this we mean non-perturbative in the matter sector, in the sense that the matter equations of motion can be fully nonlinear on a fixed background. To the order at which we work they do not couple to the graviton.  If one were to go to higher order, the matter field would also receive corrections through their coupling to the graviton field.} and, as usual, of order $O(\kappa^0)$, so that $S_M$ is taken to be \emph{a priori} devoid of $\kappa$ and thus of order $O(\kappa^0)$ too.

Let us now investigate the expansion (going back to at least \cite{Taub1})
\begin{equation}
    g_{\mu \nu}=g_{\mu \nu}^0+\kappa h_{\mu \nu}+\frac{1}{2}\kappa^2 k_{\mu \nu}+O(\kappa^3)
\end{equation} 
of the metric around some exact vacuum background solution $g_{\mu \nu}^0$; $\kappa$ is our formal expansion parameter, while the symmetric tensor $h_{\mu \nu}$ denotes the graviton field and $k_{\mu \nu}$ is the symmetric second order metric perturbation. Formally, we can think of $h_{\mu \nu},k_{\mu \nu}$ as the first and second derivative along some curve $g_{\mu \nu}(\kappa)$ in field space (not yet necessarily tangential to the space of solutions), i.e.\ $h_{\mu \nu}=\frac{\dd{g_{\mu \nu}}}{\dd{\kappa}}\big|_{\kappa=0}$ and $k_{\mu \nu}=\frac{\dd^2{g_{\mu \nu}}}{\dd{\kappa}^2}\big|_{\kappa=0}$. Including $k_{\mu \nu}$ will become important later.

\subsection{Expanding the Einstein equations}

Next, we expand the Einstein equations up to second order around the background, as this will give rise to the constraints. We will not explicitly discuss the matter equations of motion. Writing $E_{\mu \nu}=\frac{4}{\kappa^2}G_{\mu \nu}-T_{\mu \nu}$, where $G_{\mu \nu}$ is the Einstein tensor (including any cosmological constant) and $T_{\mu \nu}$ the $O(\kappa^0)$ matter stress-energy tensor (though $E_{\mu \nu}$ could in fact also be the gravity equations of motion of a general gravity theory \cite{Altas:2017fcp,Altas:2018dci}), we can expand schematically
\begin{equation}
    E_{\mu \nu}[g;\phi]=\frac{1}{\kappa^2}E^{(0)}_{\mu \nu}[g_0]+\frac{1}{\kappa}E^{(1)}_{\mu \nu}[g_0;h]+E_{\mu \nu}^{(2)}[g_0;h;k;\phi]+O(\kappa)\approx 0\,,
\end{equation}
where $\approx$ denotes on-shell equality and 
\begin{equation}\label{eoms}
\begin{aligned}
    E_{\mu \nu}^{(0)}[g_0]&=4G_{\mu \nu}^{(0)}=4G_{\mu \nu}[g_0]\,,\\
    E_{\mu \nu}^{(1)}[g_0;h]&=4G^{(1)}_{\mu \nu}(h)\,,\\
    E_{\mu \nu}^{(2)}[g_0;h;k;\phi]&=2G^{(1)}_{\mu \nu}(k)-T^{(0)}_{\mu \nu}+2G_{\mu \nu}^{(2)}(h,h)\\
    &=2G^{(1)}_{\mu \nu}(k)-T_{\mu \nu}[g_0;\phi]+2G_{\mu \nu}^{(2)}(h,h)\,,
\end{aligned}
\end{equation}
and 
\begin{equation}\label{Einsteintensor}
\begin{aligned}
    G_{\mu \nu}[g]&=G_{\mu \nu}^{(0)}+\kappa G^{(1)}_{\mu \nu}(h)+\frac{1}{2}\kappa^2\left(G_{\mu \nu}^{(2)}(h,h)+G^{(1)}_{\mu \nu}(k)\right)+O(\kappa^3)\\
    T_{\mu \nu}[g;\phi]&=T^{(0)}_{\mu \nu}+\kappa T^{(1)}_{\mu \nu}(h)+O(\kappa^2)
\end{aligned}
\end{equation}
are the $\kappa$-expansions of the Einstein and stress-energy tensors. As regards the notation: when we write square brackets, $A[g;\phi;\ldots]$, we mean that $A$ is a functional of the fields $g,\phi,\ldots$; when we write round brackets, $A^{(i)}(h),A^{(i)}(h,h)$, we mean that $A^{(i)}$ is a multilinear object in its arguments $h_{\mu \nu}$; in this latter notation, we leave the functional dependence on background metric and matter (which here is $O(\kappa^0)$) implicit. For example, $G^{(2)}_{\mu \nu}(h,h)$ is quadratic in the gravitation field $h_{\mu \nu}$ and $G^{(2)}_{\mu \nu}$ should be read as a matrix operator on the space of  metric perturbation variables. It will become important that the vector operator $G^{(1)}_{\mu \nu}$ (see \cite[Sec.\ 1]{fischer1980structure} for an explicit expression) in the linearized $\kappa G^{(1)}_{\mu \nu}(h)$ and quadratic $\kappa^2 G^{(1)}_{\mu \nu}(k)$ expansion terms of the Einstein tensor are identical.

The equations of motion are then solved order by order, $E^{(i)}_{\mu \nu}=0$. For our discussion, it will become central whether or not, given $g^0,h$ solving the two lowest order equations, $E^{(2)}_{\mu \nu}=0$ and the matter equations of motion can be solved for $k_{\mu \nu}$ and $\phi$ (given possibly some suitable gauge fixing).

\subsection{Expanding the action}

Let us for the moment focus exclusively on the linearized theory. 
Its equations of motion, $E^{(1)}_{\mu \nu}=0$, are obtained by expanding the action in Eq.~\eqref{fullaction} to second order in $h_{\mu \nu}$ (and first order in $k_{\mu \nu}$) around the background solution and keeping only the lowest non-trivial order in $\kappa$. As $S_M[g,\phi]=O(\kappa^0)$, this means expanding $S[g^0+\kappa h+\kappa^2 k,\phi]$ up to order $O(\kappa^0)$, yielding (e.g.\ see \cite{Christensen:1979iy})\footnote{The $O(\kappa^{-1})$ term vanishes on account of the (vacuum with cosmological constant) background equations of motion $E^{(0)}[g_0]=0$ and we have dropped the constant $O(\kappa^{-2})$ background term. Furthermore, the term linear in the second order metric perturbation $k_{\mu \nu}$ also drops out on account of the background equations of motion.}
\begin{eqnarray}
    S[h,\phi]&=&2\int_\mathcal{M}\epsilon^0\left[\frac{1}{4}\tilde h^{\mu \nu}\left(-g^0_{\mu \rho}g^0_{\sigma \nu}(\nabla^0)^\alpha\nabla_\alpha^0+2R^0_{\mu \rho}g^0_{\sigma \nu}-2R^0_{\mu \rho \nu \sigma}\right)h^{\rho \sigma}-\frac{1}{2}(\nabla^0)^\rho\tilde h_{\rho \mu}(\nabla^0)^\sigma\tilde h_\sigma{}^\mu\right.\nonumber\\
    &&\left. -\tilde h^{\mu \rho}\left(R^0_{\rho \sigma}-\frac{1}{4}g^0_{\rho \sigma}R^0\right)h^\sigma{}_\mu-\frac{1}{2}\Lambda\tilde h^{\mu \nu}h_{\mu \nu}\right]+S_M[g^0,\phi]+\tilde S_\partial[q^0,\delta q,\phi_\partial]+O(\kappa)\,,\label{expaction}
\end{eqnarray}
where quantities equipped with a ${}^0$ superscript refer to the background (e.g., $\epsilon^0$ is the background volume form),
\begin{equation}
    \tilde h_{\mu \nu}=h_{\mu \nu}-\frac{1}{2}g^0_{\mu \nu}h
\end{equation}
is the usual trace-reversed graviton field, $h=(g^0)^{\mu \nu}h_{\mu \nu}$ is the trace of $h_{\mu \nu}$, and $\tilde S_\partial$ is the second order expansion around the $q^0$ background with $\delta q$ the boundary fluctuation of the metric; this term too is of order $O(\kappa^0)$ and depends on the boundary conditions. Altogether, this corresponds to a theory of uncoupled fields $h,\phi$ propagating on the background spacetime $(\mathcal M,g^0)$.

\subsection{Diffeomorphisms, isometries and constraints}

Under infinitesimal diffeomorphisms, generated by a vector field $\kappa\xi$, we have the field variations
\begin{equation}
    \delta_{\kappa\xi}g_{\mu \nu}=\kappa\mathcal L_\xi g^0_{\mu \nu}+\kappa^2\mathcal L_\xi h_{\mu \nu}+O(\kappa^3)\qquad \Rightarrow \qquad \delta_{\kappa\xi}h_{\mu \nu}=\nabla^0_\mu\xi_\nu+\nabla^0_\nu\xi_\mu+\kappa\mathcal L_\xi h_{\mu \nu}+O(\kappa^2)\,,\label{variation}
\end{equation}
where indices are raised and lowered with the background metric $g^0$ and
\begin{equation}\label{variation2}
\delta_{\kappa\xi}\phi=\kappa\mathcal{L}_\xi\phi\,.
\end{equation}
In other words, at the $\kappa$-order of the expanded action in Eq.~\eqref{expaction}, only the graviton field $h_{\mu \nu}$ transforms non-trivially under infinitesimal diffeomorphisms, $\delta_{\kappa\xi}h_{\mu \nu}=\nabla^0_\mu\xi_\nu+\nabla^0_\nu\xi_\mu+O(\kappa)$. It can be checked that, for compactly supported $\xi$ with vanishing support on the boundary $\partial\mathcal M$, the action Eq.~\eqref{expaction} is invariant and a covariant phase space analysis (e.g., see \cite{Hollands:2012sf,Donnelly:2016rvo} for the asymptotically flat case) reveals that such field variations amount to degenerate directions of the presymplectic form in field space, leading to the linearized gauge constraint currents
\begin{equation}\label{lincon}
    C_{\kappa\xi}=4G_{\mu \nu}^{(1)}(h)\,\xi^\mu(\epsilon^0)^\nu+O(\kappa)\,.
\end{equation}
Here, $(\epsilon^0)_\nu=\partial_\nu\sins\epsilon^0$ is the contraction of the background volume form with a coordinate basis vector. This expression agrees with the linear term of the full theory constraint associated with $\kappa\xi$ in a $\kappa$-expansion, see Sec.~\ref{sec: Constraints in the covariant phase space} and \cite{Jensen:2023yxy}. While matter is already diffeomorphism-invariant at leading order in $\kappa$, one can construct the physical graviton observables using standard methods, such as extracting the transverse-traceless modes.

The gauge constraints of the linearized theory thus depend linearly on the graviton field and not at all on the matter (incl.\ clock QRFs). As such, they cannot correspond to the desired constraints of the form Eq.~\eqref{2clockconstraint}.

In particular, note that if $\xi$ is a compactly supported Killing vector of the background metric $g^0$, then the constraint \begin{equation}\label{vanish}
    C[\kappa\xi]=\int_\Sigma C_{\kappa\xi}=0+O(\kappa)
\end{equation}
vanishes identically (i.e.\ even off-shell) at the order under consideration \cite{fischer1980structure}.  This is because $\xi$ (if compactly supported) is a gauge symmetry of the background spacetime and consistent with Eq.~\eqref{variation} which entails that in this case $\delta_{\kappa\xi}h_{\mu \nu}=0+O(\kappa)$. For compactly supported Killing fields, the first non-trivial constraints are guaranteed to arise at second order \cite{fischer1980structure,Arms:1982ea}; these can be viewed as the first non-trivial order of the constraint $C[\xi]$ as opposed to $C[\kappa\xi]$ and such rescaled constraints are legitimate for Killing fields given that they leave the background invariant. We paraphrase the proof of Eq.~\eqref{vanish}, which is crucial in the study of linearization instabilities, in Appendix~\ref{app_isolin}; it is instructive for understanding why this method of proof is difficult to extend to non-trivial matter backgrounds and generalized gravity theories. As we will see in Sec.~\ref{sec: Constraints in the covariant phase space}, these difficulties can be easily overcome with a general covariant phase space analysis.

Compatible with this result, it is shown in \cite{Abbott:1981ff,Altas:2019qcv} that for (not necessarily compactly supported) Killing fields, the integrand $C_{\kappa\xi}$ is essentially a total divergence, on-shell of the background equations of motion $E^{(0)}[g_0]=0$ (but off-shell of $E^{(1)}[g_0;h]=0$): 
\begin{equation}\label{totald}
    \xi^\mu G^{(1)}_{\mu \nu}(h)=(\nabla^{(0)})^\mu F_{\mu \nu}(h)\,,
\end{equation}
where 
\begin{equation}\label{Feq}
    F_{\mu \nu}(h)=\xi^\rho(\nabla^{(0)})^\sigma K_{\nu \mu \rho \sigma}(h)-K_{\nu \sigma \rho \mu}(h)(\nabla^{(0)})^\sigma\xi^\rho
\end{equation}
with
\begin{equation}
    K_{\nu \mu \rho \sigma}=\frac{1}{2}\left(g^0_{\mu \rho}\tilde h_{\nu \sigma}+g^0_{\nu \sigma}\tilde{h}_{\mu \rho}-g^0_{\mu \sigma}\tilde{h}_{\nu \rho}-g^0_{\nu \rho}\tilde{h}_{\mu \sigma}\right)\,,
\end{equation}
so that, crucially, $F_{\mu \nu}=-F_{\nu \mu}$ is antisymmetric. This means that\footnote{For the pullback $(\epsilon^0)_\nu(\nabla^{(0)})_\mu F^{\mu \nu}\big|_\Sigma$ we can always choose the coordinate system such that $\partial_0$ is normal and $\partial_i$ tangential to $\Sigma$. Then, by antisymmetry, $(\epsilon^0)_\nu(\nabla^{(0)})_\mu F^{\mu \nu}\big|_\Sigma=(\epsilon^0)_0(\nabla^{(0)})_iF^{i0}\big|_\Sigma$ is a total spatial derivative associated with the induced metric, so that Stokes' theorem applies. The integral is coordinate independent, however.}
\begin{equation}\label{boundaryterm}
C[\kappa\xi]=\int_\Sigma C_{\kappa\xi}=4\int_\Sigma(\epsilon^0)_\nu(\nabla^{(0)})_\mu F^{\mu \nu}+O(\kappa)=4\int_{\partial\Sigma} (\epsilon^0)^{\mu \nu}F_{\mu \nu}(h)+O(\kappa)
\end{equation}
is a pure boundary term at the order under consideration, where $(\epsilon^0)_{\mu \nu}=\partial_\mu\sins\partial_\nu\sins \epsilon^0$. The leading boundary term is the first order contribution to the usual ADM charges.  For example, for a flat background $g^0=\eta$ with $\Lambda=0$ and flat coordinates, one obtains the ADM energy for a timelike Killing field normal to $\Sigma$,
\begin{equation}
\int_{\partial\Sigma} (\epsilon^0)^{\mu \nu}F_{\mu \nu}(h)=\int_{\partial\Sigma}\epsilon_{\partial\Sigma}n_i(\partial_j h_{ij}-\partial_i h)\,,
\end{equation}
where $n_i$ is the unit normal to $\partial\Sigma$ \cite{Abbott:1981ff}. Similarly, one obtains the other Poincar\'e generators and in (A)dS backgrounds the corresponding Killing symmetry generators from Eq.~\eqref{boundaryterm}.\footnote{In de Sitter one may choose a flat slicing and integrate over a partial Cauchy slice inside the cosmological horizon \cite{Abbott:1981ff}.}

In particular, if the spacetime is spatially  closed this constraint $C[\kappa\xi]$ vanishes identically at this order, Eq.~\eqref{vanish}. It is perhaps worthwhile to stress that this result is distinct from the statement that the Hamiltonian in generally covariant theories is \emph{on-shell} a pure boundary term that does not vanish. Indeed,  if $\xi$ is timelike, Eq.~\eqref{boundaryterm}, which is an identity \emph{off-shell} of the (linearized) equations of motion, says that the corresponding Hamiltonian \emph{constraint} is a pure boundary term that thus has to vanish on-shell. Where terms of the form Eq.~\eqref{totald} appear within higher-order constraints (evaluated on higher-order metric perturbations, e.g.\ as in Eq.~\eqref{2ndorder}) they correspond to order-by-order contributions to the ADM charge(s). The relation of the two statements will become clearer in our covariant phase space discussion in Sec.~\ref{sec: Constraints in the covariant phase space}.

The observation in Eq.~\eqref{totald} is gleaned from an explicit expansion of the Einstein tensor, rendering this line of argument tedious to generalize to include matter backgrounds or more general gravity theories. Instead, we will directly extend the identity \eqref{boundaryterm} to arbitrary generally covariant theories (and matter gauge theories) in Sec.~\ref{sec: Constraints in the covariant phase space} using simple arguments invoking the powerful covariant phase space language.

This result already suggests that the static patch boost constraint of CLPW for de Sitter must be a second order constraint, in line with the fact that Hamiltonians are usually at least quadratic in the variables they describe. However, how does one justify this constraint on the linear graviton theory, given that such higher order gauge constraints do \emph{not} arise from the quadratic (in $h_{\mu \nu}$) action \eqref{expaction} and to obtain them one would have to expand the action to higher order in $\kappa$? In particular, imposing them at the order at which we are working would be tantamount to considering the Killing vector field $\xi$ (as opposed to its infinitesimal version $\kappa\xi$ above). By Eqs.\ \eqref{variation} and~\eqref{variation2} these result in the field variations
$\delta_\xi h_{\mu \nu}=\mathcal L_\xi h_{\mu \nu}$ and $\delta_\xi\phi=\mathcal L_\xi \phi$, which typically leave the action \eqref{expaction} invariant, but do not correspond to degenerate directions of the presymplectic form. They would thus rather correspond to \emph{physical} transformations 
if we took the theory defined by \eqref{expaction} at face value. For example, suppose $(\mathcal{M},g^0=\eta)$ is Minkowski space, so that the action in Eq.~\eqref{expaction} describes fields propagating on a standard Minkowski background for which Poincar\'e symmetries are not gauge symmetries. While these are not compactly supported, the same conclusion applies to a  field theory of a non-interacting $h$ and $\phi$ on a de Sitter background, where isometries are compactly supported.

\subsection{Linearization (in)stabilities}

The justification for imposing isometry constraints on the linearized theory comes from considering the conditions under which its solutions approximate those of the untruncated theory.
Since the integral $C[\kappa\xi]$ is a boundary term even off-shell of $E^{(1)}_{\mu \nu}=0$ for Killing fields, this will also be true if we replace $h_{\mu \nu}$ in Eq.~\eqref{boundaryterm} with the second order variable $k_{\mu \nu}$, which too is symmetric. Regarding our hierarchy of equations of motion in Eq.~\eqref{eoms}, this entails for 
Killing fields
\begin{equation}\label{2ndorder}
E^{(2)}_{\mu \nu}\approx0\qquad\Rightarrow\qquad \int_\Sigma\left(-2G^{(2)}_{\mu \nu}(h,h)+T_{\mu \nu}^{(0)}\right)\xi^\mu(\epsilon^0)^\nu\approx 2\int_{\partial\Sigma}(\epsilon^0)^{\mu \nu}F_{\mu \nu}(k)+O(\kappa)\,.
\end{equation}
The bulk integrals on the left-hand side are called \emph{Taub charges} \cite{taub2011variational,Taub1,Taub2} and there is one for each independent Killing field. 

On-shell of the equations of motion following from Eq.~\eqref{expaction} (i.e.\ the linearized gravity equations of motion $E^{(1)}_{\mu \nu}\approx0$ and the matter equations of motion), the charge current is divergence-free
\begin{equation}\label{divfreecur}
(\nabla^{(0)})^\mu\left(2G^{(2)}_{\mu \nu}(h,h)-T_{\mu \nu}^{(0)}\right)\xi^\nu=0\,.
\end{equation}
This is clear for the matter stress-energy, and for the second order Einstein tensor this follows from an expansion of the Einstein equations \cite{Abbott:1981ff,Wald:1984rg}. Regarding gauge invariance of the Taub charges under linearized diffeomorphisms, it is clear from Eq.~\eqref{variation2} that the matter contribution is invariant. For the graviton contribution, this is more subtle because $G^{(2)}_{\mu \nu}(h,h)$, unlike its linear order (on-shell) counterpart $G^{(1)}_{\mu \nu}(h)$, is not invariant \cite{Wald:1984rg}. However, it can be shown that its contraction with a Killing field is invariant up to a boundary term \cite[App.~B]{Altas:2019qcv}:
\begin{equation}
\xi^\mu\delta_{\kappa\zeta}G^{(2)}_{\mu \nu}(h,h)=-(\nabla^{(0)})^\mu F_{\mu \nu}(\delta_{\kappa\zeta} h)\,,
\end{equation}
with $F_{\mu \nu}$ given in Eq.~\eqref{Feq} and $\zeta$ an arbitrary spacetime vector field. Thus, the Taub charges are gauge-invariant, i.e.\ invariant under linearized small diffeomorphisms generated by compactly supported $\zeta$. They are, however, not necessarily invariant under large diffeomorphisms.

The second order on-shell equality Eq.~\eqref{2ndorder} implies distinct repercussions regarding conservation and stability for the cases when there is or is not a boundary of spacetime.

\subsubsection{Instabilities in spatially closed spacetimes with isometries}\label{sssec_instacompact}

In the study of linearization stability, one considers the space of solutions to the Einstein equations with some matter content and aims to characterize the conditions under which a perturbative solution at linear order around an exact background solution can be integrated into another nearby exact solution. More precisely, considering the perturbative solution as a vector at the background solution, one examines the precise necessary and sufficient conditions for the linearized solution to be tangential to some curve of exact solutions \cite{Deser:1973zza,Moncrief1,Moncrief:1976un,Moncrief:1978te,Moncrief:1979bg,Arms1,Arms:1979au,fischer1980structure,Arms:1982ea,Marsden_lectures,Saraykar:1981qm,Higuchi:1991tk,Higuchi:1991tm,Losic:2006ht}; such perturbative solutions are called `integrable'. This provides a precise definition for what it means for a linearized solution to approximate an exact one. A background solution  is called linearization stable if \emph{every} linearized solution around it is integrable.

Let us first consider the case of spatially closed spacetimes, i.e.\ ones with compact Cauchy slices without boundary, before discussing spacetimes with asymptotic structures in the next subsection. In that case, it has been shown that an exact solution $g_0$ in vacuum is linearization stable if and only if $g_0$ admits no Killing symmetries \cite{Moncrief1,Moncrief:1976un,fischer1980structure,Arms:1982ea,Marsden_lectures}. The space of solutions of the Einstein equations has a smooth manifold-like structure in a neighborhood of such $g_0$ \cite{fischer1980structure,Arms:1982ea,Marsden_lectures}. When $g_0$ does admit isometries, $g_0$ is linearization unstable and a linearized solution $h$ is tangent to the space of solutions if and only if all its corresponding Taub charges vanish \cite{fischer1980structure,Arms:1982ea,Marsden_lectures,Arms:1986vk}. Such isometric solutions correspond to conical singularities in the space of solutions, see Figure~\ref{Figure: linearization instability singularities} (the term conical refers to the fact that the equations defining these singularities are quadratic). Similar results hold for spacetimes with reasonable matter content, such as Maxwell theory \cite{Arms1}, Yang-Mills theory \cite{Arms:1979au,Arms:1982ea,Altas:2021htf} and some scalar field theories \cite{Saraykar:1981qm,Saraykar2}. These latter results have been established for graviton and matter perturbations around a background that includes matter, unlike in the present section, and we will discuss this case briefly in Sec.~\ref{sec_matter}, presenting some challenges, and then separately in more detail in Sec.~\ref{sec: Constraints in the covariant phase space} at a general covariant phase space level where these challenges are overcome.

Let us provide some intuition for our case of a vacuum background and non-perturbative matter. In the absence of a boundary, Eq.~\eqref{2ndorder} tells us that the Taub charges must vanish on-shell of $E^{(2)}_{\mu \nu}=0$:
\begin{equation}\label{2ndorder3}
\int_\Sigma\left(-2G^{(2)}_{\mu \nu}(h,h)+T_{\mu \nu}^{(0)}\right)\xi^\mu(\epsilon^0)^\nu\approx0+O(\kappa)\,.
\end{equation}
Importantly, the second order metric perturbation $k_{\mu \nu}$ has dropped out. Eq.~\eqref{divfreecur} implies that the Taub charges are conserved for closed Cauchy slices, but they do not vanish automatically \cite{Arms:1986vk,Moncrief1,Moncrief:1976un,Moncrief:1978te,Moncrief:1979bg,fischer1980structure,Arms:1982ea,Marsden_lectures,Altas:2018dci,Altas:2019qcv}. For a timelike Killing field $\xi$, the corresponding Taub charge records the energy of the graviton and matter fields as measured by an observer with worldline tangential to $\xi$; indeed, $-G_{\mu \nu}^{(2)}$ is usually defined to be the graviton (pseudo) stress-energy tensor \cite{Wald:1984rg}. The modular Hamiltonian in the crossed product constructions will be one such charge. We will come back to this in Sec.~\ref{ssec_modflow}. 

The Taub charges are related to the second order part and thus first non-vanishing contribution of the diffeomorphism constraints associated with isometries as follows
\begin{equation}\label{2ndorderconstraint}
    C^{(2)}[\kappa\xi]=\int_\Sigma\left(2G^{(2)}_{\mu \nu}(h,h)-T_{\mu \nu}^{(0)}+2hG^{(1)}_{\mu \nu}(h)\right)\xi^\mu(\epsilon^0)^\nu\,,
\end{equation}
where the contribution $2hG^{(1)}_{\mu \nu}(h)$ with $h=h_{\mu \nu}(g^0)^{\mu \nu}$ comes from the variation of the volume form. In fact, this form of the constraint assumes scalar matter; for tensorial matter (including gauge fields) there are additional contributions \cite{Seifert:2006kv}. We will come back to this more systematically using the covariant phase space language in Sec.~\ref{sec: Constraints in the covariant phase space} and this will also generalize the Taub charges. For the moment, it suffices to note that, on-shell of the linearized Einstein equations $E^{(1)}_{\mu \nu}=4G^{(1)}_{\mu \nu}(h)=0$, the Taub charges coincide with (minus) the second order diffeomorphism constraints associated with Killing fields.

Thus, the Taub charges constitute non-trivial second order constraints on $h_{\mu \nu}$ and $\phi$ because $T^{(0)}_{\mu \nu}=T_{\mu \nu}[g_0;\phi]$. However, $h_{\mu \nu}$ is supposed to be the output of the linearized equations $E^{(1)}_{\mu \nu}\approx0$ and $\phi$ is determined by solving the matter equations of motion of the linearized theory in Eq.~\eqref{expaction}, both of which are oblivious to Eq.~\eqref{2ndorder3}. This is the instantiation of a linearization instability: if the linearized solution $h_{\mu \nu}$, together with $\phi$, does not already solve Eq.~\eqref{2ndorder3}, it cannot be extended to higher orders, rendering it inconsistent with exact solutions of the full theory. Hence, in order to avoid spurious solutions at linear order, one has no option but to impose the conditions in Eq.~\eqref{2ndorder3} for each linearly independent  Killing field $\xi$ already at that order. These second order \emph{linearization stability conditions} are thus \emph{necessary} at linear order. 

Are there additional second-order constraints ensuring stability -- and what if we go to even higher orders? Will further non-trivial stability conditions arise? The (in a non-linear theory) surprising feature is that, for example in vacuum general relativity, as well as general relativity coupled to electromagnetic or Yang-Mills fields, it can be shown that the second order stability conditions are, in fact, also \emph{sufficient} \cite{fischer1980structure,Arms:1982ea,Marsden_lectures}, and crucially there is thus just a finite number of them. 

To gain some intuition for the sufficiency of the second order stability constraints, we paraphrase  arguments from \cite{Arms:1986vk,Taub2}. The precise argument depends on the detailed equations of motion. Note that the second order equations of motion $E^{(2)}_{\mu \nu}\approx0$ in Eq.~\eqref{eoms} are \emph{linear} in the second order metric perturbation $k_{\mu \nu}$. When the operator $G_{\mu \nu}^{(1)}$ -- viewed as a linear map from the vector space $\mathcal{S}_2$ of symmetric two-tensor fields to itself -- is surjective, $E^{(2)}_{\mu \nu}\approx0$ can be solved for $k_{\mu \nu}$ and so $g^0$ is linearization stable. When $G_{\mu \nu}^{(1)}$ fails to be surjective, it may happen that for fixed $h$, $2G_{\mu \nu}^{(2)}(h,h)-T_{\mu \nu}^{(0)}$ in Eq.~\eqref{eoms} lies outside the image of $G_{\mu \nu}^{(1)}$. $E^{(2)}_{\mu \nu}\approx0$ can then only be solved for $k_{\mu \nu}$ provided
\begin{equation}\label{Taub}
    P\left(2G_{\mu \nu}^{(2)}(h,h)-T_{\mu \nu}^{(0)}\right)=0\,,
\end{equation}
where $P$ is the projector onto the complement of the image of $G_{\mu \nu}^{(1)}$ in $\mathcal{S}_2$. For vacuum general relativity and general relativity coupled to Yang-Mills fields it turns out that $P$ is precisely given by the Taub charges in Eq.~\eqref{2ndorder3} and thus finite-dimensional.\footnote{More precisely, $P$ is the projector onto the subspace of $\mathcal{S}_2$ spanned by $\xi_{(\mu} n_{\nu)}$, with $\xi$ varying over the Killing fields and $n_\mu$ the unit normal to $\Sigma$, and orthogonality is defined by the $L^2$-inner product constituted by the integral in Eq.~\eqref{2ndorder3}. See \cite{arms1981symmetry,Gotay:1997eg} for arguments why such a projector is quite generally given by $L^2$-inner products with background symmetry generators in non-linear field theories with gauge symmetries.} 

To address higher orders, we need to expand also the matter fields perturbatively. For only the following paragraph, we use the notation 
\begin{equation}\label{expansion2}
    g_{\mu \nu}=g_0+\sum_{n\geq1}\frac{\kappa^n}{n!} h^{(n)},\qquad\qquad \phi=\sum_{n\geq0}\frac{\kappa^n}{n!}\varphi^{(n)}\,,
\end{equation}
where $\varphi^{(0)}$ is a solution to the matter equations of motion following from Eq.~\eqref{expaction}. It will get corrected at higher orders where matter will start interacting with the metric perturbations; $\varphi^{(n)}$ shall denote a solution to the matter equations of motion obtained by expanding the matter action $S_M[g,\phi]+S_\partial[q,\phi_\partial]$ in Eq.~\eqref{fullaction} to order $n$ in $\kappa$. Note that $\varphi^{(n)}$ depends only on the metric perturbations up to order $n$ and matter perturbations up to order $n-1$.

Now, by power counting in $\kappa$, the $n$-th order Einstein equations for $n\ge2$ are of the schematic form
\begin{align}
    E^{(n)}_{\mu \nu}[g_0;\ldots;h^{(n)}; \varphi^{(0)};\ldots;\varphi^{(n-2)}] =&\frac{4}{n!}G^{(1)}_{\mu \nu}(h^{(n)})-T_{\mu \nu}^{(n-2)}[g_0;h^{(1)};\ldots; h^{(n-2)};\varphi^{(0)};\ldots;\varphi^{(n-2)}]\nonumber\\
&+\tilde{G}_{\mu \nu}^{(n)}[g_0;h^{(1)};\ldots;h^{(n-1)}]\approx0\,.\nonumber
\end{align}
It is clear that this has a solution for $h^{(n)}$, provided
\begin{equation}
    P\left(\tilde{G}_{\mu \nu}^{(n)}[g_0;h^{(1)};\ldots;h^{(n-1)}]-T_{\mu \nu}^{(n-2)}[g_0;h^{(1)};\ldots; h^{(n-2)};\varphi^{(0)};\ldots;\varphi^{(n-2)}]\right)=0\,.
\end{equation}
The linearized solution $h=h^{(1)}$ is integrable to an exact one (assuming analyticity and convergence) if these stability conditions are obeyed for all $n$. Note that $\tilde{G}^{(n)}_{\mu \nu}$ is \emph{linear} in $h^{(n-1)}$ for $n>2$. The remarkable fact for vacuum general relativity and general relativity coupled to Yang-Mills fields is that all of these conditions for $n>2$ can be solved for the higher-order metric perturbations $h^{(2)},h^{(3)},\ldots$ once $E^{(1)}_{\mu \nu}\approx0$ and Eq.~\eqref{Taub} is solved for $h=h^{(1)}$ \cite{fischer1980structure,Arms:1982ea,Marsden_lectures}.

\subsubsection{(In)stabilities in spacetimes with boundary}

In spacetimes with an asymptotic boundary, such as asymptotically flat or anti-de Sitter spacetimes, it depends on the boundary conditions whether or not instabilities arise at background solutions with isometries, and whether or not the Taub charges are conserved. If we permit an asymptotic fall-off of the second-order metric perturbations $k_{\mu \nu}=O(1/r^x)$ with $x>d-3$, then the boundary term in Eq.~\eqref{2ndorder} vanishes again (for $r\to\infty$) and we recover Eq.~\eqref{2ndorder3} (the volume form comes with a $r^{d-2}$ factor). Linearization instabilities will then arise in the same way as in the case of spatially closed spacetimes with Eq.~\eqref{2ndorder3} once again constituting necessary\footnote{They are also likely sufficient in that case, though we are not aware of a proof.} stability conditions.\footnote{The conditions discussed here do not result in instabilities when the Taub charges vanish identically. For example, for a Minkowski background $g^0=\eta$, requiring a fall-off of the gravitons as $h_{\mu \nu}=O(1/r^x)$, $x>d-3$, yields a vanishing first-order ADM mass and the only solution for $h_{\mu \nu}$ compatible with that is the zero one \cite{Brill:1968ca,Deser:1968zzf}. The Taub charges are then trivially zero and no linearization instability occurs.} Singularities of the conical type will then again arise in the space of solutions. 

When $h_{\mu \nu}=O(1/r^y)$, we have $G^{(2)}_{\mu \nu}(h,h)=O(1/r^{2(y+1)})$ (e.g., see \cite{Altas:2019qcv} for an explicit expression), and so Eq.~\eqref{divfreecur} tells us that the Taub charges are conserved provided $y>(d-4)/2$ and also the matter drops off sufficiently quickly such that $T^{(0)}_{\mu \nu}=O(1/r^z)$ with $z>d-2$. For typical matter, $T_{\mu \nu}$ is at least bilinear in $\phi$, so that conservation is guaranteed for $\phi=O(1/r^{z/2})$. A counter-example is slow roll inflation, which yields a stress-energy tensor linear in $\phi$ and conservation thus requires a faster drop-off. Without conservation one would have to impose vanishing Taub charge constraints on each Cauchy slice separately in order to avoid instabilities.

However, typically, one considers a drop-off of the metric, as well as its perturbations, of order $O(1/r^{d-3})$, in which case the ADM boundary charges in Eq.~\eqref{2ndorder} remain finite. The graviton contribution of the Taub charges remains conserved and for the matter contribution the same argument as in the previous paragraph applies.
That equation then does not constitute a constraint on the first order variables; rather, it can be solved for the second order variables $k_{\mu \nu}$ and manifests a second order correction to the charges, e.g.\ to energy, momentum and angular momentum in asymptotically flat spacetimes \cite[Sec.~6]{Moncrief:1976un}. For such boundary conditions, no linearization instabilities arise. Indeed, this has been established for flat space in \cite{Deser:1973zzb} and more generally for asymptotically flat spacetimes in \cite{choquet1979maximal}, so that no singularities arise in the space of solutions, and it assumes the structure of a manifold also around symmetric background solutions. 

We note that this analysis can also be carried out for partial Cauchy slices. For example, $\Sigma$ may be the partial slice between the asymptotics and the event horizon in a black hole spacetime. This was already suggested by Moncrief \cite[Sec.~6]{Moncrief:1976un} and this is what has been done in the construction of modular crossed products for black holes in \cite{Chandrasekaran:2022cip,Kudler-Flam:2023qfl,Faulkner:2024gst,Chen:2024rpx}, a subject we shall return to in Sec.~\ref{subsec: diffeomorphism covariant Lagrangians}. Similarly, $\Sigma$ could be the partial Cauchy slice inside the cosmological horizon of some observer in de Sitter space  \cite{Abbott:1981ff}.

\subsection{Taub charges as generators of modular flow and crossed products}\label{ssec_modflow}

Let us now connect the discussion of linearization (in)stabilities with the recent wave of works on constructing subregion gravitational algebras in perturbative quantum gravity using so-called modular crossed products \cite{Chandrasekaran:2022cip,Jensen:2023yxy,Witten:2023qsv,Witten:2023xze,Witten:2021unn,Kudler-Flam:2023hkl,Kudler-Flam:2023qfl,Kudler-Flam:2024psh,DeVuyst:2024pop,Faulkner:2024gst,Aguilar-Gutierrez:2023odp,Chen:2024rpx,Gomez:2023upk,Gomez:2023wrq,Fewster:2024pur,Kolchmeyer:2024fly}. It was already briefly noted in \cite{Jensen:2023yxy,Kaplan:2024xyk} that the generating constraints of the modular crossed product in gravity can be understood in terms of linearization instabilities. Our discussion here will be more comprehensive and make this relation explicit.

Modular crossed products have garnered significant interest because they come with an intrinsic regularization of regional entanglement entropies, leading to UV-finite entropies without the need of the introduction of an explicit UV cut-off. They resulted in new proofs \cite{Faulkner:2024gst} and more general forms \cite{KirklinGSL} of the generalized second law and in the observation that gravitational entropy is observer-dependent \cite{DeVuyst:2024pop,DVEHK}. Technically, this is rooted in a transition of the type of von Neumann algebra from the standard type $\rm{III}_1$ associated with regional algebras in quantum field theory \cite{Fredenhagen:1984dc,Buchholz:1986bg,Buchholz:1995gr,haag2012local,Yngvason_2005} to a type $\rm{II}_\infty$ or $\rm{II}_1$. This is significant because the former type $\rm{III}_1$ possesses no trace and thus no density operators or well-defined entropies, while the latter type $\rm{II}$ does posses all of those (in the case of type $\rm{II}$ algebras, traces are defined up to a multiplicative constant which results in state-independent shift ambiguity for entropies. See footnote \ref{footnote: renormalized entropy} and e.g.\ \cite{WittenRevModPhys.90.045003,Sorce:2023fdx,Takesaki1979,Witten:2021unn} for an introduction). 

The aim of this subsection is to explain how to understand these crossed product algebras from the point of view of linearization (in)stabilities and that the modular flow generator is a Taub charge as introduced above. Here, we shall restrict ourselves to spacetimes with a background possessing Killing symmetries. In Sec.~\ref{sec_generalregions}, we shall comment on proposals for modular crossed products associated with general subregions on backgrounds that do not necessarily admit isometries \cite{Jensen:2023yxy,Chen:2024rpx}.

Within this background spacetime, we consider some subregion $\mathcal U$, taken to be the causal development of some partial Cauchy slice, as well as its causal complement $\mathcal U'$. For example, $\mathcal{U}$ and $\mathcal{U}'$ could be two complementary static patches in de Sitter space associated with two complementary observers traveling along geodesic worldlines coincident with flow lines of a time like Killing field \cite{Chandrasekaran:2022cip,Kolchmeyer:2024fly,Chen:2024rpx,Kudler-Flam:2023hkl}. Similarly, $\mathcal{U},\mathcal{U}'$ could be regions split by a bifurcate Killing horizon in black hole spacetimes. For instance, maximally extended asymptotically flat Schwarzschild spacetimes were considered in \cite{Chandrasekaran:2022cip,Kudler-Flam:2023qfl,Chen:2024rpx}, Schwarzschild-de Sitter black holes in \cite{Kudler-Flam:2023qfl,Chen:2024rpx}, Schwarzschild- and Kerr-anti de Sitter black holes in \cite{Kudler-Flam:2023qfl}, and general Killing horizons in \cite{Kudler-Flam:2023qfl,Faulkner:2024gst,KirklinGSL}.

We now consider a quantum field theory including gravitons $h_{\mu \nu}$ and possibly several other species of fields (but not the second order metric perturbations $k_{\mu \nu}$) in the global spacetime, whose Hilbert space we denote by $\mathcal H_{\rm QFT}$. Working in the $\kappa\to0$ limit, the setting in the above works thus corresponds to the quantum version of the theory defined by the expanded action in Eq.~\eqref{expaction}, i.e.\ the linearized theory on the gravity side. In particular, we are interested in invariance under linearized diffeomorphisms as in Eqs.~\eqref{variation} and~\eqref{variation2}. Thus, the algebra $\mathcal{A}_\mathcal{U}\subset\mathcal{B}(\mathcal{H}_{\rm QFT})$ associated with subregion $\mathcal{U}$ is assumed to be comprised of all bounded QFT observables with exclusive support in $\mathcal{U}$ that are invariant under all linearized diffeomorphisms, as well as any additional matter gauge symmetries. In the same vein, also the Hilbert space $\mathcal{H}_{\rm QFT}$ will henceforth be assumed to solve all the linearized diffeomorphism constraints, as well as matter gauge symmetry constraints. As linearized diffeomorphisms act exclusively on the graviton field (cf.~Eq.~\eqref{variation2}), this can be achieved by standard techniques such as extracting the transverse-traceless components; furthermore, the commutator of linearized diffeomorphism constraints is of order $O(\kappa^2)$ and so they obey an Abelian algebra at first order, greatly simplifying their imposition on states also. Somewhat more precisely, $\mathcal{A}_{\mathcal{U}}$ is assumed to be a Type $\rm{III}_1$ von Neumann factor,\footnote{A factor is a subalgebra with trivial center, i.e.\ only multiples of the identity commute with all of $\mathcal{A}_\mathcal{U}$.} as generally expected for asymptotically scale invariant field theories \cite{Fredenhagen:1984dc,Buchholz:1986bg,Buchholz:1995gr,haag2012local,Yngvason_2005} and thus in particular closed under bi-commutation. Similarly, its commutant is a Type $\rm{III}_1$ factor and, assuming Haag duality \cite{haag2012local,WittenRevModPhys.90.045003}, we identify it with $\mathcal{A}_{\mathcal{U'}}$, i.e.\ the observable algebra associated with the causal complement $\mathcal U'$.

The type transition of the regional algebras now comes about by considering modular flow within them. We shall not review modular theory here (see \cite{haag2012local,WittenRevModPhys.90.045003} for an introduction). For our purposes it suffices to mention that each such modular flow is associated with a global KMS (cyclic and separating) state and constitutes an outer automorphism of both $\mathcal{A}_{\mathcal{U}}$ and $\mathcal{A}_{\mathcal{U}'}$. Outer means that its generator, the modular Hamiltonian $H$, is not contained in either of these two algebras and this is a characterizing property of type $\rm{III}_1$ algebras. The crux is that the boost Hamiltonian for each Killing horizon with boost Killing vector $\xi$,
\begin{equation}\label{boost}
    H_\xi = \int_\Sigma(\epsilon^0)^\mu\xi^\nu T_{\mu \nu}\,,
\end{equation}
where $T_{\mu \nu}$ is the stress-energy tensor for all involved fields (including the pseudo-tensor $-G^{(2)}_{\mu\nu}$ for the gravitons), is a modular Hamiltonian  for the regions separated by the bifurcation surface and associated with a KMS state \cite{Bisognano:1975ih,Sewell:1982zz,sorce2024analyticity}; indeed, the boost symmetry leaves $\mathcal{U}$ and $\mathcal{U}'$ invariant. (More generally, it is an open problem when modular flow has a spacetime geometric interpretation \cite{sorce2024analyticity}.)

Now from Eq.~\eqref{boundaryterm} it is clear that the first order of $H_\xi$ is a pure boundary term, namely the first order ADM mass and this only depends on gravitons. In particular, the first order of $H_\xi$ vanishes identically for spatially closed spacetimes, such as de Sitter and can thus not be the modular Hamiltonian.
Instead, it is clear that the modular Hamiltonian $H_\xi$ is given by the \emph{second-order} contribution constituted by the corresponding \emph{Taub charge} in Eq.~\eqref{2ndorder}, which does not vanish automatically and, as noted below Eq.~\eqref{2ndorder3}, contains the stress-energy tensor of \emph{all} involved matter and graviton fields. This object does measure the energy of all fields along the Killing flow lines.

\subsubsection{Spatially closed case}

\begin{figure}
    \centering
    \begin{tikzpicture}[scale=1.16]
        \begin{scope}[red!40!white]
            \foreach \i in {0.3,1,1.85,2.6} {
                \draw[postaction={decorate,decoration={markings,mark=at position .57 with {\arrow[scale=1.1]{stealth}}}}] (2,2) .. controls ({2+0.8*\i},{2-0.8*\i}) and ({2+0.8*\i},{0.8*\i-2}) .. (2,-2);

                \draw[postaction={decorate,decoration={markings,mark=at position .57 with {\arrow[scale=1.1]{stealth}}}}] (6,-2) .. controls ({6-0.8*\i},{0.8*\i-2}) and ({6-0.8*\i},{2-0.8*\i}) .. (6,2);
            }
        \end{scope}
        \draw (2,2) -- (6,-2) -- (6,2) -- (2,-2) -- (2,2);
        \draw (2,2) -- (6,2);
        \draw (2,-2) -- (6,-2);
        \draw[very thick, blue] (4,0) .. controls (5,-0.3) and (5.5,-0.3) .. (6,-0.3) node[midway,below,blue!60!black] {\large$\Sigma$};
        \draw[very thick, blue] (4,0) .. controls (3,0.3) and (2.5,0.3) .. (2,0.3);
        \node[right] at (6,0) {$r=0$};
    \end{tikzpicture}
    \caption{Each static patch in de Sitter spacetime has an associated boost Killing symmetry.}
    \label{Figure: de Sitter boost}
\end{figure}
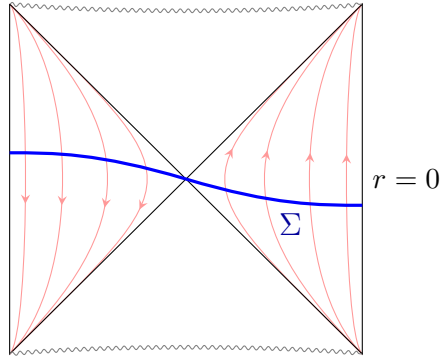

Let us for the moment focus on the spatially closed case, such as de Sitter (see Fig.~\ref{Figure: de Sitter boost}). As emphasized in \cite{Chandrasekaran:2022cip}, isometries are gauge in that case and the corresponding generators, including $H_\xi$, should thus be imposed as gauge constraints. Although $H_\xi$ cannot be seen at the purely linearized level, our above discussion makes clear why the second-order Taub charge $H_\xi$ should be imposed as a constraint on the linear graviton theory following from the action in Eq.~\eqref{expaction} nevertheless: it is one of the linearization stability conditions in Eq.~\eqref{2ndorder3} that are necessary and sufficient to eliminate spurious solutions of the linearized theory and our discussion around Eq.~\eqref{2ndorderconstraint} makes furthermore clear that it indeed is the second-order gauge constraint associated with $\xi$ on-shell of the linearized Einstein equations, $H_\xi=-C^{(2)}[\kappa\xi]$. Crucially, this provides a justification for considering only a \emph{finite number} of the infinitely many independent second-order constraints, namely one for each isometry; implementing other second-order constraints is not necessary for ensuring that the linear order is a valid approximation to an exact solution. 

Of course, the observation of necessity and sufficiency is one based on the classical theory. As we do not have a full quantum theory of gravity, it is currently not possible to rigorously prove, as in the classical case, that the quantum version of the Taub charge constraints must be imposed on the perturbative quantum theory also in order to guarantee a good approximation to the exact non-perturbative quantum theory. However, it is as strong an argument as currently feasible and, for example, in the case of scalar QED it is possible to make the case for second-order constraints on the linearized theory to avoid incorrect states in spatially closed spacetimes \cite[Sec.~6]{Higuchi:1991tk} (see also Sec.~\ref{ssec_QED}; similar arguments can be made for perturbative quantum gravity with matter around a de Sitter background \cite{Losic:2006ht}. See \cite{Moncrief:1978te,Moncrief:1979bg,Higuchi:1991tk,Higuchi:1991tm,Kaplan:2024xyk} for further discussion of linearization instabilities in quantum gravity.

There is one subtlety though: imposing the Taub charges as gauge constraints in spatially closed spacetimes, Eq.~\eqref{2ndorder3}, means also constructing the subalgebras of $\mathcal{A}_{\mathcal{U}}$ and its commutant that commute with $H_\xi$. However, as observed in \cite{Chandrasekaran:2022cip,Chen:2024rpx}, $H_\xi$ acts ergodically on the regional algebras in de Sitter space, which means that the only observables that are invariant under its action are multiples of the identity. Thus, no interesting physics is encoded in such a setup. To sidestep this issue, CLPW introduced an observer equipped with a clock in each static patch  \cite{Chandrasekaran:2022cip}; it is interesting to note that internalizing observers to properly describe thermodynamic aspects of de Sitter space was already suggested by Moncrief in \cite{Moncrief:1978te,Moncrief:1979bg} based on his linearization instability analysis. The clocks are additional degrees of freedom, constituting internal quantum reference frames \cite{Hoehn:2019fsy,delaHamette:2021oex,Hoehn:2023axh,Giacomini:2017zju,Carette:2023wpz} for the time translations relative to which the QFT degrees of freedom can be dressed \cite{DeVuyst:2024pop,Fewster:2024pur}, and this modifies the gauge-invariant content of the theory. The idea is that the observer can measure quantum field degrees of freedom in the vicinity of their worldline, which, by the timelike tube theorem \cite{Borchers1961,Araki1963AGO,Witten:2023qsv,Strohmaier:2023hhy,Strohmaier:2023opz}, entails that they can probe the quantum field everywhere in $\mathcal U$. This gives physical meaning to dressing all the QFT observables to the worldline clock.

Let us briefly analyze how the introduction of clocks affects the discussion of stability. The most natural choice is to add the clocks at the same order in the $\kappa$-expansion as all the remaining matter degrees of freedom, hence, to already include it in the action \eqref{expaction} of the linearized theory
\begin{equation}
    S_M[g^0;\phi]\to S_M[g^0;\phi]+\sum_i S_i[g^0;t_i]\,,
\end{equation}
where $i$ runs over all the observer clocks in spacetime and $t_i$ labels their internal configuration degrees of freedom. For example, similarly to \cite{Witten:2023xze,Kolchmeyer:2024fly}, a clock action could read
\begin{equation}\label{eq: clock worldline action}
S_i=\int_{\gamma_i}\dd s\left(p_i\dot t-\sqrt{-g^0_{ss}}(H_i(t_i,p_i))\right)\,,
\end{equation}
where $s$ parametrizes the observer's worldline $\gamma_i$ and $g^0_{ss}$ is the induced metric on the latter. This action is invariant under reparametrizations of $s$ and if $\gamma_i$ is one of the Killing flow lines associated with $\xi$ then this reparametrization-invariance matches the invariance under diffeomorphisms generated by $\xi$. Extremizing $S_i$ tells us that $\gamma_i$ is a geodesic and the clock's contribution to the boost Hamiltonian is
\begin{equation}\label{clockham}
    H_i=\int_\Sigma (\epsilon^0)^\mu\xi^\nu T^{(0)}_{i,\mu \nu}\,,
\end{equation}
where $T^{(0)}_{i,\mu \nu}$ is the stress-energy tensor associated with $S_i$. More generally, $\gamma_i$ need not be a geodesic (especially not when including several observers  traveling along flow lines of the same Killing field in the same static patch), but Eq.~\eqref{clockham} accommodates such cases for different choices of clock action $S_i$. The precise form will not be important here and the clock Hamiltonian $H_i$ may be bounded or not \cite{Hoehn:2019fsy,Hoehn:2020epv,DVEHK} or could even have discrete spectrum in the quantum theory \cite{Chataignier:2024eil}. The upshot is that the Taub boost charge in Eq.~\eqref{2ndorder3} that has to vanish on-shell of $E^{(2)}_{\mu \nu}\approx0$ now reads
\begin{equation}\label{taub1}
C=\int_\Sigma\left(-2G^{(2)}_{\mu \nu}(h,h)+T_{\mu \nu}^{(0)}\right)\xi^\mu(\epsilon^0)^\nu+\sum_i H_i=H_\xi+\sum_i H_i\approx0+O(\kappa)\,
\end{equation}
and on-shell of the first-order equations of motion this also coincides with the second-order gauge constraint associated with the boost vector field $\xi$. This recovers the form of the constraint in Eq.~\eqref{2clockconstraint}. Note that this still signals a linearization instability as before because all variables in Eq.~\eqref{taub1}, including the ones of the clocks, are already determined from the first-order equations of motion. Unless these first-order solutions already satisfy Eq.~\eqref{taub1}, they are spurious and not integrable to an exact one. Thus, once more it is necessary to impose Eq.~\eqref{taub1} as a stability condition. Of course, when there are additional isometries as in de Sitter, there will be further stability conditions to impose as gauge constraints.\footnote{It is likely that the insertion of the clocks does not change the fact that the vanishing of the Taub charges is also sufficient to guarantee integrability of the linearized solution, though we have not checked the conditions for this to hold.} Those will however not correspond to a modular flow and turn out not to play a role in the type transition of the regional algebras \cite{Chandrasekaran:2022cip}; invariance under those finitely many constraints can be achieved by invoking the more general quantum reference frames of \cite{delaHamette:2021oex}.  

Now what is a modular crossed product? We shall not provide details here as this goes beyond the scope of this article, instead only giving brief intuition (see \cite{Witten:2021unn,Chandrasekaran:2022cip,Jensen:2023yxy,Takesaki1979} for expositions). The addition of the clocks yields a non-trivial gauge-invariant regional algebra
\begin{equation}\label{algebra}
    \hat{\mathcal{A}}^C_{\mathcal U}:=\qty\Big(\mathcal{A}_{\mathcal{U}}\otimes\bigotimes_i\mathcal{B}(\mathcal{H}_i))^C\,,
\end{equation}
where $i$ runs over all the clocks inside $\mathcal U$, $\mathcal{H}_i$ is the kinematical Hilbert space factor associated with clock $i$ and $\mathcal{A}^C$ denotes the subalgebra of $\mathcal{A}$ that commutes with $C$. It turns out that, while $\mathcal{A}_\mathcal{U}$ was assumed to be a type $\rm{III}_1$ von Neumann factor, $\hat{\mathcal{A}}^C_\mathcal{U}$ is of type $\rm{II}$ \cite{DeVuyst:2024pop,DVEHK} (see also \cite{Chandrasekaran:2022cip,Jensen:2023yxy,Chen:2024rpx} for the case of a single clock in $\mathcal U$),\footnote{It is type $\rm{II}_1$ factor if only a single clock with bounded from below Hamiltonian resides in $\mathcal U$ and otherwise type $\rm{II}_\infty$.} which thus possesses a trace, density operators and renormalized entropies. Essentially, this happens because modular flow has become inner: while we noted before that $H_\xi$ is not contained in $\mathcal{A}_\mathcal{U}$, $U(s):=e^{is\sum_i H_i}$ is contained in $\hat{\mathcal{A}}^C_\mathcal{U}$ for any $s\in\mathbb{R}$ and because this algebra is gauge-invariant, conjugating $\hat{\mathcal{A}}^C_\mathcal{U}$ with $U(s)$ is equivalent to conjugating it with $e^{-is H_\xi}$. An algebra with inner modular flow cannot be of type $\rm{III}$ \cite{Takesaki1979}.

The algebra in Eq.~\eqref{algebra} can be written as a set of relational observables describing the involved degrees of freedom relative to any one of the clocks (together with reorientations of the clock frame) \cite{DeVuyst:2024pop,DVEHK}. Thus, the relational description ensures a naturally regularized entropy that is not available prior to imposing the linearization stability conditions. When one of the clocks in $\mathcal U$, say $j$, is an ideal one, meaning its energy spectrum is non-degenerate and takes value in the entire real line \cite{Hoehn:2019fsy}, then Eq.~\eqref{algebra} constitutes a so-called  crossed product, $\hat{\mathcal{A}}^C_\mathcal{U}=\qty\Big(\mathcal{A}_\mathcal{U}\otimes\bigotimes_{i\neq j}\mathcal{B}(\mathcal{H}_i))\rtimes_\alpha \mathbb{R}$, of $\mathcal{A}_\mathcal{U}\otimes\bigotimes_{i\neq j}\mathcal{B}(\mathcal{H}_i)$ with the modular flow $\alpha$ \cite{Witten:2021unn,Chandrasekaran:2022cip,Jensen:2023yxy,DVEHK}. Otherwise, when $H_j$ has non-degenerate spectrum, it is a subalgebra of a crossed product.\footnote{These considerations can also be extended to clocks with degenerate spectrum \cite{DVEHK}.}

As discussed above, we assumed that all first-order diffeomorphism constraints as well as matter gauge constraints are already imposed on the quantum states in $\mathcal{H}_{\rm QFT}$ and regional algebras $\mathcal{A}_\mathcal{U}$; this is analogous to how in our discussion of classical instabilities we assumed the first order equations of motion (and thus constraints) to be obeyed. We then focused on only a single second-order boost Hamiltonian constraint (or linearization stability condition), which turns out to be crucial for the type conversion of the regional von Neumann algebras. Of course, also all the other linearization stability conditions corresponding to the remaining second-order isometry constraints should be imposed on states and observables. As suggested in \cite{Chandrasekaran:2022cip}, these additional constraints should not affect the type conversion of the regional algebras and this has indeed been confirmed in \cite{Fewster:2024pur,AliAhmad:2024eun},\footnote{These works impose constraints only on the algebra, but not states, which suffices for establishing the type transition (at a kinematical level). The appropriate QRF framework for imposing constraints on both states and observables, as needed in gauge theory and gravity, can be found in \cite{delaHamette:2021oex}.} leading to an enlarged crossed product. Following \cite{Chandrasekaran:2022cip}, we have thus not further discussed these additional constraints, and we may assume that the algebra in Eq.~\eqref{algebra} already commute with those, and that quantum states solve these too.

\subsubsection{Black holes and boundaries}

\begin{figure}
    \centering
    \begin{tikzpicture}[scale=1.16]
        \begin{scope}[red!40!white]
            \foreach \i in {0.3,1,1.85,2.6} {
                \draw[postaction={decorate,decoration={markings,mark=at position .57 with {\arrow[scale=1.1]{stealth}}}}] (2,2) .. controls ({2-0.8*\i},{2-0.8*\i}) and ({2-0.8*\i},{0.8*\i-2}) .. (2,-2);
                \draw[postaction={decorate,decoration={markings,mark=at position .57 with {\arrow[scale=1.1]{stealth}}}}] (2,2) .. controls ({2+0.8*\i},{2-0.8*\i}) and ({2+0.8*\i},{0.8*\i-2}) .. (2,-2);

                \draw[postaction={decorate,decoration={markings,mark=at position .57 with {\arrow[scale=1.1]{stealth}}}}] (6,-2) .. controls ({6-0.8*\i},{0.8*\i-2}) and ({6-0.8*\i},{2-0.8*\i}) .. (6,2);
                \draw[postaction={decorate,decoration={markings,mark=at position .57 with {\arrow[scale=1.1]{stealth}}}}] (6,-2) .. controls ({6+0.8*\i},{0.8*\i-2}) and ({6+0.8*\i},{2-0.8*\i}) .. (6,2);
            }
        \end{scope}
        \draw (0,0) -- (2,2) -- (6,-2) -- (8,0) -- (6,2) -- (2,-2) -- (0,0);
        \draw[decorate,decoration={snake,amplitude=0.7pt,segment length=3pt},gray] (2,2) .. controls (3,1.9) and (5,1.9) .. (6,2);
        \draw[decorate,decoration={snake,amplitude=0.7pt,segment length=3pt},gray] (2,-2) .. controls (3,-1.9) and (5,-1.9) .. (6,-2);
        \draw[very thick, blue] (4,0) .. controls (3,0.3) and (1,0.3) .. (0,0);
        \draw[very thick, blue] (4,0) .. controls (5,-0.3) and (7,-0.3) .. (8,0) node[midway,below,blue!60!black] {\large$\Sigma$};
        \fill (8,0) circle (0.07) node[right] {$i^0$};
        \node[above right] at (7,1) {$\mathscr{I}^+$};
        \node[below right] at (7,-1) {$\mathscr{I}^-$};
    \end{tikzpicture}
    \caption{The Killing symmetry generating the horizon in the extended Schwarzschild spacetime is depicted on its Penrose diagram.}
    \label{Figure: Schwarzschild both side boost}
\end{figure}
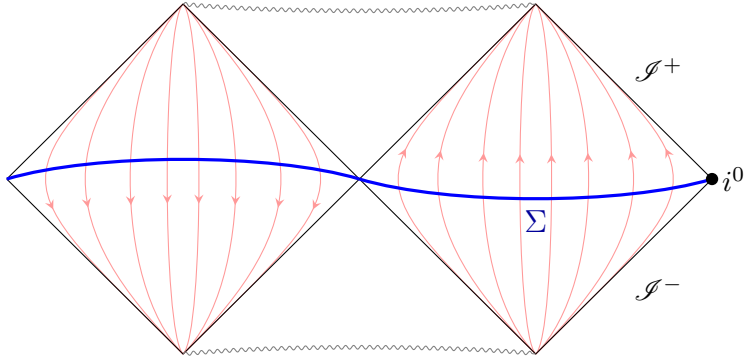

When the spacetime does have an asymptotic boundary, the situation changes somewhat. Rather than Eq.~\eqref{2ndorderconstraint}, the second-order constraint associated with a boost Killing vector field now reads
\begin{equation}\label{BHcon0}
    C^{(2)}[\kappa\xi]=\int_\Sigma\left(2G^{(2)}_{\mu \nu}(h,h)-T_{\mu \nu}^{(0)}+2hG^{(1)}_{\mu \nu}(h)\right)\xi^\mu(\epsilon^0)^\nu+2\int_{\partial\Sigma}(\epsilon^0)^{\mu \nu}F_{\mu \nu}(k)\,
\end{equation}
and so on-shell of $E^{(1)}_{\mu \nu}=0$ it equates the corresponding bulk Taub charge/boost Hamiltonian with the second-order boundary ADM energy, Eq.~\eqref{2ndorder}, i.e.
\begin{equation}\label{BHcon}
    C=C^{(2)}[\kappa\xi]=-H_\xi+H^{(2)}_{\rm ADM}\,,
\end{equation}
where $H_{\rm ADM}^{(2)}$ is the surface integral in Eq.~\eqref{BHcon0}. For vacuum general relativity, this was already noted by Moncrief in \cite[Sec.~6]{Moncrief:1976un}. Note that in the case of a maximally extended Schwarzschild spacetime, as in Fig.~\ref{Figure: Schwarzschild both side boost}, or the two-sided AdS black hole, $H_{\rm ADM}^{(2)}=H_R-H_L$ consists of two pieces, one for the left and one for the right boundary (the relative minus sign arises because the boost time goes forward in the right half and backward in the left one). For example, Eq.~\eqref{BHcon} reproduces the boost constraint of a maximally extended Schwarzschild spacetime of \cite{Chandrasekaran:2022cip}. Owing to the boundary contribution, it is now no longer necessary to explicitly insert observers with clocks into the spacetime, the second-order ADM Hamiltonians can take over this role. Constructing the gauge-invariant algebra
\begin{equation}
    \left(\mathcal{A}_\mathcal{U}\otimes\mathcal{B}(\mathcal{H}_R)\right)^C
\end{equation}
for the right region, where $\mathcal{H}_R$ denotes the Hilbert space of the second-order metric perturbations $k_{\mu \nu}$ on the right asymptotic boundary, one again finds that it is of type $\rm{II}_\infty$ because it is a crossed product since the spectrum of $H_R$ is non-degenerate and unbounded. 

However, in this case, provided all the second-order ADM charges are non-vanishing, there is no linearization instability. Thus, it is in fact not necessary to impose the second-order constraint $C$ in order to guarantee that the linearized solution is tangential to an exact solution. The justification for imposing it (but not other second-order constraints) on the linearized theory nevertheless must thus come from elsewhere. We shall come back to this issue in Sec.~\ref{sec_generalregions} when discussing general subregions in perturbative quantum gravity. Essentially, one should impose some second-order constraints if one is interested in certain second-order variables like area fluctuations. We shall also discuss black holes in some more detail in Sec.~\ref{subsec: diffeomorphism covariant Lagrangians} using the covariant phase space language.

\section{Gravitons and matter on a background with matter}\label{sec_matter}

As mentioned in Sec.~\ref{sssec_instacompact}, in most studies of linearization instabilities that involve matter \cite{Arms1,Arms:1979au,Arms:1982ea,Saraykar:1981qm,Saraykar2,Arms:1986vk,Altas:2021htf}, one perturbs around a background solution $g_0$ that solves the Einstein equations sourced by some matter configuration $\phi_0$, which itself solves the corresponding matter equations of motion with metric $g_0$. In these works, one treats gravity and matter perturbations on an equal footing and, for mathematical simplicity, $\kappa$ and all matter coupling constants are ignored (or implicitly set to $1$). 

Let us now see how we can make this consistent with our setting in the $\kappa\to0$ limit in which crossed product algebras appear. In order to treat metric and matter perturbations on an equal footing, we must also expand the matter fields
\begin{equation}\label{matterpert}
    \phi = \phi_0+\kappa\varphi+\frac{1}{2}\kappa^2\psi+O(\kappa^3)
\end{equation}
to second order around the background solution $\phi_0$. To this end, it is convenient to scale the background matter fields $\phi_0$ such that their stress-energy tensor is of order $O(\kappa^{-2})$,\footnote{Such a rescaling has also been briefly mentioned
in the crossed product discussions in \cite[Sec.~4]{Chen:2024rpx} and \cite[footnote 7]{Jensen:2023yxy}.} which also means that $S_M[g_0,\phi_0]=O(\kappa^{-2})$. In order to mimic the discussion in \cite{Arms1,Arms:1979au,Arms:1982ea,Saraykar:1981qm,Saraykar2,Arms:1986vk,Altas:2021htf} and expand everything consistently in $\kappa$, we need to ensure this is done so that:
\begin{itemize}
    \item[({\color{red}$\square$})] The background solution $(g_0,\phi_0)$ is an exact solution of the full non-linear gravity and matter theory, so that $\phi_0$ has no explicit $\kappa$-dependence. So when (possibly) rescaling the matter fields to ensure $S_M[g_0,\phi_0]=O(\kappa^{-2})$, no explicit $\kappa$-dependence should arise in $S_M$ aside from an overall $\kappa^{-2}$ factor that can be combined with the one from the Einstein-Hilbert action to a global one. 
\end{itemize}
When the matter theory is free and thus quadratic, this is easy to achieve by just rescaling $\phi\to\phi/\kappa$. However, when the matter theory contains interactions, this is more subtle but can be achieved by redefining the matter coupling constants. For example, suppose that we consider the Einstein-Yang-Mills system with \begin{equation} {S_M=-\frac{1}{2}\int_\mathcal{M}\Tr(F\wedge\star F)}
\end{equation}
where ${F=\dd A+g A\wedge A}$ and $g$ is the Yang-Mills coupling strength. In this case, we can rescale $A\to A/\kappa$ and redefine $\tilde{g}=g/\kappa$ as the new Yang-Mills coupling constant to achieve the condition in ({\color{red}$\square$}). This describes a regime where the gravitational and Yang-Mills fields contribute approximately equally to the action and energy. When we consider a scalar field with quartic interactions, 
\begin{equation} \label{lambda4}
S_M=-\frac{1}{2}\int_\mathcal{M}\epsilon\left(g_{\mu\nu}\nabla^\mu\phi\nabla^\nu\phi-m^2\phi^2-\frac{2\lambda}{4!}\phi^4\right)\,,
\end{equation}
we can scale $\phi\to\phi/\kappa$ and declare $\tilde\lambda=\lambda/\kappa^2$ as a new coupling constant to meet ({\color{red}$\square$}).\footnote{In some cases, as an alternative to ({\color{red}$\square$}) (though not mimicking \cite{Arms1,Arms:1979au,Arms:1982ea,Saraykar:1981qm,Saraykar2,Arms:1986vk,Altas:2021htf}), we could also accept an explicit $\kappa$-dependence inside $S_M$ (while ensuring $S_M=O(\kappa^{-2})$) and suitably expanding also the matter action in $\kappa$. For example, this works in slow-roll inflation, where $V(\phi)\sim\phi$. However, in most cases, this will not be useful. For instance, in Eq.~\eqref{lambda4}, we could also scale $\phi\to\phi/\sqrt{\kappa}$, so that the contribution of the leading order is $-2\lambda/4!\kappa^2\phi^4$. However, its equations of motion would just yield $\phi=0$. If we did something analogous in a theory where the highest field power is odd, $2n+1>2$, we would get fractional powers of $\kappa$, ruining our uniform treatment of matter and gravity perturbations.}

When considering a background with matter, we will henceforth assume that ({\color{red}$\square$}) is obeyed. This modifies the discussion surrounding Eq.~\eqref{eoms} in terms of where matter enters the expansion.
Equipping the rescaled stress-energy tensor with a tilde for distinction from the original one, we now have
\begin{equation}
    E_{\mu \nu}[g;\phi]=\frac{1}{\kappa^2}E^{(0)}_{\mu \nu}[g_0;\phi_0]+\frac{1}{\kappa}E^{(1)}_{\mu \nu}[g_0;\phi_0;h;\varphi]+E_{\mu \nu}^{(2)}[g_0;\phi_0;h;\varphi;k;\psi]+O(\kappa)\approx 0\,,
\end{equation}
with the Einstein equations at different order given by
\begin{equation}\label{eoms2}
\begin{aligned}
    E_{\mu \nu}^{(0)}[g_0;\phi_0]&=4G_{\mu \nu}^{(0)}-\tilde{T}^{(-2)}_{\mu \nu}=4G_{\mu \nu}[g_0]-\kappa^2\tilde{T}_{\mu \nu}[g_0,\phi_0]\,,\\
    E_{\mu \nu}^{(1)}[g_0;\phi_0;h;\varphi]&=4G^{(1)}_{\mu \nu}(h)-\tilde{T}^{(-1)}_{\mu \nu}(h;\varphi)\,,\\
    E_{\mu \nu}^{(2)}[g_0;\phi_0;h;\varphi;k;\psi]&=2G^{(1)}_{\mu \nu}(k)-\frac{1}{2}\tilde{T}^{(0)}_{\mu \nu}[g_0;\phi_0;h;\varphi;k;\psi]+2G_{\mu \nu}^{(2)}(h,h)\,,
\end{aligned}
\end{equation}
where the Einstein tensor is expanded as before in Eq.~\eqref{Einsteintensor}, but the expansion of the stress-energy tensor now becomes
\begin{equation}
    \tilde{T}_{\mu \nu}[g;\phi]=\frac{1}{\kappa^2}\tilde{T}^{(-2)}_{\mu \nu}+\frac{1}{\kappa} \tilde{T}^{(-1)}_{\mu \nu}(h;\varphi)+\frac{1}{2}\tilde{T}^{(0)}_{\mu \nu}[g_0;\phi_0;h;\varphi;k;\psi]+O(\kappa)\,
\end{equation}
with a second-order term schematically of the form
\begin{equation}
    \tilde{T}^{(0)}_{\mu \nu}[g_0;\phi_0;h;\varphi;k;\psi]= \tilde{T}^{(-1)}_{\mu \nu}(k;\psi) +\tilde{T}^{(0)}_{\mu \nu}\left((h;\varphi),(h;\varphi)\right) \,.
\end{equation}
We use the same notation as before, except that we now also have $\tilde{T}^{(-1)}_{\mu \nu}(h;\varphi)$, which is to be read as linear in the vector $(h;\varphi)$ of both the gravitons $h$ and matter perturbations $\varphi$, and $\tilde{T}^{(0)}_{\mu \nu}((h;\varphi),(h;\varphi))$, which is bilinear in that vector. Note that the vector operator $\tilde{T}^{(-1)}_{\mu \nu}$ is the same in both the $O(\kappa^{-1})$ and $O(\kappa^0)$ terms of the stress-energy expansion.

The equations of motion, $E^{(i)}_{\mu \nu}\approx0$ are once more solved order by order and the key question is whether or not, given $h,\varphi$ from the first-order Einstein and matter equations,  one can solve $E^{(2)}_{\mu \nu}\approx0$ and the second-order matter equations for $k,\psi$ (again, possibly given a suitable gauge fixing).

What about the linearized theory? Expanding $S[g,\phi]$ to order $O(\kappa^0)$, as before, yields an action that coincides with Eq.~\eqref{expaction}, except that the matter piece $S_M[g_0,\phi]$ is replaced by $S_M^{(2)}[h,\varphi]$, where the latter is the second-order expansion in the graviton and matter fluctuations of $S_M[g,\phi]$ (which is of $O(\kappa^{-2})$ due to our scaling) and that the -- for our discussion uninteresting -- boundary term is also suitably adjusted. This action will define the linearized theory of perturbations around the $(g_0,\phi_0)$ background. Note that this action will generally couple graviton and matter perturbations $h$ and $\varphi$, as already indicated in the expression for $E^{(1)}_{\mu \nu}$ above.

Notice also that the expansion in \eqref{matterpert} has the consequence that linearized diffeomorphisms now act non-trivially on the matter too at the order at which we are operating (unless $\phi_0$ is constant in spacetime). Indeed, we have in analogy to Eq.~\eqref{variation}
\begin{equation}\label{variation3}
\delta_{\kappa\xi}\phi=\kappa\mathcal L_\xi\phi_0+\kappa^2\mathcal L_\xi\varphi+O(\kappa^3)\qquad\Rightarrow\qquad \delta_{\kappa\xi}\varphi = \mathcal L_\xi\phi_0\,.
\end{equation}
Together with Eq.~\eqref{variation}, this is again a gauge symmetry of the expanded action and now leads to the constraints with matter contribution\footnote{Again, for non-trivial tensorial matter, there will be additional contributions \cite{Seifert:2006kv}, as we will discuss in App.~\ref{sect: EH with matter}.}
\begin{equation}\label{lincon2}
    C_{\kappa\xi}=\left(4G_{\mu \nu}^{(1)}(h)-\tilde{T}^{(-1)}_{\mu \nu}(h;\varphi)\right)\xi^\mu(\epsilon^0)^\nu+O(\kappa)\,.
\end{equation}
Again, this coincides with the linear expansion of the full-theory constraint. Gauge-invariant observables at leading order can be constructed as composite objects using dressing methods, e.g., by adapting constructions as in \cite{Donnelly:2015hta,Donnelly:2016rvo,Giddings:2018umg,Giddings:2022hba,Frob:2022ciq,Frob:2023gng}.\footnote{For example, it can be easily checked that inserting Eq.~\eqref{matterpert} (for scalar fields) into the dressed observables of \cite{Donnelly:2015hta,Donnelly:2016rvo,Giddings:2018umg} yields observables invariant under the variations in Eqs.~\eqref{variation} and \eqref{variation3} at order $\kappa$. }

Arguing for instabilities in this case requires some modifications. We explain in App.~\ref{app_isolin} why the arguments of \cite{fischer1980structure} establishing Eq.~\eqref{vanish} cannot be easily extended to backgrounds that include matter. Similarly, the derivation of Eq.~\eqref{boundaryterm} relies on an explicit $\kappa$-expansion of the vacuum Einstein equations and so this line of argument too is difficult to generalize to include matter backgrounds. Both of these challenges can be easily overcome using the modern covariant phase space language, as we shall see next.

\section{Constraints and instabilities in the covariant phase space formalism}
\label{sec: Constraints in the covariant phase space}

While the original literature on linearization instabilities was motivated by understanding the solution space and the consistency of perturbative expansions in general relativity, the topic applies to any gauge-covariant field theory with non-linear equations of motion. To highlight this, it is useful to reframe the preceding discussion of linearization instability using the modern covariant phase space formalism \cite{Witten_Crnkovic,Lee:1990nz, Wald:1993nt, Jacobson:1993vj, Wald:1999wa, Iyer:1994ys, Iyer:1995kg}, as opposed to the original formulation in the canonical phase space setting \cite{fischer1980structure,Arms:1982ea,Marsden_lectures,Deser:1973zza,Moncrief1,Moncrief:1976un,Arms:1982ea,Deser:1973zzb,choquet1979maximal,Arms1,Saraykar:1981qm,Saraykar2,Arms:1979au,Arms:1982ea}.   This provides a unifying framework in which to consider local gauge transformations of any gauge-covariant Lagrangian, viewing diffeomorphism covariance as a special case. In the case of gravitational theories, this also allows us to sidestep the difficulties discussed in Sec.~\ref{sec_matter} regarding matter-sourced backgrounds.  We will see in complete generality that the first-order constraint density associated with any background symmetry of a gauge-covariant Lagrangian is spacetime exact.  This is at the heart of linearization instability: if the spatial slices are closed, or if fall-off conditions entail that the relevant boundary terms vanish, it may become necessary to include higher-order constraints as restrictions on first-order fluctuations to ensure a valid truncation of the global theory.  We will not undertake a complete analysis in terms of existence and uniqueness of solutions to the relevant differential equations (for discussions of these aspects, see the original literature in the canonical formalism \cite{Moncrief1,Moncrief:1976un,fischer1980structure,Arms:1982ea,Arms1,Arms:1979au,Saraykar:1981qm,Saraykar2,Marsden_lectures,Deser:1973zzb,choquet1979maximal,Moncrief:1977hy,Arms_1981}) but identify the universal structure of these scenarios.

There are two previous works that translate the study of linearization instabilities into a covariant setting with the aim to generalize it to a general class of field theories. Reference \cite{Arms:1986vk} employs covariant equations at the level of our preceding discussion (though for more general theories), while \cite{Khavkine:2013iei} is based on de Rham cohomology. Neither invokes the covariant phase space formalism and, as far as we are aware, the present work appears to be the first to do so in the context of linearization (in)stabilities.

\subsection{Notations}
\label{subsec: notations}
We begin by establishing some notation.  As is common with modern covariant phase space computations, we consider differential forms to be part of the variational bicomplex \cite{anderson1992introduction}, on which a spacetime $p$-form and field-space $q$-form is referred to as a $(p,q)$-form.  The exterior derivatives on spacetime and on field space will be denoted $\sd$ and $\fd$, respectively, and we take these to commute.  We let $\sins$ and $\fins$ represent the interior product on spacetime and field space, respectively.  We will use $\Phi$ to collectively represent all the fundamental fields, both metric and matter.  A `hat' will designate vectors on field space, and as a special case $\hat{\xi}$ will represent the field-space lift of a spacetime vector field $\xi$, which is defined to induce the same infinitesimal diffeomorphism flow on fundamental fields:
\begin{equation}
    \hat{\xi}:=
    \int \extd^d x~ \sL_{\xi}\Phi^a\frac{\delta}{\delta \Phi^a(x)}.
\end{equation} 
The field-space Lie derivative and spacetime Lie derivative of forms can now both be written using Cartan's magic formula, which is employed liberally in covariant phase space manipulations: $\fL_{\hat{\xi}}:=\fd \hat{\xi}\fins+\hat{\xi}\fins \fd$ and $\sL_{\xi}:= \sd \xi\sins+\xi\sins \sd$.

We will sometimes explicitly indicate the functional dependence of certain quantities using the notation
\begin{equation}\label{eq: functional dependence notation}
    \mathcal{F}:=\mathcal{F}\left[A_1,...,A_N; B_1,...,B_M\right]
\end{equation}
where the semicolon conveys that $\mathcal{F}$ is both linear and completely antisymmetric in all arguments beyond it, a dependence which naturally arises from forms of field-space degree $M$. 

\subsection{General gauge-covariant Lagrangians}
\label{subsec: general gauge constraints}
Given a Lagrangian top form $L$ in $d$ dimensions, the usual variational principle can be expressed in terms of its field-space exterior derivative, leading to the fundamental relation
\begin{equation}\label{eq: variation L general}
    \fd L = E+\sd \theta,
\end{equation}
which identifies the equation of motion $(d,1)$-form, $E:=E_a[\Phi] \delta \Phi^a$, and the symplectic potential current $(d-1,1)$-form, $\theta:=\theta[\Phi;\delta\Phi]$.\footnote{The ambiguities associated with shifting $\theta$ by a closed form or shifting $L$ by a boundary term can be collectively resolved through proper consideration of the boundary conditions of the theory \cite{Harlow:2019yfa}.  We will discuss such boundary terms later in the case of diffeomorphism-covariant Lagrangians, but for the moment we only require $\theta$ to satisfy \eqref{eq: variation L general}.}  
Let $\hat{X}$ represent the field-space vector generating an infinitesimal gauge transformation corresponding to a local gauge parameter $X$:
\begin{equation}\label{eq: gauge action}
\hat{X}\fins\fd\Phi = \delta_X\Phi.
\end{equation}
For this to be a gauge symmetry it must leave the Lagrangian invariant up to a spacetime exact term, which we denote using $\alpha_X$:
\begin{equation}\label{eq: covariant L general}
    \hat{X}\fins\fd L =\sd \alpha_X.
\end{equation}
For such Lagrangians one can write down an associated  Noether current $(d-1,0)$ form, $J_X$, which is closed on-shell (i.e.\ closed when evaluated on field configurations satisfying $E=0$):\footnote{\label{footnote: J conserved}Currents such as $J_X$ are sometimes referred to as ``conserved'' due to their on-shell closure.  Under suitable fall-off conditions this is enough to entail an exactly conserved integral on spacelike slices via Stokes' theorem, but more generally deriving a conserved charge associated with $J_X$ can require the inclusion of boundary terms.}
\begin{align}
    \label{eq: define J general}
    J_X &:=\hat{X}\fins \theta-\alpha_X,\\
    \label{eq: closed J general}
    \extd J_X &= -\hat{X}\cdot E.
\end{align}
Equation \eqref{eq: closed J general} follows from the definition $\eqref{eq: define J general}$ by employing \eqref{eq: variation L general} and \eqref{eq: covariant L general}. A classic result due to Wald \cite{Wald:1990mme} states that any identically closed form locally constructed from fundamental and background fields and finitely many of their derivatives, and which vanishes for vanishing fundamental fields, can be written as an exact form.  This implies a standard decomposition as
\begin{equation}\label{eq: decompose J general}
    J_X=C_X + \extd Q_X,
\end{equation}
where $Q_X$ is a $(d-2,0)$ form and $C_X$ is a constraint piece that vanishes on-shell. Such a decomposition is ambiguous under shifts of the form $C_X\to C_X+\dd{Y}_X$, $Q_X \to Q_X-Y_X$, where $Y_X$ vanishes on-shell. Any such choice is sufficient for purposes of establishing the identities below, but $C_X$ can be fixed by requiring it to depend  linearly on $X$ and not on its derivatives.
A field-space exterior derivative of this decomposition leads to
\begin{equation}\begin{split}\label{eq: delta C general 0}
    \delta C_{X} &=\delta(\hat{X}\cdot \theta) - \delta \alpha_{X} -\extd \delta Q_{X} \\
    &=-\hat{X}\cdot \omega + \fL_{\hat{X}}\theta-\delta \alpha_{X}-\extd\delta Q_{X}.
\end{split}\end{equation}
The second equality introduces the symplectic current $(d-1,2)$ form $\omega:=\delta \theta$ and uses Cartan's magic formula for field-space Lie derivatives.
The combination $\fL_{\hat{X}}\theta-\delta \alpha_{X}$ appearing in \eqref{eq: delta C general 0} can be seen to be on-shell closed.
\begin{equation}\begin{split}\label{eq: on-shell closed combination}
    \extd(\fL_{\hat{X}}\theta-\delta \alpha_{X})
    =\fL_{\hat{X}}(\extd\theta - \delta L) = -\fL_{\hat{X}} E.
\end{split}\end{equation}
This vanishes on-shell because $\hat{X}$ represents a gauge symmetry, so it necessarily preserves the on-shell condition.
Once again invoking \cite{Wald:1990mme}, we can decompose\footnote{See comments below equation \eqref{eq: decompose J general} regarding the ambiguities associated with such decompositions.} this on-shell closed form as a piece which vanishes on-shell and a spacetime exact piece:
\begin{equation}\label{eq: define tilde C, tilde q general}
    \fL_{\hat{X}}\theta-\delta \alpha_{X} = \mathcal{C}_X + \extd \mathcal{Q}_X.
\end{equation}
Note that while $C_{X}$ and $Q_{X}$ from \eqref{eq: decompose J general} are a $(d-1,0)$-form and $(d-2,0)$-form, respectively, $\mathcal{C}_{X}$ and $\mathcal{Q}_{X}$ are a $(d-1,1)$-form and $(d-2,1)$ form, respectively. Inserting \eqref{eq: define tilde C, tilde q general} into equation \eqref{eq: delta C general 0} and defining the $(d-2,1)$-form
\begin{equation}\label{eq: boundary term general}
    B_{X}:=\mathcal{Q}_{X}-\fd Q_{X}
\end{equation}
gives
\begin{equation}\begin{split}\label{eq: delta C general 2}
    \fd C_{X} 
    &=\mathcal{C}_{X}-\hat{X}\cdot \omega +\sd B_{X}\\
    &=\mathcal{C}_{X}[\Phi;\fd\Phi]- \omega[\Phi;\delta_X\Phi,\fd\Phi] +\extd B_{X}[\Phi;\fd\Phi].\\
\end{split}\end{equation}
Equation \eqref{eq: delta C general 2} is a key relation that we will manipulate.  When evaluated on solutions to the equations of motion, the $\mathcal{C}_X$ term vanishes, and for any field configuration that is symmetric under some gauge transformation $\hat{X}_s$ (meaning that $\hat{X}_s\fins\fd\Phi=\delta_{X_s}\Phi=0$ for all fields) then we also have $\omega[\Phi;\delta_{X_s}\Phi,\fd\Phi]=\omega[\Phi;0,\fd\Phi]=0$, leaving only the exact term $\sd B_{X_s}$.  If the field-space variations satisfy the linearized equations of motion, then $\extd B_{X_s}=0$.  In the context of diffeomorphism-covariant Lagrangians, Wald \cite{Wald:1993nt} used such an identity to prove the first law for perturbations around stationary black holes.  For our purposes, it is important that even for \textit{arbitrary} variations (around solutions where $\hat{X_s}\fins \fd\Phi=0$), $\fd C_{X_S}$ is always spacetime exact.  This fact plays an important role in our analysis of the perturbative constraints in the next subsection.

\subsection{Perturbative Constraints}
\label{subsec: perturbative constraints}
We are interested in a perturbative expansion of the gauge constraints\footnote{What we refer to as the gauge constraint $C_X$, defined in equation \eqref{eq: decompose J general}, is more appropriately called a constraint \textit{density} from the point of view of the pre-phase space structure, because its integral on a Cauchy slice is what generates a gauge transformation.  We will only use the term ``density'' for $C_X$ when we wish to emphasize this distinction.} $C_X$ around a fixed background solution.  Let $\Phi_\varepsilon$ represent some one-parameter smooth curve of field-space configurations parametrized by $\varepsilon$, and let $\hat{V}_\varepsilon$ be a field-space vector flowing tangential to this curve:
\begin{equation}\label{eq: define V}
\hat{V}_\varepsilon:=\frac{\extd \Phi^{a}_\varepsilon}{\extd \varepsilon}\frac{\delta}{\delta \Phi^a}.
\end{equation}
Field-space Lie derivatives along $\hat{V}_\varepsilon$ evaluated at $\varepsilon=0$ can be used to establish the terms of a perturbative expansion in $\varepsilon$. 
For the fundamental fields we write 
\begin{equation}\label{eq:fundamentalFieldExpansion}
\begin{split}
    &\Phi^{(n)}:=\left(\fL^{n}_{\hat{V}_\varepsilon}\Phi_\varepsilon\right)\big|_{\varepsilon=0},\\
    &\Phi_\varepsilon = \Phi^{(0)}+\varepsilon \Phi^{(1)}+\frac{1}{2} \eps^2\Phi^{(2)}+O(\varepsilon^3).
\end{split}\end{equation}
For $n>0$, the field perturbations are taken to be arbitrary (off-shell the $n$th order equation of motion) unless explicitly stated below, but we stipulate that the $\Phi^{(0)}$ configurations always satisfy the equations of motion, constituting the perturbative background solution:
\begin{equation}\begin{split}\label{eq: background EOM}
    E[\Phi^{(0)};\delta\Phi]=0.\\
\end{split}\end{equation}
For any field-space $0$-form $\mathcal{F}$ we write $\mathcal{F}:=\mathcal{F}[\Phi_\varepsilon,\varepsilon]$ and
\begin{equation}\label{eq:fieldExpansion}
\begin{split}
    &\mathcal{F}^{(n)}:=\left(\frac{\sd}{\sd\varepsilon}\mathcal{F}[\Phi_\varepsilon,\varepsilon]\right)\big|_{\varepsilon=0}=\left(\fL_{\hat{V}_\varepsilon}+\frac{\partial}{\partial\varepsilon}\right)^{n}\mathcal{F}[\Phi_\varepsilon,\varepsilon]\big|_{\varepsilon=0},\\
    &\mathcal{F} = \mathcal{F}^{(0)}+\varepsilon \mathcal{F}^{(1)}+\frac{1}{2}\eps^2\mathcal{F}^{(2)}+O(\varepsilon^3).
\end{split}\end{equation}
We have here allowed for explicit dependence on $\varepsilon$ in addition to the implicit dependence through $\Phi_\varepsilon$. For example, this will occur in general relativity coupled to matter on a vacuum background, where $\kappa$ will assume the role of $\varepsilon$. To lighten notation, we will hereafter leave off the subscript $\varepsilon$ on both $\hat{V}_\varepsilon$ and $\Phi_\varepsilon$.  It is important to note in what follows that while $\Phi$ will implicitly depend on $\varepsilon$ along some configuration-space curve as described above, the $\Phi^{(n)}$ for all integers $n\geq0$ do \textit{not} depend on $\varepsilon$, even implicitly.  In the case that the constraint carries no explicit dependence on $\varepsilon$, with the help of equation \eqref{eq: delta C general 2} we can write 
\begin{equation}\begin{split}\label{eq: first order constraint, general}
     C_{X}^{(1)}&:=\left(\fL_{\hat{V}_\varepsilon}C_{X}\right)\big|_{\varepsilon=0}
     =(\hat{V}_\varepsilon\fins\fd C_{X})\big|_{\varepsilon=0}
     = \left(-\omega\left[\Phi;\delta_X \Phi,\frac{\extd \Phi}{\extd \varepsilon}\right]+ \sd B_{X}\left[\Phi; \frac{\extd \Phi}{\extd \varepsilon}\right]\right)\bigg|_{\varepsilon=0}\\
     &= -\omega\left[\Phi^{(0)};\delta_X\Phi^{(0)}, \Phi^{(1)}\right]+ \sd B_{X}\left[\Phi^{(0)}; \Phi^{(1)}\right].
\end{split}\end{equation}
$C_{X}^{(1)}$ is the constraint density associated with $X$ in the linearized theory.
For the second-order constraint, we have\footnote{Here we take advantage of the linearity of $\omega$ and $\extd B$ in their latter arguments (those after the semicolons, in the notation of  \eqref{eq: functional dependence notation}) to write
\begin{equation*}\begin{split}\label{eq: useful identity}
    \left(\frac{\extd}{\extd\varepsilon}\mathcal{F}[A_1...,A_n;B_1,...,B_m]\right)\bigg|_{\varepsilon=0}
    &=
    \left(\frac{\extd}{\extd\varepsilon}\mathcal{F}[A_1...,A_n;B^{(0)}_1,...,B^{(0)}_m]\right)\big|_{\varepsilon=0}\\
    &+\sum_{r=1}^{m}\mathcal{F}\left[A_1...,A_n;B^{(0)}_1,...,B^{(1)}_r,...,B^{(0)}_m\right].
\end{split}\end{equation*}
Recall that $B^{(n)}:=\frac{\extd^n B}{\extd\varepsilon^n}\big|_{\varepsilon=0}$ for any field-space $0$-form $B$, and so $\frac{\extd}{\extd\varepsilon}(B^{(n)})=0$ for all non-negative integers $n$.}
\begin{equation}\begin{split}\label{eq: second order constraint, general}
     &C_X^{(2)}:=(\fL^2_{\hat{V}_\varepsilon}C_{X})|_{\varepsilon=0} 
      =\left(\frac{\extd}{\extd\varepsilon}\left(\mathcal{C}_X\left[\Phi;\frac{\extd\Phi}{\extd\varepsilon}\right]-\omega\left[\Phi;\delta_X\Phi,\frac{\extd\Phi}{\extd \varepsilon}\right]
      +\sd B_X\left[\Phi;\frac{\extd\Phi}{\extd\varepsilon}\right]\right)\right)\bigg|_{\varepsilon=0}\\
      =&
      \left(\frac{\extd}{\extd\varepsilon}\left(\mathcal{C}_X\left[\Phi;\Phi^{(1)}\right]
      - \omega\left[\Phi;\delta_X\Phi^{(0)},\Phi^{(1)}\right]
      +\sd B_X\left[\Phi;\Phi^{(1)}\right]
      \right)\right)\bigg|_{\varepsilon=0}
      -\omega\left[\Phi^{(0)};(\delta_X\Phi)^{(1)},\Phi^{(1)}\right]\\
      &~ -\omega\left[\Phi^{(0)};\delta_X\Phi^{(0)},\Phi^{(2)}\right]
      +\sd B_X\left[\Phi^{(0)};\Phi^{(2)}\right].
\end{split}\end{equation}
The last line isolates the dependence on the second-order perturbations $\Phi^{(2)}$.  These terms appear in the same form as the first-order constraint evaluated on $\Phi^{(2)}$ instead of $\Phi^{(1)}$.  This pattern repeats itself at higher orders; each `new' (highest-order) perturbation enters the constraint in precisely this form, which we codify by defining 
\begin{equation}\label{eq: universal piece}
    \tilde{C}_X[\Phi^{(0)},\mathcal{V}]:=\left(\mathcal{V}^a\frac{\delta}{\delta \Phi^a} \fins \delta C_X\right)\big|_{\varepsilon=0}
    =-\omega\left[\Phi^{(0)};\delta_X\Phi^{(0)},\mathcal{V}\right]
      +\sd B_X\left[\Phi^{(0)};\mathcal{V}\right]
\end{equation}
    and writing the $n$th-order constraint density as
\begin{equation}\begin{split}\label{eq: nth order constraint, general}
   C^{n}_X[\Phi^{(0)},...,\Phi^{(n)}]= \tilde{C}_X[\Phi^{(0)},\Phi^{(n)}] -\mathcal{F}^{(n)}_X[\Phi^{(0)},...,\Phi^{(n-1)}],
\end{split}\end{equation}
for some functional $\mathcal{F}^{(n)}_{X}$ of lower-order perturbations (with $\mathcal{F}_X^{(1)}=0$).  An `instability' occurs if at some order $n$ all preceding constraints can be satisfied by some set $\Phi^{(0)},...,\Phi^{(n-1)}$, but the $n$th-order constraint cannot be solved for any $\Phi^{(n)}$.  This would imply that additional restrictions must be placed on the solution space of the $(n-1)$-order theory to ensure that its solutions are integrable to solutions of the full theory.  

A simple condition that could ensure this never occurs (at any order) is if the operator $\tilde{C}_X[\Phi^{(0)}, \cdot]$ has trivial cokernel at all points, implying that the vanishing of \eqref{eq: nth order constraint, general} can always be solved for $\Phi^{(n)}$ given any source function $\mathcal{F}_X^{(n)}$.

At the opposite extreme, an instability occurs immediately (at 1st order) if 
$\tilde{C}_X[\Phi^{(0)}, \cdot]$ is a total derivative term, and if the spatial slices are closed (or if relevant fall-off conditions are imposed).  In such scenarios, the integral of the first-order constraint density on a Cauchy slice vanishes identically.  For any non-trivial $\mathcal{F}_X^{(2)}$, integrating the second-order constraint density on a Cauchy slice then results in a nontrivial restriction on the first-order solution space.  This is a linearization instability.

More generally the system might avoid either of the above two scenarios, but the question of whether the range of any $\mathcal{F}^{(n)}$ intersects the cokernel of $\tilde{C}_X[\Phi^{(0)}, \cdot]$ requires more detailed analysis, which we avoid.  However, the case of linearization instability is particularly relevant when considering perturbations around a background solution with a symmetry, as we now explain.

Suppose that the background solutions $\Phi^{(0)}$ are symmetric under some particular gauge transformation generated by $X_s$, meaning that $(\hat{X_s}\fins \delta\Phi)|_{\varepsilon=0}=\delta_{X_s}\Phi^{(0)}=0$ for all fields.   In this case, because $\omega[\Phi^{(0)};\delta_{X_s}\Phi^{(0)},\mathcal{V}]=\omega[\Phi^{(0)};0,\mathcal{V}]=0$, the first- and second-order constraint densities \eqref{eq: first order constraint, general} and \eqref{eq: second order constraint, general} become
\begin{align}
\label{eq: first order constraint, general symmetry}
     C_{X_s}^{(1)}&=\sd B_{X_s}\left[\Phi^{(0)}; \Phi^{(1)}\right],
\end{align}
and
\begin{align}
\label{eq: second order constraint, general symmetry}
     C_{X_s}^{(2)}&=
      \left(\frac{\extd}{\extd\varepsilon}\left(\mathcal{C}_{X_s}\left[\Phi;\Phi^{(1)}\right]
      +\sd B_{X_s}\left[\Phi;\Phi^{(1)}\right]
      \right)\right)\bigg|_{\varepsilon=0}
      -\omega\left[\Phi^{(0)};(\delta_{X_s}\Phi)^{(1)},\Phi^{(1)}\right]\\ \nonumber
      &~ +\sd B_{X_s}\left[\Phi^{(0)};\Phi^{(2)}\right].
\end{align}
Crucially, $\tilde{C}_{X_s}$ of \eqref{eq: universal piece} simplifies to an exact term, and we see that linearization instability will occur if the spatial slices are closed or if the solution space is restricted by boundary or fall-off conditions such that $B_{X_s}$ vanishes at the boundary.

We can also consider the case that the Lagrangian, and therefore constraints, depend explicitly on the perturbative parameter $\varepsilon$ (as in the case of gravity when the matter Lagrangian and gravitational Lagrangian differ by an explicit power of $\kappa^2$). To establish a perturbative expansion of the constraint, we assume the Lagrangian has been rescaled by a multiple of $\varepsilon$ such that the $\varepsilon\rightarrow 0$ limit gives a finite result.  The general first- and second-order constraint densities are shown for such scenarios in appendix \ref{App: perturbative constraints with explicit dependence}.  When perturbing around a background with symmetry $X_s$, the modified constraints are relatively simple:
\begin{align}
\label{eq: first order constraint, general symmetry, varepsilon}
     C_{X_s}^{(1)}&=
     \sd B_{X_s}\left[\Phi^{(0)},\varepsilon=0; \Phi^{(1)}\right]
     +\frac{\partial C_{X_s}}{\partial\varepsilon}\bigg|_{\varepsilon=0},
\end{align}
and
\begin{equation}\begin{split}
\label{eq: second order constraint, general symmetry, varepsilon}
     &C_{X_s}^{(2)}=
     +\left(\hat{V}_{X_s}\fins\fd
     \left(\mathcal{C}_{X_s}\left[\Phi,\varepsilon=0;\Phi^{(1)}\right]
      +\sd B_{X_s}\left[\Phi,\varepsilon=0;\Phi^{(1)}\right]\right)\right)\big|_{\varepsilon=0}
      -\omega\left[\Phi^{(0)},\varepsilon=0;(\delta_{X_s}\Phi)^{(1)},\Phi^{(1)}\right]\\
      &+2\frac{\partial}{\partial\varepsilon}\left(\mathcal{C}_{X_s}\left[\Phi^{(0)},\varepsilon;\Phi^{(1)}\right]+\sd B_{X_s}\left[\Phi^{(0)},\varepsilon;\Phi^{(1)}\right]\right)\big|_{\varepsilon=0}
       +\sd B_{X_s}\left[\Phi^{(0)},\varepsilon=0;\Phi^{(2)}\right]
       +\frac{\partial^2 C_{X_s}}{\partial\varepsilon^2}\bigg|_{\varepsilon=0}.
\end{split}\end{equation}
In principle, this explicit dependence on the perturbative parameter can alter the conclusion that the first-order constraint density is spacetime exact when associated with a background symmetry.  However, if the explicit dependence is at least quadratic (as in the case of gravity when there is an explicit $\kappa^2$ difference between the Einstein-Hilbert Lagrangian and matter Lagrangian), \eqref{eq: first order constraint, general symmetry, varepsilon} is again spacetime exact.  Regardless, the second-order fields $\Phi^{(2)}$ again enter the second-order constraint only in an exact term.\footnote{One should keep in mind that when there is explicit $\varepsilon$ dependence in the action, the order (in $\varepsilon$) at which field equations are imposed is no longer simply indicated by the superscript on the various field $\Phi^{(n)}$.  For instance the field equation for $g^{(1)}$ is at a lower order than the field equation for a scalar field $\phi^{(1)}$ when the latter enters only in a matter Lagrangian at $\kappa^2$ higher order than the gravity Lagrangian.}

\section{Diffeomorphism-covariant Lagrangians}

We will now apply this formalism to generally covariant theories. Our perturbative expansions follow in the same spirit as Hollands and Wald \cite{Hollands:2012sf} in the context of black hole and black brane stability in vacuum general relativity, though it should be noted that they were considering notions of dynamical and thermodynamic stability, neither of which is identical to the notion of linearization instability we are considering here. Our discussion is also more general by covering arbitrary generally covariant theories with or without matter on vacuum or non-trivial matter backgrounds.

\subsection{Diffeomorphism constraints}\label{subsec: diffeomorphism covariant Lagrangians}

We now consider diffeomorphisms as a special instance of a gauge symmetry under the formalism just established (sections \ref{subsec: general gauge constraints} and \ref{subsec: perturbative constraints}); this subsection puts standard results into the context of our framework.  In this case, for any spacetime vector field $\xi$ we write the associated field-space vector as $\hat{X}=\hat{\xi}$ such that
\begin{equation}
    \hat{\xi}\fins \fd \Phi = \sL_\xi\Phi.
\end{equation}
We start with a diffeomorphism-covariant Lagrangian, which entails that 
\begin{equation}\begin{split}
    \hat{\xi}\fins\fd L &= \sd(\xi\sins L).\\
\end{split}\end{equation}
We can therefore take
\begin{equation}\begin{split}
    \alpha_\xi&:=\xi\sins L,
\end{split}\end{equation}
where $\alpha_{X}$ was defined for general gauge transformations in \eqref{eq: covariant L general}. Referring to equations \eqref{eq: define J general} and \eqref{eq: decompose J general}, we write the gauge current and its decomposition as 
\begin{equation}\begin{split}\label{eq: define J diffeos}
    J_\xi&:=\hat{\xi}\fins\theta-\xi\sins L
    =C_\xi+\sd Q_\xi.\\
\end{split}\end{equation}
Since we have not specified the Lagrangian (only that it is diffeomorphism covariant), we must likewise leave the form of $C_\xi$ and $Q_\xi$ unspecified for now, though we will give explicit examples (including tensorial matter) in App.~\ref{sect: EH with matter} (see \eqref{eq:resultsScalar} and \eqref{eq:resultsEM}). 
We next seek an analogous decomposition of the terms $\fL_{\hat{X}}\theta-\fd\alpha_{\hat{X}}$ (referring to equation \eqref{eq: define tilde C, tilde q general} for the general gauge case).
The covariance of $\theta$ implies that\footnote{The extra piece $\widehat{\delta \xi}\fins\theta$, which only appears if we allow for field-dependent spacetime vector fields $\xi$, is necessary because $\theta$ is a field-space 1-form.  By contrast, covariance of the Lagrangian (a field-space 0-form) is always the statement that $\fL_\xi L=\sL_\xi L$.}
\begin{equation}
    \fL_\xi \theta = \sL_\xi \theta + \widehat{\delta \xi}\fins\theta.
\end{equation}
This means we can write
\begin{equation}\begin{split}
\label{eq: define mathcal Cxi, Qxi}
    \fL_{\hat{\xi}}\theta-\delta\alpha_{\xi}
    &=\sL_{\xi}\theta +\widehat{\fd\xi}\fins\theta - \delta (\xi\sins L)\\
    &=\sd(\xi\sins\theta) +\widehat{\fd\xi}\fins\theta - \delta\xi\sins L-\xi\sins E\\
    &=\sd(\xi\sins\theta)+ J_{\fd\xi}-\xi\sins E\\
    &\implies\\
    \mathcal{C}_{\xi}&:=C_{\fd\xi}-\xi\sins E,
    \qquad
    \mathcal{Q}_{\xi}:=Q_{\fd\xi}+\xi\sins\theta.
\end{split}\end{equation}
In the second equality we used $\fd L = E + \sd \theta$ and Cartan's magic formula for the spacetime Lie derivative of $\theta$.
Thereafter, we used the definition and decomposition of $J_{\delta\xi}$ from \eqref{eq: define J diffeos} with $\delta\xi$ in place of $\xi$.  Defining
\begin{equation}\label{eq: boundary term diffeos}
    B_\xi:=\mathcal{Q}_\xi-\fd Q_\xi
    =\xi\sins\theta+Q_{\fd\xi}-\fd Q_\xi,
\end{equation}
the variation of the diffeomorphism constraint $C_\xi$ then decomposes according to
\begin{equation}\begin{split}\label{eq: delta C diffeos}
    \delta C_{\xi} 
    &=-\hat{\xi}\cdot \omega + \mathcal{C}_{\xi}+\extd B_\xi\\
    &=-\hat{\xi}\cdot \omega +C_{\fd\xi}-\xi\sins E 
    +\sd(\xi\sins\theta+Q_{\fd\xi}-\delta Q_\xi).
\end{split}\end{equation}
This is the form of equation \eqref{eq: delta C general 2} in the case of diffeomorphisms. Upon integrating on a (partial or complete) Cauchy slice $\Sigma$ this becomes
\begin{equation}\label{eq: integrated constraint}
    \fd \int_\Sigma C_\xi =\int_{\Sigma}\left(-\hat{\xi}\fins\omega +C_{\fd\xi}-\xi\sins E\right)
    +\int_{\partial\Sigma}\left(\xi\sins\theta+Q_{\delta\xi}-\fd Q_\xi\right).
\end{equation}
The role played by the boundary integral $\int_{\partial\Sigma}\xi\sins\theta$ requires some explanation.  As shown in \cite{Harlow:2019yfa}, in order to have a well-defined variational principle on regions with a (possibly asymptotic) boundary $\Gamma$, the total action might require a boundary term $\int_\Gamma \ell$.  The pullback of $\theta$ to $\Gamma$, when combined with $\fd \ell$, must then be spacetime exact up to a term which vanishes under imposition of the boundary conditions.  This entails the following decomposition of $\theta|_{\Gamma}$:
\begin{equation}\label{eq: theta decomposition}
    \theta|_\Gamma=-\fd\ell +\sd \beta -\mathcal{E},
\end{equation}
where $\mathcal{E}$ must vanish if boundary conditions are imposed.  When the boundary manifold $\partial\Sigma$ appearing in \eqref{eq: integrated constraint} is a cut of this boundary $\Gamma$, this decomposition can be employed to write\footnote{Here we assume that $\beta$ is `anomaly-free' along $\xi$, meaning that $\fL_\xi\beta = \sL_\xi \beta+\widehat{\fd\xi}\fins\beta$.}
\begin{equation}\label{eq: boundary integral decomposition}
    \int_{\partial\Sigma}\xi\sins \theta=
     \int_{\partial\Sigma}\left(-\fd(\xi\sins\ell-\hat{\xi}\fins\beta)+(\fd\xi\sins\ell-\widehat{\fd\xi}\fins\beta)+\hat{\xi}\fins\fd\beta-\xi\sins\mathcal{E}\right).
\end{equation}
Restricting to field-independent vector fields ($\delta\xi=0$), equation \eqref{eq: integrated constraint} then becomes
\begin{equation}\begin{split}\label{eq: Hamilton's equation}
    \fd \int_\Sigma C_\xi =
    -\hat{\xi}\fins\Omega_\Sigma
    -\fd \mathcal{H}_\xi^{\partial \Sigma}
    -\int_{\Sigma}\xi\sins E 
    -\int_{\partial\Sigma}\xi\sins\mathcal{E}\, ,
\end{split}\end{equation}
where
\begin{equation}\begin{split}\label{eq: full symplectic form}
    \Omega_\Sigma :=\int_\Sigma\omega - \int_{\partial\Sigma}\delta\beta,
\end{split}\end{equation}
is the presymplectic form of the theory, which can be shown to be independent of the Cauchy slice $\Sigma$ when evaluated on shell (in which case we may leave off the subscript $\Sigma$), and
\begin{equation}\begin{split}\label{eq: full charge}
    \mathcal{H}_\xi^{\partial \Sigma} :=\int_{\partial \Sigma}\left(Q_\xi+\xi\sins\ell-\hat{\xi}\fins\beta\right)
\end{split}\end{equation}
is the charge associated with the flow along $\xi$.  On shell, the equations of motion and boundary conditions, equation \eqref{eq: Hamilton's equation} shows that $\mathcal{H}_\xi$ generates this flow on the (pre)phase space.

\subsection{Perturbative constraints with matter-free background}

Suppose we want to develop a perturbative expansion of the diffeomorphism constraints where $\kappa=\sqrt{32\pi G_N}$ is treated as the perturbative parameter, taking the place of $\varepsilon$ in section \ref{subsec: perturbative constraints}.  We can use the formalism and notation of that section, except that in this case the Lagrangian, and the constraints themselves, depend \textit{explicitly} on the perturbative parameter. This means that the constraints at each order are obtained as\footnote{We always rescale the constraint by a power of $\kappa$ such that the $\kappa\rightarrow 0$ limit gives a finite contribution.}
\begin{equation}
    C^{(n)}_{\xi}:=\left[\left(\hat{V}\fins\fd + \frac{\partial}{\partial\kappa}\right)^n C_\xi\right]\big|_{\kappa=0}.
\end{equation}
We write expressions for the first- and second-order diffeomorphism constraints in this case of general dependence in App.~\ref{App: perturbative constraints with explicit dependence}.  For a standard gravity action at $O(\kappa^{-2})$, coupled to an $O(\kappa^{-0})$ matter Lagrangian, things are fairly straightforward, as we now explain by outlining how to recover the results reported for that case in Sec.~\ref{sec:gravitonsandmatter}.  After rescaling the Lagrangian by $\kappa^2$, the explicit $\kappa$ dependence in the constraints is now quadratic, and the first- and second-order constraints are then simply
\begin{equation}
    C^{(1)}_{\xi}:=\left(\hat{V}\fins\fd C_\xi\right)\big|_{\kappa=0}
\end{equation}
and 
\begin{equation}
    C^{(2)}_{\xi}:=
    \left[(\hat{V}\fins\fd)^2 C_\xi\right]\big|_{\kappa=0}+\left(\frac{\partial^2 C_\xi}{\partial\kappa^2}\right)\big|_{\kappa=0}.
\end{equation}
Evaluation at $\kappa=0$ drops any dependence on the matter fields from both $\hat{V}\fins\fd C_\xi$ and $(\hat{V}\fins\fd)^2 C_\xi$.  These terms can thus be evaluated by considering only the gravity action.  The first-order term is always exact when evaluated on shell of a background Killing symmetry $\sL_{\xi_s} g^{(0)}_{\mu\nu}=0$, as can be inferred from 
\eqref{eq: first order constraint, general symmetry}.

In the second-order constraint, the $(\hat{V}\fins\fd)^2C_\xi|_{\kappa=0}$ can also be evaluated by considering only the part of the constraint associated with the gravity action, whereas the only role of the matter is in the $\left(\frac{\partial^2 C_\xi}{\partial\kappa^2}\right)\big|_{\kappa=0}$ term of the second-order constraint.  This can then be rewritten as
\begin{equation}
    C^{(2)}_{\xi}:=
    \left[\frac{\sd^2}{\sd\kappa^2} C^{(\text{gravity})}_\xi\right]\big|_{\kappa=0}+\left[\frac{\partial^2}{\partial\kappa^2} C^{(\text{matter})}_\xi\right]\big|_{\kappa=0}.
\end{equation}
It is shown explicitly in appendix \ref{app_lalala} how these terms combine to recover the standard expressions we derived already in Sec.~\ref{sec:gravitonsandmatter}, such as Eq.~\eqref{2ndorder}, yielding:
\begin{equation}
    C^{(2)}_{\xi_s}=\left(4G^{\mu(2)}_{~\nu}(h,h)-2 T^{\mu(0)}_{~\nu}\right)\xi_s^\nu\epsilon^{(0)}_{\mu}
      +\sd B_{\xi_s,\text{EH}}\left[g^{(0)},\Phi^{(2)}\right].
\end{equation}

\subsection{Perturbative constraints with matter background}

Next, we derive the (in)stability conditions for gravitons and matter on a background with non-trivial matter sources, treating $\kappa$ again as a formal parameter (taking the place of $\varepsilon$ in section \ref{subsec: perturbative constraints}). That is, we revisit the setting of Sec.~\ref{sec_matter} and generalize it to arbitrary generally covariant theories using the covariant phase space formalism. This overcomes the challenges mentioned in Sec.~\ref{sec_matter} regarding the derivation of the statement that the first-order contribution of any diffeomorphism constraint associated with a Killing field is a pure boundary term, and hence vanishes identically in spatially closed spacetimes. This will assume that the isometries are also symmetries of the matter background.

We recall condition ({\color{red}$\square$}) in Sec.~\ref{sec_matter} on the rescaling of the matter fields, which now means that, aside from an overall $\kappa^{-2}$ prefactor, there is no further explicit $\kappa$ dependence inside the Lagrangian $L$ (encompassing all field species).
This means that the constraint can be rescaled so that its $\kappa\rightarrow 0$ contribution is finite and no explicit $\kappa$ dependence remains.  Retaining the notation of \eqref{eq:fieldExpansion} but for the moment allowing the background configuration $\Phi^{(0)}:=\Phi|_{\kappa=0}$ to be off-shell, we can return to the expression \eqref{eq: delta C diffeos} to write the first-order constraint density as
\begin{equation}\begin{split}\label{eq: first order constraint, diffeo}
     C_{\xi}^{(1)}
     &= -\xi\sins E\left[\Phi^{(0)};\Phi^{(1)}\right] -\omega\left[\Phi^{(0)};\sL_\xi\Phi^{(0)}, \Phi^{(1)}\right]+ \sd B_{\xi}\left[\Phi^{(0)}; \Phi^{(1)}\right]\,,
\end{split}\end{equation}
while the second-order constraint density is
\begin{equation}\begin{split}\label{eq: second order constraint, diffeo}
     C_\xi^{(2)}
      =&
      \left(\frac{\extd}{\extd\kappa}\left(-\xi\sins E\left[\Phi;\Phi^{(1)}\right]
      - \omega\left[\Phi;\sL_\xi\Phi^{(0)},\Phi^{(1)}\right]
      +\sd B_\xi\left[\Phi;\Phi^{(1)}\right]
      \right)\right)\bigg|_{\kappa=0}\\
      &~
      -\xi\sins E\left[\Phi^{(0)};\Phi^{(2)}\right]
      -\omega\left[\Phi^{(0)};\sL_\xi\Phi^{(1)},\Phi^{(1)}\right]
      -\omega\left[\Phi^{(0)};\sL_\xi\Phi^{(0)},\Phi^{(2)}\right]
      +\sd B_\xi\left[\Phi^{(0)};\Phi^{(2)}\right].
\end{split}\end{equation}
Evaluated on a background solution satisfying a symmetry $\sL_{\xi_s}\Phi^{(0)}=0$, these become
\begin{equation}\begin{split}\label{eq: first order constraint, diffeo with sym}
     C_{\xi_s}^{(1)}
     &= \sd B_{\xi_s}\left[\Phi^{(0)}; \Phi^{(1)}\right],
\end{split}\end{equation}
and
\begin{equation}\begin{split}\label{eq: second order constraint, diffeo with sym1}
     C_{\xi_s}^{(2)}
      =&
      -\xi_s\sins\left(E^{(1)}_a \Phi^{a(1)}\right)
      -\omega\left[\Phi^{(0)};\sL_{\xi_s}\Phi^{(1)},\Phi^{(1)}\right]
      +\left(\frac{\extd}{\extd\kappa}\sd B_{\xi_s}\left[\Phi;\Phi^{(1)}\right]
      \right)\bigg|_{\kappa=0}
      +\sd B_{\xi_s}\left[\Phi^{(0)};\Phi^{(2)}\right].
\end{split}\end{equation}
In this form it is easy to see that when integrated on a closed slice, the second-order constraint \eqref{eq: second order constraint, diffeo with sym1} may result in a non-trivial condition on first-order solutions $\Phi^{(1)}$, and a linearization instability. This produces a result analogous to Eqs.~\eqref{2ndorder} and \eqref{2ndorderconstraint}, however, now for arbitrary generally covariant theories on a possibly matter-sourced background, thus in particular encompassing the situation of Sec.~\ref{sec_matter} and resolving the challenges mentioned there.

For purposes of relating to the cross product construction, it is perhaps more informative to work directly with the integrated form \eqref{eq: Hamilton's equation} for a perturbative expansion.  Defining an expansion of the integrated constraint via
\begin{equation}\label{eq: perturbative integrated constraints}
C^{(n)}_{\xi,\Sigma}:=\left(\frac{\sd^n}{\sd\kappa^n}\int_\Sigma C_\xi\right)\big|_{\kappa=0}\,,
\end{equation}
we can write\footnote{Note that in these expressions, the boundary terms on $\partial\Sigma$ may contain multiple contributions if the boundary is disconnected, as will be the case in many important cases (e.g.\ two asymptotic boundaries, or one boundary and a bifurcate Killing horizon).} (off-shell at all orders, for generality)
\begin{equation}\begin{split}\label{eq: first order integrated}
    C_{\xi,\Sigma}^{(1)}
    =&-\Omega_{\Sigma}\left[\Phi^{(0)};\sL_\xi\Phi^{(0)},\Phi^{(1)}\right]-(\delta\mathcal{H}_{\xi,\partial\Sigma})\left[\Phi^{(0)};\Phi^{(1)}\right]\\
    &-\int_\Sigma \xi\sins E\left[\Phi^{(0)};\Phi^{(1)}\right]
    -\int_{\partial\Sigma}\xi\sins\mathcal{E}\left[\Phi^{(0)};\Phi^{(1)}\right],
\end{split}\end{equation}
and
\begin{equation}\begin{split}\label{eq: first C2 expression}
    &C^{(2)}_{\xi,\Sigma}
    =-\frac{\sd}{\sd\kappa}(\delta\mathcal{H}_{\xi,\partial\Sigma})\left[\Phi;\Phi^{(1)}\right]\big|_{\kappa=0}-(\delta\mathcal{H}_{\xi,\partial\Sigma})\left[\Phi;\Phi^{(2)}\right]\\
    &-\frac{\sd}{\sd\kappa}\Omega_{\Sigma}\left[\Phi;\sL_\xi\Phi^{(0)},\Phi^{(1)}\right]
    \big|_{\kappa=0}
    -\Omega_{\Sigma}\left[\Phi^{(0)};\sL_\xi\Phi^{(1)},\Phi^{(1)}\right]
    -\Omega_{\Sigma}\left[\Phi^{(0)};\sL_\xi\Phi^{(0)},\Phi^{(2)}\right]\\
    &-\int_\Sigma \xi\sins \frac{\sd}{\sd\kappa}E\left[\Phi;\Phi^{(1)}\right]\big|_{\kappa=0}
    -\int_{\partial\Sigma} \xi\sins\frac{\sd}{\sd\kappa}\mathcal{E} \left[\Phi;\Phi^{(1)}\right]\big|_{\kappa=0}\\
    &-\int_\Sigma \xi\sins E\left[\Phi^{(0)};\Phi^{(2)}\right]
    -\int_{\partial\Sigma} \xi\sins\mathcal{E} \left[\Phi^{(0)};\Phi^{(2)}\right].
\end{split}\end{equation}
If we work on-shell of the equations of motion at the preceding  order for each constraint and boundary conditions at the current order, the last line of \eqref{eq: first order integrated} and the last two lines of \eqref{eq: first C2 expression} can be dropped, and we do this henceforth. For the boundary conditions this means that we impose the same type of boundary conditions at all orders, in line with the fact that the entire exact solution space is defined via some fixed set of boundary conditions.

The first two terms of \eqref{eq: first C2 expression} may be grouped as (minus) the full second-order charge variation $\frac{\sd^2}{\sd\kappa^2}\mathcal{H}_{\xi,\Sigma}\big|_{\kappa=0}$, including terms both quadratic in $\Phi^{(1)}$ and linear in $\Phi^{(2)}$. Alternatively, we may choose to group terms quadratic in $\Phi^{(1)}$ with the other terms and isolate the boundary charge variation $(\delta\mathcal{H}_{\xi,\partial\Sigma})\left[\Phi;\Phi^{(2)}\right]$.  This is natural from the following perspective. First define
\begin{equation}\label{eq: C^[1]}
    C^{[1]}_{\xi,\Sigma}\left[\Phi;\frac{\sd\Phi}{\sd\kappa}\right]:=\frac{\sd}{\sd\kappa}\int_\Sigma C_\xi=\hat{V}\fins \fd\int_\Sigma C_\xi.
\end{equation}
Here, the superscript square bracket reflects the fact that this is just the first-order variation from \eqref{eq: perturbative integrated constraints} prior to evaluating at $\kappa=0$: $C^{(1)}_{\xi,\Sigma}=C^{[1]}_{\xi,\Sigma}\left[\Phi;\frac{\sd\Phi}{\sd\kappa}\right]\big|_{\kappa=0}=C^{[1]}_{\xi,\Sigma}\left[\Phi^{(0)};\Phi^{(1)}\right]$. The second-order integrated constraint can then be expressed as
\begin{equation}\begin{split}\label{eq: alternate C2 expression}
    C^{(2)}_{\xi;\Sigma}&=\frac{\sd}{\sd\kappa}C^{[1]}_{\xi;\Sigma}\left[\Phi;\Phi^{(1)}\right]\big|_{\kappa=0}+
    C^{[1]}_{\xi;\Sigma}\left[\Phi^{(0)};\Phi^{(2)}\right]\\
    &=\frac{\sd}{\sd\kappa}C^{[1]}_{\xi;\Sigma}\left[\Phi;\Phi^{(1)}\right]\big|_{\kappa=0}
    -\Omega_{\Sigma}\left[\Phi^{(0)};\sL_\xi\Phi^{(0)},\Phi^{(2)}\right]-(\delta\mathcal{H}_{\xi,\partial\Sigma})\left[\Phi^{(0)};\Phi^{(2)}\right]\\
    &\qquad\qquad-\int_\Sigma \xi\sins E\left[\Phi^{(0)};\Phi^{(2)}\right]
    -\int_{\partial\Sigma}\xi\sins\mathcal{E}\left[\Phi^{(0)};\Phi^{(2)}\right].
\end{split}\end{equation}
In the second equality we used the fact that $C^{[1]}_{\xi,\Sigma}\left[\Phi^{(0)};\Phi^{(2)}\right]$ is just the first-order constraint evaluated on second-order fields $\Phi^{(2)}$ in place of $\Phi^{(1)}$, the form of which can be taken from equation \eqref{eq: first order integrated}.
Comparing equations \eqref{eq: first C2 expression} and \eqref{eq: alternate C2 expression} fully on-shell leaves us with two ways to express (the vanishing of) the second-order constraint:
\begin{equation}\label{eq: vanishing C2, v1}
    \frac{\sd^2}{\sd\kappa^2}\mathcal{H}_{\xi_s,\partial\Sigma}\big|_{\kappa=0} = -\Omega\left[\Phi^{(0)};\sL_{\xi_s}\Phi^{(1)},\Phi^{(1)}\right]\,,
\end{equation}
or
\begin{equation}\label{eq: vanishing C2, v2}
    (\fd\mathcal{H}_{\xi_s,\partial\Sigma})\left[\Phi^{(0)};\Phi^{(2)}\right]=\frac{\sd}{\sd\kappa}C^{[1]}_{\xi_s,\Sigma}\left[\Phi;\Phi^{(1)}\right]\big|_{\kappa=0}\,.
\end{equation}
The charge variations on the left hand side of these expressions differ in that the first is the full second-order variation of the boundary charge (containing terms quadratic in first-order perturbations as well as linear in second-order perturbations), while the second is only the variation depending on second-order fluctuations.   The full content of these two equalities is the same, of course; it is merely a matter of moving quadratic charge variations between the left and the right hand side.\footnote{This is confirmed by looking at the off-shell difference between the two right hand sides:  
\begin{equation}\begin{split}
    &\frac{\sd}{\sd\kappa}C^{[1]}_{\xi;\Sigma}\left[\Phi;\Phi^{(1)}\right]\big|_{\kappa=0}
    =-\Omega_\Sigma\left[\Phi;\sL_\xi\Phi^{(1)},\Phi^{(1)}\right]
    \\
    &\quad-\frac{\sd}{\sd\kappa}\left(\Omega_\Sigma\left[\Phi;\sL_\xi\Phi^{(0)},\Phi^{(1)}\right]+\int_\Sigma \xi\sins E\left[\Phi;\Phi^{(1)}\right]+\delta \mathcal{H}_{\xi,\partial\Sigma}\left[\Phi;\Phi^{(1)}\right]-\int_{\partial\Sigma}\xi\sins \mathcal{E}\left[\Phi;\Phi^{(1)}\right]\right)\big|_{\kappa=0}.
\end{split}\end{equation}
On some boundaries and for some boundary conditions, the quadratic contribution to the charge variation may vanish identically on shell, in which case there is literally no difference.} 
The expression \eqref{eq: vanishing C2, v1} is in some sense more natural in that  the \textit{full} charge variation does not care about the splitting between orders, which could be parametrized differently, while the right hand side is directly expressed in terms of symplectic flow.  The second expression clearly divides a boundary charge variation in $\Phi^{(2)}$ from a bulk integral that is quadratic in $\Phi^{(1)}$, and in some contexts provides an easy way to write down the form of that bulk integral explicitly. This discussion will become relevant when contemplating modular flow in such settings in Secs.~\ref{ssec_overall2ndorder} and~\ref{ssec_modularflowagain}. 

It is interesting to consider what happens when $\xi$ is \textit{not} a background symmetry. Then $\Phi^{(2)}$ also appears in a term $-\Omega_\Sigma\left[\Phi^{(0)};\sL_\xi\Phi^{(0)},\Phi^{(2)}\right]$, which generically ruins the known relationship between the bulk integral and a modular flow generator, and also the fact that the second-order fields only enter at the boundary.

Finally we briefly mention as an example scenario, a particularly crucial case which was treated in the context of establishing first- and second-order constraints in \cite{Hollands:2012sf} and in the context of careful crossed product constructions in \cite{Kudler-Flam:2023qfl}: take $\Sigma$ to be a partial Cauchy surface extending from a bifurcate Killing horizon generated by $\xi_s$, to asymptotic spatial infinity.  In such scenarios, when $\xi_s$ is the horizon-generating Killing vector, the contribution to $\mathcal{H}^{\partial\Sigma}_{\xi_s}$ at spatial infinity is a linear combination of the ADM mass $M$ and black hole  angular momentum $J$.  The contribution at the bifurcate horizon (where $\xi_s=0$) is simply proportional to the black hole area $A$ (or in higher derivative gravity the ``Wald entropy'' \cite{Wald:1993nt}).  The first-order constraint becomes an expression of the first law of black hole mechanics, while the second-order constraint can be expressed either as
\begin{equation}\label{eq: BH first law second order 1}
    \Omega[\Phi^{(0)}; \sL_{\xi_s}\Phi^{(1)},\Phi^{(1)}]
    =\frac{\sd^2 M}{\sd\kappa^2}-\Omega_H\frac{\sd^2 J}{\sd\kappa^2} - 4 K \frac{\sd^2 A}{\sd\kappa^2},
\end{equation}
or, equivalently\footnote{See discussion below equation \eqref{eq: vanishing C2, v2} for discussion of the difference between these charge variations.}
\begin{equation}\label{eq: BH first law second order 2}
    \frac{\sd}{\sd\kappa}C^{[1]}_{\xi_s,\Sigma}\left[\Phi;\Phi^{(1)}\right]\big|_{\kappa=0}
    =M^{(2)}-\Omega_H J^{(2)} - 4 K A^{(2)}\,.
\end{equation}

\subsection{Boundary preserving Killing symmetries}
\label{sec: boundrary preserving}

Let us next consider a variety of forms that the boundary charges can take. Under general conditions on the form of the Lagrangian,~\cite{Iyer:1994ys} showed (see also~\cite{Fursaev:1998hr}) that the charge $Q_\xi$ may be written in the form
\begin{equation}
    Q_\xi = \xi_\mu W^\mu[\Phi] - \fdv{L}{R_{\mu\nu\rho\sigma}}\epsilon_{\mu\nu}\nabla_\rho\xi_\sigma,
\end{equation}
for some $W(\Phi)$ that depends locally on the fields $\Phi$ and their derivatives. Here $L=L_{\text{EH}}+L_{\text{M}}$ is the total Lagrangian, which one may always write in the form
\begin{equation}
    L = L[g_{\mu\nu}, \nabla_\mu R_{\nu\rho\sigma\tau}, \dots,\nabla_{(\mu}\dots\nabla_{\nu)}R_{\rho\sigma\tau\epsilon},\phi,\nabla_\mu\phi,\dots,\nabla_{(\mu}\dots\nabla_{\nu)}\phi],
\end{equation}
i.e.\ as a local function of the metric $g$, the matter fields $\phi$, the Riemann tensor $R_{\mu\nu\rho\sigma}$, and the \emph{symmetrized} covariant derivatives of these objects. One can then take a kind of functional derivative of $L$ with respect to $R_{\mu\nu\rho\sigma}$, as if it were independent of $g$ and $\phi$; this defines
\begin{equation}
    \fdv{L}{R_{\mu\nu\rho\sigma}} = \pdv{L}{R_{\mu\nu\rho\sigma}}-\nabla_\tau\pdv{L}{\nabla_\tau R_{\mu\nu\rho\sigma}} + \dots + (-1)^m \nabla_{(\tau_1}\dots\nabla_{\tau_m)}\pdv{L}{\nabla_{(\tau_1}\dots\nabla_{\tau_m)}R_{\mu\nu\rho\sigma}}.
\end{equation}
For pure gravity, we have $\fdv{L}{R_{\mu\nu\rho\sigma}}= \frac2{\kappa^2}g^{\mu\rho}g^{\nu\sigma}$ and $W^\mu(\Phi)=0$, which reproduces~\eqref{eq: gravitational charge contribution}. A minimally coupled scalar field does not change this, as shown above. In Einstein-Maxwell theory we also have $\fdv{L}{R_{\mu\nu\rho\sigma}}= \frac2{\kappa^2}g^{\mu\rho}g^{\nu\sigma}$, but now we have $W^\mu(\phi)= -\frac12 F^{\alpha\beta}A^\mu\epsilon_{\alpha\beta}$ (as in~\eqref{eq:resultsEM}).

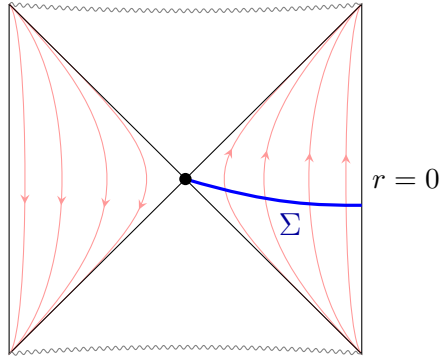
\begin{figure}
    \centering
    \begin{tikzpicture}[scale=1.16]
        \begin{scope}[red!40!white]
            \foreach \i in {0.3,1,1.85,2.6} {
                \draw[postaction={decorate,decoration={markings,mark=at position .57 with {\arrow[scale=1.1]{stealth}}}}] (2,2) .. controls ({2+0.8*\i},{2-0.8*\i}) and ({2+0.8*\i},{0.8*\i-2}) .. (2,-2);

                \draw[postaction={decorate,decoration={markings,mark=at position .57 with {\arrow[scale=1.1]{stealth}}}}] (6,-2) .. controls ({6-0.8*\i},{0.8*\i-2}) and ({6-0.8*\i},{2-0.8*\i}) .. (6,2);
            }
        \end{scope}
        \draw (2,2) -- (6,-2) -- (6,2) -- (2,-2) -- (2,2);
        \draw (2,2) -- (6,2);
        \draw (2,-2) -- (6,-2);
        \draw[very thick, blue] (4,0) .. controls (5,-0.3) and (5.5,-0.3) .. (6,-0.3) node[midway,below,blue!60!black] {\large$\Sigma$};
        \fill (4,0) circle (0.07);
        \node[right] at (6,0) {$r=0$};
    \end{tikzpicture}
    \caption{The boost in de Sitter spacetime preserves the boundary of any of the static patch's Cauchy surfaces $\Sigma$.}
    \label{Figure: static patch boost}
\end{figure}

If $\xi$ vanishes on $\partial\Sigma$, then $W$ does not contribute there, so we have
\begin{equation}
    \mathcal{H}_\xi = A_\xi - \int \hat{\xi}\cdot\beta,
\end{equation}
where
\begin{equation}
    A_\xi = \int_{\partial\Sigma} Q_\xi = \int_{\partial\Sigma}\fdv{L}{R_{\mu\nu\rho\sigma}}\epsilon_{\mu\nu}\nabla_\rho\xi_\sigma.
\end{equation}
This happens for example on the boundary of the partial Cauchy surface whose domain of dependence is a de Sitter static patch, as shown in Fig.~\ref{Figure: static patch boost}.

\subsubsection{Minimal vs non-minimal coupling, and higher-curvature theories}

Let us suppose $\xi$ is a constant surface gravity boost (see~\cite{Jensen:2023yxy}), and $\fdv{L}{R_{\mu\nu\rho\sigma}}=\frac2{\kappa^2}g^{\mu\rho}g^{\nu\sigma}$, as in the case of pure gravity and Einstein-Maxwell theory. Then $A_\xi$ is proportional to the area of $\partial\Sigma$. If we are in pure higher curvature gravity, $\fdv{L}{R_{\mu\nu\rho\sigma}}$ will contain extra contributions from the higher curvature terms in the gravitational action, so $A_\xi$ will not be proportional to the area but some other geometrical quantity. Let us focus for now on Einstein gravity.

For the case of a minimally coupled scalar field, and Einstein-Maxwell gravity, $\fdv{L}{R_{\mu\nu\rho\sigma}}$ is the same as in the pure gravity case, so $A$ is the area. It is reasonable to ask whether this is always the case. In fact, the answer is yes, \emph{so long as} the matter is minimally coupled. Let us be precise about what it means for matter to be minimally coupled to gravity. The Lagrangian of matter on a fixed background geometry may be written as a function of the matter fields and their partial derivatives:
\begin{equation}
    L_{\text{M}} = L_{\text{M}}[\phi,\nabla^{(0)}_\mu\phi,\dots,\nabla^{(0)}_{(\mu}\dots\nabla^{(0)}_{\nu)}\phi],
\end{equation}
where $\nabla^{(0)}$ is the background covariant derivative. Minimal coupling to gravity follows from replacing the symmetrized background covariant derivatives by symmetrized covariant derivatives of the dynamical metric $g$:
\begin{equation}
    \nabla^{(0)}_{(\mu}\dots\nabla^{(0)}_{\nu)}\phi \to \nabla_{(\mu}\dots\nabla_{\nu)}\phi,
\end{equation}
and also using the volume form of $g$ rather than the background volume form. With this \emph{minimal} coupling, one observes that the functional derivative of the matter Lagrangian with respect to $R_{\mu\nu\rho\sigma}$ vanishes, so we have
\begin{equation}
    A_\xi = -\int_{\partial\Sigma} Q_\xi = \int_{\partial\Sigma}\fdv{L_\text{EH}}{R_{\mu\nu\rho\sigma}}\epsilon_{\mu\nu}\nabla_\rho\xi_\sigma.
\end{equation}
Thus, $A_\xi$ is the same as in pure Einstein-Hilbert gravity, when matter is minimally coupled.

By contrast, when matter is not minimally coupled, $A_\xi$ will in general have extra contributions. For example, consider a scalar field conformally coupled to gravity, with Lagrangian
\begin{equation}
    L_{\text{M}} = \qty(-\frac12\nabla_\mu\phi\nabla^\mu\phi-\frac{D-2}{8(D-1)} R\phi^2)\epsilon.
\end{equation}
Then we have
\begin{equation}
    \fdv{L_\text{M}}{R_{\mu\nu\rho\sigma}} = - \frac{D-2}{8(D-1)}\phi^2 g^{\mu\rho} g^{\nu\sigma},
\end{equation}
so $A_\xi$ contains contributions from the scalar field on $\partial\Sigma$. 

\subsection{Stress tensors of the perturbative fields}

In the second-order constraint~\eqref{eq: alternate C2 expression}, some terms are integrals evaluated over the boundary $\partial\Sigma$, while by contrast the term $-\Omega_\Sigma[\Phi^{(0)};\lie_{\xi_s}\Phi^{(1)},\Phi^{(1)}]$ involves a bulk integral over $\Sigma$. This term is the canonical generator of the Killing symmetry $\xi_s$ in the linearized phase space.

For various reasons it can be useful to write this in terms of the integral of a stress tensor, as is often done in field theories. There is not a unique way to do this~\cite{Iyer:1994ys,Fursaev:1998hr}. One possibility involves the `gravitational' stress tensor $T_{\mu\nu}$, defined from the metric equations of motion via
\begin{equation}
    T^{\mu\nu} = 2E^{\mu\nu}_{(g)}.
\end{equation}
Another possibility, which we shall give a bit more detail on now, is what is known as a `canonical stress tensor'. The form $\omega[\Phi^{(0)};\lie_{\xi_s}\Phi^{(1)},\Phi^{(1)}]$ is linear in $\xi_s$ and its derivatives. Since we are assuming that $\xi_s$ is a Killing vector of the background metric, its components obey the first order differential equation
\begin{equation}
    \lie_\xi g^{(0)}_{\mu \nu} = \nabla^{(0)}_\mu\xi_{s\,\nu} + \nabla^{(0)}_\nu\xi_{s\,\mu} = 0,
\end{equation}
where $\xi_{s\,\mu} = g_{\mu \nu}^{(0)}\xi_s^\nu$. Using this equation we can always rewrite any derivatives of $\xi_s$ normal to $\Sigma$ as linear combinations of $\xi_s$ and its derivatives tangent to $\Sigma$. Then, we can integrate 
\begin{equation}
    \int_\Sigma\omega[\Phi^{(0)};\lie_{\xi_s}\Phi^{(1)},\Phi^{(1)}]
\end{equation}
by parts, in order to write 
\begin{equation}
    -\Omega_\Sigma[\Phi^{(0)};\lie_{\xi_s}\Phi^{(1)},\Phi^{(1)}] = \int_{\Sigma} t_{\mu\nu}\xi_s^\mu \epsilon^\nu + \int_{\partial\Sigma} k_{\xi_s}
\end{equation}
for some $t_{\mu\nu}$, which is the canonical stress tensor, and some $k_\xi$, which is a boundary term resulting from the integration by parts.

\subsection{The overall second-order constraint as a sum of energies}\label{ssec_overall2ndorder}

We will now write in a physically instructive form the overall second-order constraint~\eqref{eq: alternate C2 expression} associated with a Killing symmetry $\xi_s$ of the background configuration.

In~\eqref{eq: alternate C2 expression} there exist various boundary terms; there is one for each connected component of the boundary. Let us label the different such components by $a=1,2,\dots$, and use $H_a$ to denote the sum of the boundary terms with support on component $a$. If $a$ is fixed by the Killing symmetry, then the corresponding boundary term $H_a$ will simplify as explained in Sec.~\ref{sec: boundrary preserving}. If on the other hand $a$ is not fixed, $H_a$ will take a more complicated form -- but such components of the boundary are typically asymptotic, and global boundary conditions will be imposed at such components, thus simplifying $H_a$ in a separate, theory-dependent way. The partial Cauchy surface for a general subregion will have both kinds of components. For example, partial Cauchy surfaces for one side of the Schwarzschild black hole have one boundary component at the bifurcate horizon, which is fixed by the horizon generating Killing symmetry, and another  component at spacelike infinity, which is not. In general relativity, $H_a$ at the bifurcate horizon will be proportional to its area, and $H_a$ at spacelike infinity will be the ADM Hamiltonian.

Suppose some clocks exist in the interior of the subregion as described in Sec.~\ref{ssec_modflow}, which are only very weakly interacting with the fields, such that we can decompose
\begin{equation}
    -\Omega_\Sigma[\Phi^{(0)};\lie_{\xi_s}\Phi^{(1)},\Phi^{(1)}] = -\tilde\Omega_\Sigma[\Phi^{(0)};\lie_{\xi_s}\Phi^{(1)},\Phi^{(1)}]
    -\sum_i\Omega_i[\Phi^{(0)};\lie_{\xi_s}\Phi^{(1)},\Phi^{(1)}],
\end{equation}
where $\tilde\Omega_\Sigma$ is the contribution of the matter fields and gravitons, and $\Omega_i$ is the contribution of clock $i$. If we write things in terms of a stress tensor $T_{\mu\nu}$, then this corresponds to the decomposition
\begin{equation}
    T_{\mu\nu} = \tilde{T}_{\mu\nu} + \sum_i T^i_{\mu\nu},
\end{equation}
where $T^i_{\mu\nu}$ is the stress tensor of clock $i$ (localized at its worldline if we use the action~\eqref{eq: clock worldline action}), and $\tilde{T}_{\mu\nu}$ is the stress tensor of the rest of the matter fields and gravitons. In any case, one may write the bulk integral contribution to the second-order constraint as $\tilde{H}+\sum_i H_i$, where $\tilde{H}$ is the contribution of the matter/gravitons, and $H_i$ is the contribution of clock $i$.

Thus, one may write the total second-order constraint as a sum of energies
\begin{equation}
    \int_\Sigma C_{\xi_s}^{(2)} = \tilde{H} + \sum_i H_i + \sum_a H_a.
    \label{eq: energy decomposition}
\end{equation}
We may view each $H_a$ as the energy of an additional clock associated with the boundary component $a$. Because $H_a$ is formed using the second-order fields (unlike the other contributions to the constraint), the boundary clocks are effectively independent of the linearized degrees of freedom. So this constraint can really be viewed as the sum of the field energy $\tilde{H}$ with the energies of some clocks labeled by $i$ and $a$.

Note that there is some ambiguity in what we have so far said, because we have not specified which boundary terms are to be included in $\tilde{H}$, and which boundary terms go into the boundary energies $H_a$. One would get different decompositions, depending on how one chooses to define $\tilde{H}$. This is exactly analogous to what happened with Eqs.~\eqref{eq: vanishing C2, v1} and~\eqref{eq: vanishing C2, v2}; neither of these are \emph{a priori} better expressions of the second-order constraint, and both have their advantages. It is hard to say whether there is a \emph{correct} decomposition, in the general case.

For example, if one were to set it to be the canonical generator of $\xi_s$ on the fields, i.e.
\begin{equation}
    \tilde{H} = -\tilde\Omega_\Sigma[\Phi^{(0)};\lie_{\xi_s}\Phi^{(1)},\Phi^{(1)}],
\end{equation}
then each boundary energy would be given by
\begin{equation}
    H_a = -\frac{\sd}{\sd\kappa}(\fd \mathcal{H}^{\partial\Sigma_a}_{\xi_s})[\Phi;\Phi^{(1)}]|_{\kappa=0}
    -(\fd \mathcal{H}^{\partial\Sigma_a}_{\xi_s})[\Phi^{(0)};\Phi^{(2)}],
\end{equation}
where $\partial\Sigma_a$ is component $a$ of $\partial\Sigma$, and 
\begin{equation}\begin{split}\label{eq: full charge one boundary component}
    \mathcal{H}^{\partial\Sigma_a}_{\xi_s} :=\int_{\partial \Sigma_a}\left(Q_\xi+\xi\sins\ell-\hat{\xi}\fins\beta\right).
\end{split}\end{equation}
If, on the other hand, one chose for example to define $\tilde{H}$ in terms of the canonical stress energy:
\begin{equation}
    \tilde{H} = \int_\Sigma t_{\mu\nu}\xi_s^\mu \epsilon^\nu,
\end{equation}
one would have 
\begin{equation}
    H_a = -\frac{\sd}{\sd\kappa}(\fd \mathcal{H}^{\partial\Sigma_a}_{\xi_s})[\Phi;\Phi^{(1)}]|_{\kappa=0}
    -(\fd \mathcal{H}^{\partial\Sigma_a}_{\xi_s})[\Phi^{(0)};\Phi^{(2)}] + \int_{\partial\Sigma_a}k_{\xi_s}.
\end{equation}
In any case, the bulk term $\tilde H$ will only depend on the linear perturbations $\Phi^{(1)}$, whereas the boundary pieces $H_a$ will also include the second-order perturbations $\Phi^{(2)}$.

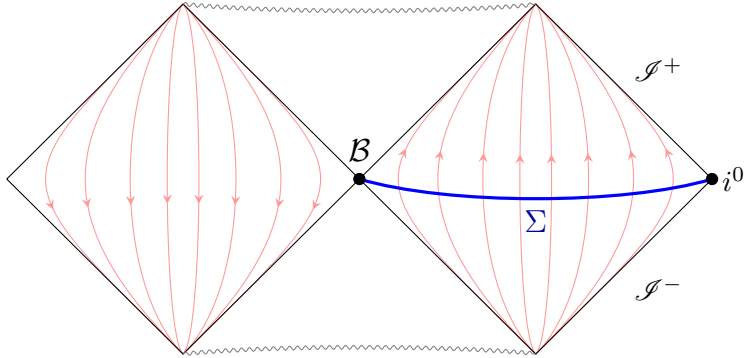
\begin{figure}
    \centering
    \begin{tikzpicture}[scale=1.16]
        \begin{scope}[red!40!white]
            \foreach \i in {0.3,1,1.85,2.6} {
                \draw[postaction={decorate,decoration={markings,mark=at position .57 with {\arrow[scale=1.1]{stealth}}}}] (2,2) .. controls ({2-0.8*\i},{2-0.8*\i}) and ({2-0.8*\i},{0.8*\i-2}) .. (2,-2);
                \draw[postaction={decorate,decoration={markings,mark=at position .57 with {\arrow[scale=1.1]{stealth}}}}] (2,2) .. controls ({2+0.8*\i},{2-0.8*\i}) and ({2+0.8*\i},{0.8*\i-2}) .. (2,-2);

                \draw[postaction={decorate,decoration={markings,mark=at position .57 with {\arrow[scale=1.1]{stealth}}}}] (6,-2) .. controls ({6-0.8*\i},{0.8*\i-2}) and ({6-0.8*\i},{2-0.8*\i}) .. (6,2);
                \draw[postaction={decorate,decoration={markings,mark=at position .57 with {\arrow[scale=1.1]{stealth}}}}] (6,-2) .. controls ({6+0.8*\i},{0.8*\i-2}) and ({6+0.8*\i},{2-0.8*\i}) .. (6,2);
            }
        \end{scope}
        \draw (0,0) -- (2,2) -- (6,-2) -- (8,0) -- (6,2) -- (2,-2) -- (0,0);
        \draw[decorate,decoration={snake,amplitude=0.7pt,segment length=3pt},gray] (2,2) .. controls (3,1.9) and (5,1.9) .. (6,2);
        \draw[decorate,decoration={snake,amplitude=0.7pt,segment length=3pt},gray] (2,-2) .. controls (3,-1.9) and (5,-1.9) .. (6,-2);
        \draw[very thick, blue] (4,0) .. controls (5,-0.3) and (7,-0.3) .. (8,0) node[midway,below,blue!60!black] {\large$\Sigma$};
        \fill (4,0) circle (0.07) node[above,shift={(0,0.1)}] {\large$\mathcal{B}$};
        \fill (8,0) circle (0.07) node[right] {$i^0$};
        \node[above right] at (7,1) {$\mathscr{I}^+$};
        \node[below right] at (7,-1) {$\mathscr{I}^-$};
    \end{tikzpicture}
    \caption{In the extended Schwarzschild spacetime, any Cauchy surface $\Sigma$ for one side is bounded by spacelike infinity $i^0$ and the bifurcate horizon $\mathcal{B}$. The bifurcate horizon is fixed by the action of the Killing symmetry, where it acts as a boost. But the action is non-trivial at $i^0$ (it only appears so on the Penrose diagram due to the conformal compactification).}
    \label{Figure: Schwarzschild boost}
\end{figure}

\subsection{Modular flow}\label{ssec_modularflowagain}

This sum of energies~\eqref{eq: energy decomposition} is the form of constraint used in~\cite{Chandrasekaran:2022cip,Jensen:2023yxy,Kudler-Flam:2023hkl,Kudler-Flam:2023qfl,Kudler-Flam:2024psh,DeVuyst:2024pop,DVEHK,Faulkner:2024gst,Aguilar-Gutierrez:2023odp,Chen:2024rpx,Gomez:2023upk,Gomez:2023wrq,Fewster:2024pur,Kolchmeyer:2024fly}. As described in Sec.~\ref{ssec_modflow}, when $\xi_s$ is appropriately chosen, $\tilde{H}$ may be identified with a modular Hamiltonian of the fields~\cite{Sewell:1982zz} (excluding the second-order perturbation $\Phi^{(2)}$). The algebra of operators invariant under this constraint then has the structure of the crossed product with respect to modular flow (or a subalgebra thereof in the case of a non-ideal clock), and such an algebra is Type II, and thus has well-defined traces and renormalized entropies. In a semiclassical regime, with an ultraviolet regulator, the methods of those papers allow one to roughly speaking approximate the entropy as
\begin{equation}
    S \approx \sum_a\expval{H_a} + S_{\text{interior}},
\end{equation}
where $S_{\text{interior}}$ is the von Neumann entropy of the UV-regulated degrees of freedom inside the subregion. This is essentially a generalized entropy formula, with $\sum_a H_a$ taking the role of the area term.\footnote{See \cite{AliAhmad:2024saq} for related discussion of how the area term may be interpreted as a quantum conditional entropy using a non-commutative version of a quantum Bayes' law.}

In a general system, $\sum_a H_a$ will not always agree with the boundary area. Indeed, for the reasons discussed above, its form depends on, for example, whether there are higher-curvature terms in the action and whether there is non-minimally coupled matter. Its form also depends on the choice of boundary conditions imposed on the linearized fields at $\partial\Sigma$. 

Without boundary conditions on the linearized fields, $\sum_a H_a$ is subject to the usual JKM ambiguities in the covariant phase space formalism~\cite{Iyer:1994ys,Jacobson:1993vj}, so we really do need to fix the boundary conditions in order to get an unambiguous answer for the entropy. Indeed, without appropriate boundary conditions, it is not clear whether one has even defined the subsystem whose degrees of freedom one wishes to study.

Some extra physical principles are required to choose the appropriate boundary conditions. For example, in general relativity with minimally coupled matter, we expect $\sum_a H_a$ to give an area term, so we might design our boundary conditions such that this is the case. By contrast, in higher-derivative theories, or with non-minimally coupled matter, we do not necessarily expect an area term. We need a different way to pick the right boundary conditions. This can be motivated by previous results. For example, if we want the entropy formula to be consistent with AdS/CFT, we might choose boundary conditions such that $\sum_a H_a$ agrees with the Camps-Dong entropy~\cite{Dong_2014,Camps:2013zua}. Alternatively, we may want the entropy formula to be consistent with a generalized second law; then we might choose boundary conditions such that $\sum_a H_a$ agrees with the Wall entropy~\cite{Wall_2015} (the Camps-Dong entropy and Wall entropy do not necessarily disagree). Of course, these would be somewhat indirect methods. There is probably a more fundamental principle which would enable one to pick good boundary conditions; we leave exploration of this for future work.

\section{Gauge theory examples}
\label{sec: gauge theory examples}

We will now illustrate the formalism of section \ref{sec: Constraints in the covariant phase space} through explicit non-gravitational gauge theories. The results will be discussed here, but explicit calculations we leave for App.~\ref{app: gauge theory examples}.

We will start with Yang-Mills theory, Maxwell, and then 3-dimensional non-Abelian Chern-Simons theory. In both cases, the sole dynamical field is the Lie algebra-valued one-form $A$. We define the curvature two-form as
\begin{equation}\begin{split}\label{eq: define F}
    F&:=[\DD,\DD]=\sd A + A\wedge A,
\end{split}\end{equation}
where $\DD:=\sd + A$ denotes the gauge-covariant derivative and $[\cdot,\cdot]$ is the Lie bracket on Lie algebra-valued forms, defined as $[A,\eta]:=A\wedge \eta -(-1)^p\eta\wedge A$ for $p$-forms $\eta$. 
The infinitesimal gauge transformation associated with a (possibly field-dependent) function $\lambda$ is\footnote{This choice of sign amounts to defining a gauge transformation to be the gauge group element $U = e^{-\lambda}.$}
\begin{equation}
    \delta_\lambda A = \DD\lambda.
\end{equation}
We will denote the field-space vector field generating this transformation as $\hat{X}=\hat{\lambda}$ such that
\begin{equation}\label{eq: infinitesimal gauge action}
    \hat{\lambda}\fins \fd A = \DD\lambda.
\end{equation}

For the perturbative constraint analysis, we will expand the one-form $A$ as in \eqref{eq:fundamentalFieldExpansion}
\begin{equation}
    A = A^{(0)} + \eps A^{(1)} + \frac{1}{2} \eps^2 A^{(2)}  + O(\eps^3).
\end{equation}

In the coming analysis it will be important that $\eps$ is not set equal to the couplings of the non-Abelian theories, i.e.~the Yang--Mills coupling $g$, and $(k/2\pi)^{-1/2}$ with $k$ the level of Chern--Simons theory. In fact, we will set these couplings to unity, in line with the literature \cite{Moncrief:1977hy,Arms1,Arms:1979au,Altas:2021htf,Khavkine:2013iei}. The reason being that unlike Einstein--Hilbert theory with global scaling $\kappa^{-2}$, this does not happen in these theories. The kinetic terms will be zero'th order in their coupling constant while the self-interaction vertices come with powers of the coupling constant. This is to be expected, as in the limit that the coupling is turned off the gauge connection $A$ should become a free field. As a consequence, setting $\eps$ to one of these couplings would mean the lowest order term $A^{(0)}$ would be a free field. Hence, it can not be an exact solution to the full interacting theory. This remains true when combined with Einstein--Hilbert theory. There we can do the rescalings $A \to A / \kappa$ and $g \to \tilde{g} = g / \kappa$ to pull out an overall $\kappa^{-2}$ which then becomes global, as argued in Sect.~\ref{sec_matter}. In that case, setting $\eps = \kappa$ poses no problem, but for the same reasons as above it will be a problem when $\eps = \tilde{g}$ as those will still be present and not result in an overall scaling in front of both actions.

In its Abelian versions, one can set $\eps$ equal to the coupling constant. Because just as in gravity, one can pull an overall scaling out of the action, this is due to the lack of self-interaction of the gauge field. It even remains true when coupled to matter, such as in scalar QED where one can pull out an overall $e^{-2}$ through appropriate rescalings, and Einstein--Hilbert gravity where one rescales the metric with some power of the coupling.

\subsection{Yang--Mills}

We start with the Yang--Mills Lagrangian form
\begin{equation}
    \begin{split}
        L^{(\text{YM})} &= -\frac{1}{2} \Tr(F\wedge\star F),\\
    \end{split}
\end{equation}
with $\star$ the usual Hodge dual for a non-dynamical metric.
The constraint density of this theory, computed in App.~\ref{app:YM}, is 
\begin{equation}
    C^{(\text{YM})}_\lambda = \Tr(\lambda \DD \star F),
\end{equation}
with straightforward variation
\begin{equation}
    \delta C_{\lambda}^{\text{YM}}= \Tr\left((\delta\lambda) \DD\star F+\lambda\delta \DD\star F\right).
\end{equation}
Nevertheless, its form in terms of the decomposition \eqref{eq: delta C general 2} is more elucidating as it explicitly separates terms which vanish when the background is symmetric, those which vanish on background solutions, and a boundary term
\begin{equation}
    \begin{split}
        \delta C^{(\text{YM})}_{\lambda} 
    &=-\hat{\lambda}\cdot \omega^{(\text{YM})} + \mathcal{C}^{(\text{YM})}_{\lambda}+\sd B^{(\text{YM})}_\lambda\\
    &=-\hat{\lambda}\cdot \omega^{(\text{YM})} + \Tr(\delta\lambda ~\DD\star F)+\sd\Tr(\lambda \fd \star F)\,,
    \end{split}
\end{equation}
with presymplectic current
\begin{equation}
    \omega^{(\text{YM})} = \Tr(\delta A \wedge \star F).
\end{equation}
With the variation $\delta C^{(\text{YM})}_{\lambda}$ known, we can perform the perturbative constraint analysis of Sec.~\ref{subsec: perturbative constraints} by acting with the field-space Lie derivative along $\hat{V}$ as defined in \eqref{eq: define V} and \eqref{eq:fieldExpansion}. In particular, we will restrict ourselves to the case where the background field is a solution to the equations of motion and is symmetric, and additionally where the gauge transformation parameter is field independent. This amounts to the following three conditions:
\begin{equation}
    \DD^{(0)} \star F^{(0)} = 0, \quad \DD^{(0)}\lambda_s = 0, \quad \fdv{\lambda_s}{A} = 0.
\end{equation}
As shown in detail in appendix \ref{app:YM},
this leads to the following first- and second-order constraint densities
\begin{align}
    \label{eq:YMConstraintFirst}
    C_{\lambda_s}^{(\text{YM},1)} &= \dd \Tr(\lambda_s \star \DD^{(0)} A^{(1)}), \\
    C_{\lambda_s}^{(\text{YM},2)} &= 2\Tr(A^{(1)} \wedge \comm{\star \DD^{(0)} A^{(1)}}{\lambda_s} ) 
    + \dd \Tr( \lambda_s \star \left(\DD^{(0)} A^{(2)} + \comm{A^{(1)}}{A^{(1)}} \right)).
    \label{eq:YMConstraintSecond}
\end{align}
As expected from (\ref{eq: first order constraint, general symmetry}, \ref{eq: second order constraint, general symmetry}) the first-order constraint density is a pure boundary term, and the second-order constraint density has the second-order perturbation $A^{(2)}$ appearing only in the boundary term. The latter means that when the boundary term vanishes after integrating $C_\lambda^{(\text{YM},2)}$ over a spatial slice, we are left with a non-trivial condition on the first-order perturbation $A^{(1)}$. This signals a linearization instability.

A similar analysis of linearization instabilities in Yang--Mills theory has been performed in \cite{Moncrief:1977hy} where they made use of the canonical formalism. Although we used the covariant phase space formalism, the result \eqref{eq:YMConstraintSecond} is consistent with \cite[Eq.~(4.10)]{Moncrief:1977hy} when written in component form. Other works exploring instabilities in the case of Yang--Mills theory are \cite{Arms1,Arms:1979au,Altas:2021htf,Khavkine:2013iei,Arms_1981,arms1981symmetry,Arms:1986vk} with conclusions compatible with ours.

\subsection{Maxwell theory and scalar  electrodynamics}\label{ssec_QED}

Pure Maxwell theory is characterized by the Lagrangian form
\begin{equation}
    L^{(\text{M})} = - \frac{1}{2} F \wedge \star F,
\end{equation}
where $F := \dd A$. Since this is just the Abelian version of the Yang--Mills case, we can straightforwardly use the results computed in the previous subsection. For instance, the constraint density and its variation are simply
\begin{equation}
    C_{\lambda}^{(\text{M})} = \lambda \dd \star F, \quad \delta C_{\lambda}^{(\text{M})} = \delta \lambda \dd \star F + \lambda \delta \dd \star F.
\end{equation}
Similarly, the variation respects the decomposition \eqref{eq: delta C general 2} and is given by 
\begin{equation}
    \begin{split}
        \delta C^{(\text{M})}_{\lambda} 
    &=-\hat{\lambda}\cdot \omega^{(\text{M})} + \mathcal{C}^{(\text{M})}_{\lambda}+\sd B^{(\text{M})}_\lambda\\
    &=-\hat{\lambda}\cdot \omega^{(\text{M})} + \delta\lambda ~\dd \star F +\sd(\lambda \fd \star F)\,,
    \end{split}
\end{equation}
where the Maxwell presymplectic current now takes the form
\begin{equation}
    \omega^{(\text{M})} = \delta A \wedge \star F.
\end{equation}
Taking further Lie derivatives along $\hat{V}$ and imposing the following three conditions
\begin{equation}
    \dd \star F^{(0)} = 0, \quad \dd \lambda_s = 0, \quad \fdv{\lambda_s}{A} = 0,
    \label{eq:maxwellConditions}
\end{equation}
leads us to the perturbative constraint densities
\begin{align}
    C_{\lambda_s}^{(\text{M},1)} &= \dd(\lambda_s \star F^{(1)}), \\
    C_{\lambda_s}^{(\text{M},2)} &= \dd(\lambda_s \star F^{(2)}).
\end{align}
It is not surprising that the second-order constraint density depends only on second-order fields (pure Maxwell theory is linear after all), and in fact is simply proportional to the second-order equation of motion. Even in the absence of a boundary, the integrated constraint imposes no new condition on first-order fields and there is no instability. 

This changes when we add charged matter to the Lagrangian. In that case, it was argued by Higuchi \cite{Higuchi:1991tk} that the charge only appears at second-order in perturbation theory. Let us recast this statement using covariant phase space language.  We show some calculational details in the Appendix \ref{app:sED}. The Lagrangian form of scalar electrodynamics (sED) in an arbitrary, fixed spacetime is
\begin{equation}
    L = -\frac{1}{2} F \wedge \star F - ((\DD \phi)^* \wedge \star \DD \phi - m^2 \phi^* \phi \eps), \quad \DD = \dd + iA,
\end{equation}
The current for this theory can be defined covariantly through
\begin{equation}
    \star j = \frac{1}{\sqrt{-g}} \fdv{S_\text{matter}}{A} = i (\phi^* \star \DD \phi - (\star \DD \phi)^* \phi),
    \label{eq:maxwellCurrent}
\end{equation}
which enters the equation of motion as 
\begin{equation}\label{eq: EOM  sED}
    \dd \star F = \star j.
\end{equation}
The addition of this current makes the theory non-linear, even in flat space.
Performing the usual analysis for the Noether current, it turns out to be 
\begin{equation}
    J_\lambda^{(\text{sED})} = \lambda (\dd \star F - \star j) - \dd(\lambda \star F),
\end{equation}
which leads to the identifications
\begin{equation}
    \label{eq:gaussLaw}
    C_\lambda^{(\text{sED})} = \lambda (\dd \star F - \star j), \quad Q_\lambda^{(\text{sED})} = -\lambda \star F.
\end{equation}
When we a consider perturbative analysis, the only background solution for the matter field which is left invariant under gauge transformations  is 
\begin{equation}
    \phi^{(0)} = 0.
\end{equation}
Consequently, the current \eqref{eq:maxwellCurrent} is only non-zero at second-order $j^{(2)}$ since every term is quadratic in the matter field.  The perturbative constraints become
\begin{align}
    C_{\lambda_s}^{(\text{sED},1)} &= \lambda_s ( \dd\star F^{(1)}), \\
    C_{\lambda_s}^{(\text{sED},2)} &= \lambda_s ( \dd\star F^{(2)}-\star j^{(2)}),
\end{align}
where $j^{(2)}$ is quadratic in $\phi^{(1)}$.  In the absence of a spatial boundary, the integrated second-order constraint then results in a nontrivial condition on the first-order solutions $\phi^{(1)}$, indicating a linearization instability.

\subsection{Chern--Simons}

For our last example, consider Chern--Simons theory in 3d locally being specified by
\begin{equation}
    L^{(\text{CS})} = \frac{1}{2}\Tr(A \wedge \dd{A} + \frac{2}{3} A \wedge A \wedge A).
\end{equation}
The constraint density that appears in this theory, see App.~\ref{app:CS}, is 
\begin{equation}
    C_{\lambda}^{(\text{CS})} = -  \Tr(\lambda F),
\end{equation}
which leads to its variation
\begin{equation}
    \begin{split}
        \delta C^{(\text{CS})}_{\lambda} &= - \Tr(\delta \lambda F + \lambda \delta F) \\
        &=-\hat{\lambda}\cdot \omega^{(\text{CS})} + \mathcal{C}^{(\text{CS})}_{\lambda}+\extd B^{(\text{CS})}_\lambda\\
        &=- \hat{\lambda}\cdot \omega^{(\text{CS})} - \Tr(\delta\lambda~ F)-\sd\Tr(\lambda~ \fd A),
    \end{split}
\end{equation}
with Chern--Simons presymplectic current
\begin{equation}
    \label{eq:CSpresymplcurrent}
    \omega^{(\text{CS})} = - \frac{1}{2} \Tr(\delta A \wedge \delta A).
\end{equation}
As in the Yang-Mills case, we can perturb around a background solution with a symmetry $\hat{\lambda}_s$, requiring
\begin{equation}
    F^{(0)} = 0, \quad \DD^{(0)}\lambda_s = 0, \quad \fdv{\lambda_s}{A} = 0.
\end{equation}
This results in the following two perturbative constraint densities
\begin{align}
    \label{eq:CSconstraint1}
    C_{\lambda_s}^{(\text{CS,1})} &= -  \dd \Tr(\lambda_s A^{(1)}), \\
    C_{\lambda_s}^{(\text{CS,2})} &=  \Tr( \comm{A^{(1)}}{\lambda_s} \wedge A^{(1)})
    -  \dd \Tr(\lambda_s A^{(2)}).
    \label{eq:CSconstraint2}
\end{align}
Hence, we again have a pure boundary term for the first-order constraint density, and the second-order perturbation only appearing in the boundary term of the second-order constraint density which signals a linearization instability for spatially closed Cauchy slices or suitable boundary conditions. This is consistent with the observation in \cite{Khavkine:2013iei}.

Likewise, we can look at the Abelian version of this theory and find that the second-order constraint density
\begin{equation}
    C_{\lambda_s}^{(\text{ACS,2})} = -  \dd(\lambda_s A^{(2)}) = -  \lambda_s F^{(2)}
\end{equation}
is again simply proportional to the second-order equation of motion which now depends only on the second-order field, thus avoiding a linearization instability. This is not surprising as Abelian Chern--Simons theory can be represented as a linear theory.

\section{General subregions}\label{sec_generalregions}

Let us briefly comment on imposing second-order constraints and crossed algebras in the generic case of subregions in spacetimes without isometries, which has been considered in \cite{Jensen:2023yxy,Chen:2024rpx}. Our discussion applies, however, also to spacetimes with isometries and boundaries, as instabilities arise in neither of these cases (except for special boundary conditions in the boundary case).

\subsection{Second-order constraints without isometries}\label{ssec_2connoiso}

A main proposal in \cite{Jensen:2023yxy} is that crossed product algebras and the ensuing von Neumann type conversion occur not only for Killing boost constraints but also for any second-order diffeomorphism constraint associated with a timelike vector that appears boost-like in the vicinity of the entangling surface of interest. This is based on a `geometric modular flow conjecture' saying that the action of such a boost-like constraint will agree infinitesimally with the modular flow of some KMS state (see also \cite{sorce2024analyticity}). It was further argued that one should impose such constraints for consistently embedding the perturbative solution into the full non-linear theory because any higher-order constraint contains contributions that will not commute with the observables of the linearized theory. Linearization instabilities would thereby provide a misleading justification for only imposing second-order constraints associated with Killing symmetries.

Our exposition of linearization instabilities leads to a slightly different picture. In the absence of Killing fields or in the presence of boundaries, a consistent embedding of linearized solutions into the full theory is guaranteed without \emph{any} second-order constraint. There is no \emph{need} to impose higher-order constraints to ensure stability of the solutions and therefore that it is a valid approximation to an exact solution. When no Killing fields are present, conical singularities in the solutions space do not occur in the vicinity of such a solution and in the presence of boundaries they do not occur at all for reasonable boundary conditions \cite{fischer1980structure,Arms:1982ea,Deser:1973zzb,choquet1979maximal}. This is very different from the situation of spatially closed backgrounds with Killing fields, where the imposition of the second-order Taub charge constraints is a \emph{necessary and sufficient} condition for a consistent embedding. While it is true that higher-order constraints contain a dependence on the linearized variables, their non-commutation with linearized observables is suppressed by higher powers of $\kappa$ and so this does not cause problems. Rescaling second-order constraints by $1/\kappa$, and imposing them for vector fields that are not Killing, on the other hand, is asking too much of the linearized theory. 

One may of course impose the second-order boost-like constraints nevertheless. But then one has to justify why it suffices to consider only a single or finite number of second-order constraints and one does not effectively open Pandora's box, given that there are infinitely many second-order constraints, and it is \emph{a priori} unclear why it is consistent to ignore those. In which sense is the entropy discussion obtained by implementing only a finite number of constraints physically relevant in relation to a full diffeomorphism invariance at the given order? The hope is that additional second-order constraints will not affect the crossed product transition from a type $\rm{III}$ to a type $\rm{II}$ algebra \cite{Jensen:2023yxy}. While this appears reasonable, it is not obvious that this will always work. For example, the current generalized entropy discussions based on crossed product constructions invoke Haag duality for the region and its complement to fix the relevant subsystems. However, dressed observables lead to challenges for microcausality when a local dressing is not feasible \cite{Donnelly:2015hta,Donnelly:2016rvo,HK} and this might become an issue when taking care of diffeomorphisms that move or deform the region and its complement, and it is not feasible to fully dress within the region.

By contrast, there is a clear answer to why it is consistent to ignore all but a finite number of second-order constraints for spatially closed spacetimes with Killing symmetries. In the next subsection, we shall provide a partial one also for the present case without Killing fields, based on whether or not one may be interested in higher-order observables. Nevertheless, this case remains on somewhat less firm grounds than the spatially closed case with isometries.

\subsection{Outline of an algebraic argument for second-order constraints}

Let us give an outline of an operational and algebraic argument for why one may want to impose second- and  higher-order constraints on more general backgrounds.

In the exact theory, all the constraints $C_\xi[\Phi]=0$ have to be imposed. Let us expand this equation order by order in the field perturbations, like we have been doing throughout the paper:
\begin{equation}
    C_\xi[\Phi^{(0)}] + \kappa C^{(1)}[\Phi^{(0)},\Phi^{(1)}] + \frac12\kappa^2 C^{(2)}[\Phi^{(0)},\Phi^{(1)},\Phi^{(2)}] + \dots = 0.
\end{equation}
Thus, looking order by order in $\kappa$, which we treat as a formal parameter, we have a series of constraints $C^{(n)}=0$. Each constraint $C^{(n)}$ depends on $\Phi^{(i)}$ up to at most $i=n$. Let us fix the background to some on-shell configuration, $C_\xi[\Phi^{(0)}]=0$.

Previously in the paper we have mostly been concerned with solving the constraints up to $n=2$. By contrast, here let us imagine that we solve the constraints to all orders. If we do so, we do not need to treat $\xi$ in a $\kappa$-dependent way. Indeed, if we were to use $\xi = \sum_k \kappa^k \xi^{(k)}$, then Taylor expanding $C_\xi[\Phi]=0$ would just result in a series of constraints which are linear combinations of the original ones $C^{(n)}=0$ above -- so imposing $C^{(n)}=0$ for all $n$ suffices to also satisfy the constraints for $\kappa$-dependent $\xi$. Simultaneously, we do not need to worry about whether the action of the diffeomorphism $\xi$ preserves the background -- since, if we wanted, we could just exchange $\xi$ for $\kappa\xi$, and get the same set of constraints $C^{(n)}=0$.

Let us comment on the fact that we are treating $\kappa$ as a formal parameter, rather than just a small number. This is an appropriate thing to do if we want to construct the strict $\kappa\to 0$ theory. As we vary $\kappa$, we change through different finite $\kappa$ theories, and for fixed $\Phi^{(i)}$, we get a trajectory through states in different theories (with different $\kappa$). Then, taking $\kappa\to 0$, and following along to the endpoint of this trajectory, we get a state in the strict $\kappa\to 0$ theory. For this to be consistent, we need the states to satisfy $C=0$ at all points along the trajectory, which is what requires us to impose $C^{(n)}=0$.

Suppose we are only interested in observables formed out of the leading-order perturbations of the fields $\Phi^{(1)}$. Let us call the algebra of such observables $\mathcal{A}^{(1)}$. We need to impose the constraints on such observables. However, the way in which we do this differs for each constraint, depending on whether that constraint depends on the higher-order fields $\Phi^{(i)}$ or not. In particular, the constraint $C^{(1)}$ only depends on $\Phi^{(1)}$, so in the algebra $\mathcal{A}^{(1)}$ we should impose this constraint by restricting to those observables which commute with $C^{(1)}$. On the other hand, $C^{(i)}$ for $i>1$ (generically) depends on the higher-order fields also, and, so long as there is sufficient kinematical freedom in the higher-order fields (and there are no linearization instabilities), we do not need to directly impose this constraint on the algebra $\mathcal{A}^{(1)}$. Indeed, we can simply just assume that the higher-order fields solve this constraint, for any possible configuration of $\Phi^{(1)}$. Thus, we (generically) only need to directly impose $C^{(1)}=0$, if we are only concerned with observables depending on $\Phi^{(1)}$.

In special cases, the higher-order constraints may simplify in such a way that they only depend on the lower-order fields. This is exactly what happens when the background $\Phi^{(0)}$ has isometries in a spatially closed universe; then, as explained previously in the paper, for each isometry the corresponding $C^{(2)}$ only depends on the first-order perturbations $\Phi^{(1)}$. Now we can no longer satisfy $C^{(2)}=0$ by assuming that the higher-order fields solve it. We must instead impose it directly on the algebra $\mathcal{A}^{(1)}$, by restricting to observables which commute with it. Thus, in the case of backgrounds with isometries, observables in $\mathcal{A}^{(1)}$ should commute with $C^{(1)}$ for all $\xi$, and they should also commute with $C^{(2)}$ for $\xi$ which are isometries. This is an algebraic version of linearization instability.

To reiterate, when there are no linearization instabilities, $C^{(2)}$ need not be directly imposed on $\mathcal{A}^{(1)}$. However, we can make $C^{(2)}$ relevant again, for all $\xi$ and for all backgrounds by \emph{enlarging} the set of observables in which we are interested. In particular, suppose we now wish to consider observables which are allowed to depend on the \emph{second-order} parts of the fields $\Phi^{(2)}$, as well as the first-order parts $\Phi^{(1)}$. Let us call this algebra $\mathcal{A}^{(2)}$. Then, $C^{(1)}$ and $C^{(2)}$ now always depend on all the fields under consideration, and should hence always be imposed directly: we must restrict to observables in $\mathcal{A}^{(2)}$ which commute with both $C^{(1)}$ and $C^{(2)}$, for all $\xi$. Generically, the higher-order constraints $C^{(i)}=0$, $i>2$, need not be imposed directly (although, by the same reasoning as above, in special cases they may need to be imposed directly, if they simplify such that they only depend on the fields up to second order).

This appears to be a partial justification for imposing second-order constraints in general subregions, as for example is done in~\cite{Jensen:2023yxy,Faulkner:2024gst,Kudler-Flam:2023qfl,KirklinGSL}. There, the second-order fluctuations of the area of the boundary of a region are explicitly included in the algebra of observables -- meaning one should therefore impose $C^{(2)}$ directly, regardless of the presence of linearization instabilities. Ultimately, this is responsible for the crossed product structure of the algebras they considered. 

But there are still many things which are not completely clear: in particular, only one second-order observable is included (the area), and $C^{(2)}$ for only one $\xi$ is imposed (a boost). These two facts are no doubt related to each other. But more generally, it would be well motivated to include \emph{all} second-order observables, and to impose $C^{(2)}$ for all $\xi$. It remains to be seen whether this can be consistently done, as indicated in Sec.~\ref{ssec_2connoiso}. As an intermediate case, one might be interested in observables which already identically commute with $C^{(2)}$ for all but finitely many $\xi$. Then one would only have to directly impose $C^{(2)}$ for the finitely many left over $\xi$. This would somewhat simulate what happens in the case of linearization instabilities, where finitely many second-order constraints must be imposed on the first order variation $\Phi^{(1)}$. Again, it is not clear that there is a general systematic way to achieve this kind of setup.

\section{Conclusion}\label{sec_conc}

In this paper, we have explored the potential justifications for imposing the kinds of constraints which were employed in recent studies of the properties of von Neumann algebras of gauge-invariant observables in gravitational subregions -- namely that they are Type II, and thus have well-defined von Neumann entropies (up to a state-independent constant). These studies were performed within the setting of perturbative gravity, but, as we have argued, the relevant constraints only appear at second order in perturbation theory. One needs to justify carefully why one should nevertheless impose them on the first-order fields.

In particular, we have pointed out that `linearization instabilities' are a fairly robust justification for the imposition of these constraints. These occur in spatially closed spacetimes, when perturbing around backgrounds with symmetries. They are extra conditions that must be imposed on the linear-order perturbations of the fields, in order for the perturbative theory to be consistent with non-perturbative physics. More precisely, for each Killing symmetry there is an associated second-order constraint which is quadratic in the first-order field perturbations. There can only be finitely many independent Killing symmetries, so there are finitely many independent such constraints. Imposing these second-order constraints is typically a necessary and sufficient condition for the first-order field perturbations to be integrable to some exact field configuration.

We examined two cases: when the background matter is in a vacuum configuration, and when the background matter is non-trivial. We developed a uniform and systematic formalism for understanding the instabilities, using the powerful covariant phase space framework. Our methods reproduce known results in the case of general relativity minimally coupled to matter. But they are also capable of describing what happens in much more general theories. Our treatment applies equally well to diffeomorphisms and local Abelian or non-Abelian gauge symmetries, and we have described various examples of this. We have thus shed light on the universal structures underlying linearization instabilities in non-linear gauge theories and gravity.

Our results also make clear that linearization instabilities are not present if spacetime is not spatially closed (and reasonable boundary conditions are imposed), or the background configuration does not have symmetries. Then the first-order constraints are all that needs to be imposed on the linearized fields. So in this case the theory is linearization stable, meaning linearization instabilities cannot be used as a justification for imposing the second-order constraint on the first order fields. 

Since linearization instabilities are only present in the spatially closed, symmetric background case, it is natural to ask what the justification would be for imposing these constraints in more general subregions. We provided some initial exploratory discussion on this issue. But it remains an important open question, and one which must be answered in order to obtain a full understanding of the scope of gravitational crossed product constructions.

In any case, we have also shown how second-order constraints may in general be written as a sum of several contributions: an integral over each component of the boundary $\partial\Sigma$ of the relevant Cauchy surface $\Sigma$, a contribution from any clocks or more general QRFs which may be present in $\Sigma$, and a bulk integral over $\Sigma$ containing contributions from the remaining fields. The boundary contributions are the only ones which depend on second-order perturbations. Imposing invariance under the full second-order constraint is what then leads to a crossed product algebra. This may be done for general gauge symmetries as well as gravitational ones, i.e.\ diffeomorphisms. But diffeomorphisms have a special status, because the modular flow of the fields is, in certain states, a diffeomorphism. Thus, in these cases the crossed product construction causes modular flow to be inner, which is what is responsible for the Type III to Type II conversion of the algebra of observables. It is unlikely that a similar type reduction will occur when imposing invariance under the constraints associated with general gauge symmetries. However, an interesting intermediate case is that of tetrad gravity, which has both diffeomorphisms and local Lorentz transformations as gauge symmetries. Modular flow, which typically acts as some kind of boost, would presumably have a non-trivial component within both of these kinds of gauge symmetry. We leave exploration of this to future work.

\phantomsection
\addcontentsline{toc}{section}{\numberline{}Acknowledgements}
\section*{Acknowledgments}

We thank Krishna Jalan and Antony Speranza for helpful discussions.
This work was supported in part by funding from the Okinawa Institute of Science and Technology Graduate University.~It was also made possible through the support of the ID\# 62312 grant from the John Templeton Foundation, as part of the \href{https://www.templeton.org/grant/the-quantum-information-structure-of-spacetime-qiss-second-phase}{\textit{`The Quantum Information Structure of Spacetime'} Project (QISS)} from the John Templeton Foundation.~The opinions expressed in this project/publication are those of the author(s) and do not necessarily reflect the views of the John Templeton Foundation. Research at Perimeter Institute is supported in part by the Government of Canada through the Department of Innovation, Science and Economic Development and by the Province of Ontario through the Ministry of Colleges and Universities.

\appendix

\section{Isometries and linearized constraints}\label{app_isolin}

Here, we shall briefly paraphrase \cite[Lemmas 1.1 \& 1.5]{fischer1980structure}, establishing Eq.~\eqref{vanish}. This will be instructive for understanding why this line of argument is difficult to generalize to non-trivial matter backgrounds and generalized gravity theories. These challenges are sidestepped by our general covariant phase space analysis in Sec.~\ref{sec: Constraints in the covariant phase space}.

Expanding the contracted Bianchi identities $\nabla^\mu G_{\mu \nu}[g]=0$ in $\kappa$, we have
\begin{equation}\label{A1}
\kappa\left(\left(\nabla^{(1)}(h)\right)^\mu\,G_{\mu \nu}[g_0] +(\nabla^{(0)})^\mu G^{(1)}_{\mu \nu}(h)\right)+O(\kappa^2)=0\,,
\end{equation}
where $\nabla^{(1)}(h)$ is the linear term in a $\kappa$-expansion of $\nabla$.  Thus, the background equations of motion $E^{(0)}[g_0]=0$ imply that the linearized Einstein tensor is divergence-free with respect to the background as $\kappa\to0$, \emph{off-shell} of the linearized equations of motion $E^{(1)}[g_0;h]=0$:
\begin{equation}\label{A2}
   (\nabla^{(0)})^\mu G^{(1)}_{\mu \nu}(h) =O(\kappa)\,.
\end{equation}
Therefore, for a background Killing field $\xi$, also $G_{\mu \nu}^{(1)}\xi^\nu$ is divergence-free with respect to the background as $\kappa\to0$. Supposing that $\xi$ is compactly supported and invoking Stokes' theorem yields
\begin{equation}
    \int_{\Sigma_1}C_{\kappa\xi}=\int_{\Sigma_2}C_{\kappa\xi}+O(\kappa)
\end{equation}
for two disjoint Cauchy slices $\Sigma_1,\Sigma_2$. Now this holds off-shell of $E^{(1)}[g_0;h]=0$. For any off-shell graviton field $h$, we can thus choose $h'$ such that $h=h'$ on $\Sigma_1$ and  $h'=0$ on $\Sigma_2$. This yields the claim
\begin{equation}
    \int_{\Sigma_1}C_{\kappa\xi}(h)=\int_{\Sigma_1}C_{\kappa\xi}(h')=\int_{\Sigma_2}C_{\kappa\xi}(h')+O(\kappa)=0+O(\kappa)\,.
\end{equation}

It is unfortunately not feasible to extend this argument to backgrounds with matter because $\nabla^a T_{\mu \nu}=0$ typically only holds on-shell of the matter equations of motion or on-shell of the Einstein equations. That is, the analog of Eq.~\eqref{A1} with $G_{\mu \nu}\to G_{\mu \nu}-\frac{1}{4}T_{\mu \nu}$ only holds partially or fully on-shell and for field variations tangential to the space of solutions. This again means, however, that Eq.~\eqref{A2} with $G_{\mu \nu}\to G_{\mu \nu}-\frac{1}{4}T_{\mu \nu}$ would only hold partially or fully on-shell of the linearized equations of motion. This would defeat the argument leading to the analog of Eq.~\eqref{2ndorder3}, which relied on Eq.~\eqref{A2} being an off-shell identity. The covariant phase space will overcome this issue.

\section{Einstein-Hilbert gravity with matter} \label{sect: EH with matter}
For the gravitational case, we restrict our explicit analysis to the identification of the constraints themselves, illustrating that the presence of various forms of matter can alter the usual form of the diffeomorphism constraint nontrivially.  This is certainly well known to experts in the field, but to the novice, seeing how it occurs is informative.  The first and second-order constraint densities are in principle determined by equations \eqref{eq: first order constraint, general} and \eqref{eq: second order constraint, general}, which we avoid writing out in full detail, given that they are rather convoluted.

Suppose we take the gravitational action to be the Einstein-Hilbert (EH) action, possibly extended to include a cosmological constant, and some matter Lagrangian $L_{\text{M}}$ for non-metric fields which we collectively denote $\phi$.  We assume the matter can be rescaled such that it contributes as a source of the background solutions. We then work with a total Lagrangian that has been stripped of a $1/\kappa^2$ factor.
\begin{equation}\begin{split}\label{eq: Einstein Hilbert with matter system}
    L[g, \phi]&:=L_{\text{EH}}[g]+L_{\text{M}}[g,\phi],\\
    L_{\text{EH}}&:=2\left(R-2\Lambda \right)\epsilon,\\ L_{\text{M}}&:=\mathcal{L}_{\text{M}}\epsilon,
\end{split}\end{equation}
with $\epsilon$ being the volume form. The field space variation of the Einstein-Hilbert Lagrangian gives
\begin{equation}\begin{split}
    \fd L_{\text{EH}} &= -2G_{(\Lambda)}^{\mu\nu} \fd g_{\mu\nu}\epsilon +\sd \theta_{\text{EH}},\\
    G_{(\Lambda)\mu\nu}&=R_{\mu\nu}-\frac{1}{2}(R-2\Lambda)g_{\mu\nu},\\
    \theta_{\text{EH}}&=2\left(g^{\alpha\mu}g^{\nu\beta}-g^{\mu\nu}g^{\alpha\beta}\right)\nabla_{\beta}\delta g_{\mu\nu}\epsilon_\alpha,\\
    \omega &:=\fd\theta_{\text{EH}}= 2P^{\alpha\beta\gamma\lambda\mu\nu}\epsilon_\alpha \fd g_{\beta\gamma}\nabla_\lambda \fd g_{\mu\nu}\,,
\end{split}\end{equation}
where $\epsilon_\alpha:=\partial_\alpha\sins \epsilon$, and
\begin{equation}\begin{split}\label{eq: omega projector for GR}
    P^{\alpha\beta\gamma\lambda\mu\nu}&:=
    \left(g^{\alpha\nu}g^{\lambda\gamma}g^{\beta\mu}-\frac{1}{2}g^{\alpha\beta}g^{\lambda\gamma}g^{\mu\nu}-\frac{1}{2}g^{\alpha\mu}g^{\lambda\nu}g^{\beta\gamma}+\frac{1}{2}g^{\alpha\lambda}g^{\mu\nu}g^{\beta\gamma}-\frac{1}{2}g^{\alpha\lambda}g^{\nu\gamma}g^{\beta\mu}\right)\,.
\end{split}\end{equation}
Without specifying the form of the matter Lagrangian we can write
\begin{equation}\begin{split}
    \fd L_{\text{M}} &= \frac{1}{2}T^{\mu\nu}\fd g_{\mu\nu}\epsilon+ E_{\phi^a}  \fd\phi^a \epsilon +\sd\theta_{\text{M}},\\
    T_{\mu\nu}&=g_{\mu\nu}\mathcal{L}_{\text{M}} -2\frac{\partial\mathcal{L}_{\text{M}}}{\partial g^{\mu\nu}},\\
    \omega_{\text{M}}&:=\fd\theta_{\text{M}}\,.
\end{split}\end{equation}
All together, the variation of the action becomes:
\begin{equation}\begin{split}
    \fd L&=
    -2E_{g}^{\mu\nu}\delta g_{\mu\nu}\epsilon+E_{\phi^a}\delta\phi^a\epsilon+\extd\left(\theta_{\text{EH}}+\theta_M\right),\\
    E_{g}^{\mu\nu}&:=G_{(\Lambda)}^{\mu\nu}-\frac{1}{4} T^{\mu\nu}.\\
\end{split}\end{equation}
The Noether current $J_\xi$ associated to a diffeomorphism generating vector $\xi$ can be divided into contributions coming from the Einstein-Hilbert piece (plus cosmological constant) and a matter piece:
\begin{equation}\begin{split}
    J_\xi:=\left(\hat{\xi}\cdot\theta_{\text{EH}} -\xi\sins L_{\text{EH}}\right)+ \left(\hat{\xi}\cdot\theta_{\text{M}} -\xi\sins L_{\text{M}}\right).
\end{split}\end{equation}
After some manipulation, the first two terms can be written
\begin{equation}\begin{split}
    \left(\hat{\xi}\cdot\theta_{\text{EH}} -\xi\sins L_{\text{EH}}\right)
    &=  4\epsilon_\alpha \left(\nabla_\beta \nabla^{[\beta}\xi^{\alpha]}+\xi_\beta G_{(\Lambda)}^{\alpha \beta}\right)\\
    &=  \sd q_\xi+ \left(4E_{g}^{\alpha\beta}+T^{\alpha\beta}\right)\xi_\beta\epsilon_\alpha,
\end{split}\end{equation}
with
\begin{equation}\begin{split}\label{eq: gravitational charge contribution}
    q_\xi&:=-2\nabla^{\alpha}\xi^\beta \epsilon_{\alpha\beta}\,.
\end{split}\end{equation}
Here we have used the identity $\nabla_\beta V^{[\beta\alpha]}\epsilon_\beta = -\frac{1}{2}\sd (V^{\alpha\beta}\epsilon_{\alpha\beta})$ for any two-index tensor $V^{\alpha\beta}$, with $\epsilon_{\alpha\beta}:=
    \partial_\beta\sins\partial_\alpha\sins \epsilon$.
We cannot fully decompose the matter contribution to the current $J_\xi$ without further specifying the matter Lagrangian, but we can use
\begin{equation}
\xi\sins L_{\text{M}}
=\mathcal{L}_{\text{M}}g^{\alpha\beta}\xi_\beta\epsilon_\alpha
=\left(T^{\alpha\beta}+2\frac{\partial \mathcal{L}_{\text{M}}}{\partial g_{\alpha\beta}}\right)\xi_\beta\epsilon_\alpha
\end{equation}
to write the full diffeomorphism current as
\begin{equation}\begin{split}\label{eq: general current}
    J_{\xi}
    &=\extd q_\xi + \left(4E^{\alpha\beta}_{g}-2\frac{\partial \mathcal{L}_{\text{M}}}{\partial g_{\alpha\beta}}\right)\xi_\beta\epsilon_\alpha
    +\hat{\xi}\cdot\theta_{\text{M}}\,.
\end{split}\end{equation} 
In the absence of matter fields, it is already apparent from \eqref{eq: general current} that the diffeomorphism current decomposes simply as
\begin{equation}\begin{split}\label{eq: diffeo current vacuum}
    J_\xi^{\text{vac}}&:=\sd Q^{\text{vac}}_\xi+C^{\text{vac}}_\xi\\
    Q^{\text{vac}}_\xi&:=q_\xi,\\
    C_\xi^{\text{vac}}&:=4G^{\alpha\beta}_{(\Lambda)}\xi_\beta\epsilon_\alpha,
\end{split}\end{equation}
with $q_\xi$ given defined as the pure gravity contribution in \eqref{eq: gravitational charge contribution}.
More generally, the decomposition will depend on the nature of the matter fields.

To complete the decomposition into constraint and charge pieces, we must specify the form of matter and consider the last two terms of \eqref{eq: general current}. We here give two examples, with the matter Lagrangian being that of either a single scalar field, or an electromagnetic gauge field.  In the former case, we have
\begin{equation}\begin{split}
L_{\text{M}}&=\mathcal{L}_\phi\epsilon:= \left(-\frac{1}{2}\nabla_\mu \phi \nabla^\mu\phi-V(\phi)\right)\epsilon,\\
\fd L_{\text{M}}&=\left(\nabla_\mu\nabla^\mu\phi-V'(\phi)\right)\fd\phi\epsilon - \sd(\nabla^\mu\phi\epsilon_\mu \fd\phi),\\
&\implies\\
-2\frac{\partial \mathcal{L}_\phi}{\partial g_{\alpha\beta}}\xi_\alpha\epsilon_\beta
+\hat{\xi}\cdot\theta_\phi
&=\nabla^\alpha \phi \nabla^\beta \phi\xi_\alpha\epsilon_\beta
-\epsilon_\beta\nabla^\beta\phi \delta_\xi \phi.
\end{split}\end{equation}
This is seen to vanish identically upon replacing $\delta_\xi\phi = \sL_\xi\phi=\xi^\alpha\nabla_\alpha \phi$.
We conclude that for scalar fields, the full diffeomorphism current is
\begin{equation}\begin{split}\label{eq:resultsScalar}
 J^{\phi}_\xi &= \extd Q^{\phi}_\xi + C^{\phi}_\xi,\\
 Q^{\phi}_\xi &= q_\xi,\\
 C^{\phi}_\xi &= 4 E^{\alpha\beta}_{(g)}\xi_\beta \epsilon_\alpha .
\end{split}\end{equation}
In this case, the charge piece $Q^\phi_\xi$ is unchanged from that of vacuum gravity, while the constraint $C_\xi^\phi$ is modified only by the inclusion of the stress tensor in the gravity equation of motion.

For an EM action, we have
\begin{equation}\begin{split}
L_{\text{M}}&=\mathcal{L}_{\text{EM}}\epsilon:= -\frac{1}{4}F_{\mu\nu}F^{\mu\nu} \epsilon,\\
\fd L_{\text{M}}&=\nabla_\mu F^{\mu\nu}\fd A_\nu \epsilon
-\sd\left(F^{\mu\nu}\fd A_\nu \epsilon_\mu\right),
\end{split}\end{equation}
with $F_{\mu\nu}:=\nabla_\mu A_\nu-\nabla_\mu A_\nu$.
This leads to
\begin{equation}\begin{split}\label{eq: EM field current contribution, app}
-2\frac{\partial \mathcal{L}_{\text{EM}}}{\partial g_{\alpha\beta}}\xi_\alpha\epsilon_\beta
+\hat{\xi}\cdot\theta_{\text{EM}}
&=F^{\alpha\mu}F^{\beta}_{~\mu}\xi_\alpha\epsilon_\beta
-\epsilon_\beta F^{\beta\mu}\delta_\xi A_\mu\\
&=F^{\alpha\mu}F^{\beta}_{~\mu} \xi_\alpha\epsilon_\beta
-\epsilon_\beta F^{\beta\mu}(\xi^\alpha \nabla_\alpha A_\mu + A_\alpha\nabla_\mu \xi^\alpha )\\
&=(\nabla_\mu F^{\beta\mu}A^\alpha)\xi_\alpha\epsilon_\beta
-\frac{1}{2}\extd(F^{\beta\mu}A^\alpha \xi_\alpha \epsilon_{\beta\mu} ).
\end{split}\end{equation}
Here we have used the diffeomorphism action on the gauge field, $\delta_\xi A_\mu = \sL_\xi A_\mu = \xi^\alpha\nabla_\alpha A_\mu +\nabla_\mu \xi^\alpha A_\alpha$. The final equality follows from an integration by parts on the $\nabla_\mu\xi^\alpha$ term, using the definition $F_{\mu\alpha}:=\nabla_\mu A_\alpha-\nabla_\mu A_\alpha$, and employing the identity $\nabla_\alpha V^{[\beta\alpha]}\epsilon_\beta = \frac{1}{2}\extd (V^{\alpha\beta}\epsilon_{\alpha\beta})$ for any two-index tensor $V^{\alpha\beta}$.  Notice that the first term of the final expression vanishes on-shell of the gauge field equation of motion, $\nabla_\mu F^{\mu\nu}=0$.
Having identified the charge and constraint contributions in \eqref{eq: EM field current contribution, app}, we see that for EM fields
\begin{equation}\begin{split}\label{eq:resultsEM}
 j^{\text{EM}}_\xi &= \extd q^{\text{EM}}_\xi + C^{\text{EM}}_\xi,\\
 Q^{\text{EM}}_\xi &=q_\xi  -\frac{1}{2}F^{\alpha\beta}A^\mu\xi_\mu\epsilon_{\alpha\beta},\\
 C^{\text{EM}}_\xi &= \left(4 E^{\alpha\beta}_{(g)}+\nabla_\mu F^{\alpha\mu}A^\beta\right)\xi_\beta \epsilon_\alpha.
\end{split}\end{equation}
Thus, in the presence of gauge fields, both the charge piece $Q^{\text{EM}}_\xi$ and the constraint piece $C_\xi^{\text{EM}}$ include non-trivial extensions to the vacuum gravity expressions.  For general expressions of the constraint in the presence of matter fields of arbitrary spin, see \cite{Seifert:2006kv} for example.

\section{Gauge theory computations}
\label{app: gauge theory examples}

This appendix gathers the explicit calculations of the perturbative constraint densities for gauge theories discussed in Section \ref{sec: gauge theory examples}. As discussed there, we will set all couplings to unity.

\subsection{Yang-Mills theory}
\label{app:YM}
We start from a non-Abelian Yang-Mills Lagrangian form
\begin{equation}\begin{split}\label{eq: YM action}
    L^{(\text{YM})} &= -\frac{1}{2} \Tr(F\wedge\star F).\\
\end{split}\end{equation}
This Lagrangian is completely invariant under the gauge transformation \eqref{eq: infinitesimal gauge action}:
\begin{equation}\begin{split}
    &\hat{\lambda}\fins\fd L^{(\text{YM})} = 0,
\end{split}\end{equation}
so we can take 
\begin{equation}\begin{split}
    &\alpha^{(\text{YM})}_{\lambda}=0,
\end{split}\end{equation}
where $\alpha_X$ was defined for general gauge transformations through equation \eqref{eq: covariant L general}.  The equations of motion and symplectic potential form follow from
\begin{equation}\begin{split}
    \delta L^{(\text{YM})} 
    &=-\Tr\left(\delta A\wedge \DD\star F\right)-\sd\Tr\left(\delta A\wedge \star F\right)\\
    &\implies\\
    E^{(\text{YM})}:=-\Tr(\fd A\wedge \DD\star F), \qquad \theta^{(\text{YM})}&:=-\Tr(\fd A\wedge\star F),\qquad \omega^{(\text{YM})}:=\Tr(\delta A\wedge \star \DD \delta A)
\end{split}\end{equation}
We can write the gauge current as (referring to equations \eqref{eq: define J general} and \eqref{eq: decompose J general}) 
\begin{equation}\begin{split}
    J^{(\text{YM})}_{\lambda}
    &:=\hat{\lambda}\fins\theta^{(\text{YM})} 
    = -\Tr(\DD\lambda\wedge\star F)\\
    &=\Tr(\lambda\DD\star F)-\sd\Tr(\lambda\star F)\\
    &\implies\\
    C^{(\text{YM})}_\lambda&:=\Tr(\lambda\DD\star F), \qquad Q^{(\text{YM})}_\lambda:=-\Tr(\lambda\star F).
\end{split}\end{equation}
We likewise need to decompose $\fL_{\hat{\lambda}}\theta^{(\text{YM})}$ (which corresponds to $\fL_{\hat{X}}\theta-\fd\alpha_X$ of equation\eqref{eq: define tilde C, tilde q general} for general gauge transformations) into a term which vanishes on-shell and a spacetime exact piece:\footnote{More explicit details of similar computations are shown in section 7 of \cite{Carrozza:2021gju}, for example.  Note that in that work the gauge transformation rule differs by a sign and the gauge parameter is denoted $\alpha$, so to compare with conventions here one should replace $\alpha\rightarrow-\lambda$.} 
\begin{equation}\begin{split}
    \fL_{\hat{\lambda}}\theta^{(\text{YM})}
    &=\Tr(\delta\lambda ~\DD\star F)-\sd\Tr(\delta\lambda\star F)\\
    &\implies\\
    \mathcal{C}^{(\text{YM})}_{\lambda}
    &:=\Tr(\delta\lambda~\DD\star F),\qquad
    \mathcal{Q}^{(\text{YM})}_{\lambda}
    :=-\Tr(\delta\lambda\star F),
\end{split}\end{equation}
where for generality we have allowed for field-dependent gauge parameters $\lambda$. We can now write the boundary $(d-2,1)$-form defined in \eqref{eq: boundary term general} for Yang-Mills theory as
\begin{equation}\label{eq: boundary piece, YM}
    B^{(\text{YM})}_{\lambda}:=\mathcal{Q}^{(\text{YM})}_{\lambda}-\fd Q^{(\text{YM})}_{\lambda}
    =\Tr(\lambda~\fd\star F)\,.
\end{equation}
One can check that the variation of the constraint density respects the decomposition of equation \eqref{eq: delta C general 2}, giving
\begin{equation}\begin{split}\label{eq: delta C YM}
    \delta C^{(\text{YM})}_{\lambda} 
    &=-\hat{\lambda}\cdot \omega^{(\text{YM})} + \mathcal{C}^{(\text{YM})}_{\lambda}+\sd B^{(\text{YM})}_\lambda\\
    &=-\hat{\lambda}\cdot \omega^{(\text{YM})} + \Tr(\delta\lambda ~\DD\star F)+\sd\Tr(\lambda \fd \star F)\,.
\end{split}\end{equation}
One could employ an explicit form of $\omega^{(\text{YM})}$ to re-express the right-hand side and write
\begin{equation}\label{eq: delta C YM alternate expression}
    \delta C_{\lambda}^{\text{(YM)}}=\Tr\left((\delta\lambda) D_A\star F+\lambda\delta D_A\star F\right).
\end{equation}
However, the decomposition \eqref{eq: delta C YM} is advantageous for our purposes because it explicitly separates terms which vanish on background solutions ($\Tr(\delta\lambda ~\DD \star F)$), terms which vanish when the fields satisfy a symmetry ($-\hat{\lambda}\fins \omega^{\text{(YM)}}$), and a boundary term which remains in both circumstances ($\extd \Tr(\lambda \fd \star F)$).

Let us now perform the perturbative analysis by acting with the Lie derivative \eqref{eq:fieldExpansion} along $\hat{V}$ \eqref{eq: define V}. For the first-order constraint density it is fairly straightforward given expression \eqref{eq: delta C YM alternate expression} as 
\begin{align}
    \delta C_\lambda^{(\text{YM)}} &= \Tr(\delta \lambda \DD \star F + \lambda \DD \star \DD \delta A + \lambda \comm{\delta A}{\star F}) \nonumber \\
    &= \Tr(\delta \lambda \DD \star F - \DD \lambda \wedge \star \DD \delta A + \lambda \comm{\delta A}{\star F}) + \dd \Tr(\lambda \star \DD \delta A),
\end{align}
where we made use of $\delta \DD \eta = \DD \delta \eta + \comm{\delta A}{\eta}$ for an arbitrary form $\eta$, the relation $\delta \DD F = \DD \delta A$, and that the Hodge star $\star$ commutes with $\delta$ since the metric is taken to be a background field. The first-order constraint density can then be computed from this expression as 
\begin{equation}
    \begin{split}
        \label{eq:YMfirstConstraintCalc}
        C_\lambda^{(\text{YM},1)}
        &= \Tr(-\DD \lambda \wedge \star \DD \dv{A}{\eps} + \dv{\lambda}{\eps} \star F + \lambda \comm{\dv{A}{\eps}}{\star F} ) \eval_{\eps = 0} + \dd \Tr(\lambda \star \DD \dv{A}{\eps}) \eval_{\eps = 0}
        \\
        &= \Tr(-\DD^{(0)} \lambda^{(0)} \wedge \star \DD^{(0)} A^{(1)} + \lambda^{(1)}\DD^{(0)} \star F^{(0)} + A^{(1)} \wedge \comm{\star F^{(0)}}{\lambda^{(0)}} ) \\ 
        &\hspace{10pt} + \dd \Tr(\lambda^{(0)} \star \DD^{(0)} A^{(1)}),
    \end{split}
\end{equation}
where we also used the following identity
\begin{equation}
    \Tr(\lambda^{(0)} \comm{A^{(1)}}{\star F^{(0)}}) = \Tr(A^{(1)} \wedge \comm{\star F^{(0)}}{\lambda^{(0)}}),
    \label{eq:traceSwitch}
\end{equation}
which follows straightforwardly from the general statement that Lie algebra-valued forms behave as ordinary forms within the trace. When the background field is left invariant by the gauge transformation, i.e.~$\DD^{(0)} \lambda_s^{(0)} = 0$, and the background equation of motion is satisfied, i.e.~$\DD^{(0)} \star F^{(0)} = 0$, or the gauge parameter is field independent $(\lambda_s=\lambda_s^{(0)})$, then we are left with the result quoted in the main body \eqref{eq:YMConstraintFirst} which is a pure boundary term
\begin{equation}
    C_{\lambda_s}^{(\text{YM},1)} = \dd \Tr(\lambda_s \star \DD^{(0)} A^{(1)}).
\end{equation}
To realise this, note that the third term vanishes because $\comm*{\star F^{(0)}}{\lambda}$ does. It is the gauge transformation of $\star F^{(0)}$ which is zero, since we assumed it leaves $A^{(0)}$ invariant as well.

For the second-order constraint density, we act with another $\delta$ on the first line of \eqref{eq:YMfirstConstraintCalc}, use the same manipulations as before, and then insert the vector $\hat{V}$, so 
\begin{equation}
    \begin{split}
        C_\lambda^{(\text{YM},2)}
        &= -\Tr(\left(\DD \dv{\lambda}{\eps} + \comm{\dv{A}{\eps}}{\lambda} \right) \wedge \star \DD \dv{A}{\eps} + \DD \lambda \wedge \star \left(\DD \dv[2]{A}{\eps} + \comm{\dv{A}{\eps}}{\dv{A}{\eps}} \right) ) \eval_{\eps = 0}
        \\
        &\hspace{10pt} + \Tr(\dv[2]{\lambda}{\eps}\DD \star F + \dv{\lambda}{\eps} \left(\DD \star \DD \dv{A}{\eps} + \comm{\dv{A}{\eps}}{\star F} \right)) \eval_{\eps = 0}
        \\
        &\hspace{10pt} +\Tr( \dv[2]{A}{\eps} \wedge \comm{\star F}{\lambda} + \dv{A}{\eps} \wedge \comm{\star \DD \dv{A}{\eps}}{\lambda} + \dv{A}{\eps} \wedge \comm{\star F}{\dv{\lambda}{\eps}} ) \eval_{\eps = 0}
        \\
        &\hspace{10pt} + \dd \Tr(\dv{\lambda}{\eps} \star \DD \dv{A}{\eps}+ \lambda\star \left(\DD \dv[2]{A}{\eps} + \comm{\dv{A}{\eps}}{\dv{A}{\eps}} \right)) \eval_{\eps = 0}
        \\
        &= -\Tr(\left(\DD^{(0)} \lambda^{(1)} + \comm{A^{(1)}}{\lambda^{(0)}} \right) \wedge \star \DD^{(0)} A^{(1)} + \DD^{(0)} \lambda^{(0)} \wedge \star \left(\DD^{(0)} A^{(2)} + \comm{A^{(1)}}{A^{(1)}} \right) )
        \\
        &\hspace{10pt} + \Tr(\lambda^{(2)}\DD^{(0)} \star F^{(0)} + \lambda^{(1)} \left(\DD^{(0)} \star \DD^{(0)} A^{(1)} + \comm{A^{(1)}}{\star F^{(0)}} \right)) 
        \\
        &\hspace{10pt} +\Tr( A^{(2)} \wedge \comm{\star F^{(0)}}{\lambda^{(0)}} + A^{(1)} \wedge \comm{\star \DD^{(0)} A^{(1)}}{\lambda^{(0)}} + A^{(1)} \wedge \comm{\star F^{(0)}}{\lambda^{(1)}} ) 
        \\
        &\hspace{10pt} + \dd \Tr(\lambda^{(1)} \star \DD^{(0)} A^{(1)} + \lambda^{(0)} \star\left(\DD^{(0)} A^{(2)} + \comm{A^{(1)}}{A^{(1)}} \right)),
    \end{split}
\end{equation}
Under the same assumptions as before (including field-independent $\lambda_s=\lambda_s^{(0)}$), this reduces to \eqref{eq:YMConstraintSecond}
\begin{equation}
\begin{split}
    C_{\lambda_s}^{(\text{YM},2)} 
    &= \Tr(-\comm{A^{(1)}}{\lambda_s}  \wedge \star \DD^{(0)} A^{(1)} + A^{(1)} \wedge \comm{\star \DD^{(0)} A^{(1)}}{\lambda_s} ) 
    + \dd \Tr( \lambda_s\star \left(\DD^{(0)} A^{(2)} + \comm{A^{(1)}}{A^{(1)}} \right)) \\
    &= 2\Tr(A^{(1)} \wedge \comm{\star \DD^{(0)} A^{(1)}}{\lambda_s} ) 
    + \dd \Tr( \lambda_s \star \left(\DD^{(0)} A^{(2)} + \comm{A^{(1)}}{A^{(1)}} \right)),
\end{split}
\end{equation}
where for the first two terms we made use of a similar identity as \eqref{eq:traceSwitch}.
The same results can be confirmed by directly using equations \eqref{eq: first order constraint, general} and \eqref{eq: second order constraint, general} in terms of $\mathcal{C}_\lambda^{(\text{YM})}$,  $\omega^{(\text{YM})}$, and $B_\lambda^{(\text{YM})}$ computed earlier in this appendix.

\subsection{Scalar electrodynamics}
\label{app:sED}

Let us derive the results from Subsect.~\ref{ssec_QED} where we look at scalar electrodynamics (sED) on a fixed background with Lagrangian form
\begin{equation}
    L^{(\text{sED})} = -\frac{1}{2} F \wedge \star F - ((\DD \phi)^* \wedge \star \DD \phi - m^2 \phi^* \phi \eps), \quad \DD = \dd + iA.
\end{equation}
Variation of the Lagrangian form leads to 
\begin{align}
    \delta L^{(\text{sED})} =~&- \delta A \wedge \dd \star F - \dd(\delta A \wedge \star F) + \delta A \wedge \star j \nonumber \\
    &-  (\DD \delta \phi)^* \wedge \star \DD \phi - \DD \delta \phi \wedge \star (\DD \phi)^*  + m^2 (\delta \phi^* \phi + \phi^* \delta \phi) \eps, 
\end{align}
with the covariant definition of the $U(1)$ current
\begin{equation}
    \star j = \frac{1}{\sqrt{-g}} \fdv{S_\text{matter}}{A} = i (\phi^* \star \DD \phi - (\star \DD \phi)^* \phi).
\end{equation}
To massage it into a form from which we can read off the equations of motion and the symplectic prepotential, let us focus on the $\delta \phi$ part first and use the Leibniz rule on the exterior derivative $\dd$, then
\begin{equation}
    -\DD \delta \phi \wedge \star (\DD \phi)^* + m^2  \phi^* \delta\phi \eps 
    = -\dd(\delta \phi \wedge \star (\DD \phi)^*) + (\DD^* \star (\DD \phi)^* + m^2\phi^* \eps) \delta \phi. 
\end{equation}
A similar result holds for the $\delta \phi^*$ part such that we can rewrite $\delta L^{(\text{sED})}$ as 
\begin{equation}
    \begin{split}
        \delta L^{(\text{sED})} = &-\delta A \wedge (\dd \star F - \star j) - \dd(\delta A \wedge \star F) \nonumber \\
    &+ (\DD \star \DD \phi + m^2 \phi \eps) \delta\phi^* + ((\DD \star \DD \phi)^* + m^2 \phi^* \eps) \delta \phi \\
    &- \dd(\delta \phi \wedge \star (\DD \phi)^* + \delta \phi^* \wedge \star \DD \phi).
    \end{split}
\end{equation}
We can thus read off the equations of motion
\begin{equation}
    \dd \star F = \star j, \quad (\DD \star \DD + m^2 \eps) \phi = 0,
\end{equation}
and the presympletic potential and current 
\begin{align}
    \theta^{(\text{sED})} &= - \delta A \wedge \star F - \delta \phi \wedge \star (\DD \phi)^* - \delta \phi^* \wedge \star \DD \phi \\ 
    &\implies\\ 
    \omega^{(\text{sED})} &= \delta A \wedge \delta \star F  + \delta \phi \wedge \star \delta (\DD \phi)^* + \delta \phi^* \wedge \star \delta \DD \phi.
\end{align}

To find the associated Noether current \eqref{eq: define J general} we use the infinitesimal gauge transformation
\begin{equation}
    \hat{\lambda}\fins \delta \phi = - i\lambda \phi, \quad\hat{\lambda}\fins\delta A = \dd \lambda,
\end{equation}
such that
\begin{equation}
    \hat{\lambda} \cdot \theta^{(\text{sED})} = - \dd (\lambda \star F) +  \lambda (\dd \star F -  \star j).
\end{equation}
We also require
\begin{align}
    \hat{\lambda} \cdot \delta L^{(\text{sED})} = i \lambda \phi^* \DD \star \DD \phi - i \lambda \phi(\DD \star \DD)^*  \phi^*  - \lambda \dd \star j,
\end{align}
in which the last term can be written as 
\begin{align}
    d \star j  
    = i (\phi^* \DD \star \DD \phi - (\DD \star \DD \phi)^* \phi),
\end{align}
This then leads to 
\begin{equation}
    \hat{\lambda} \cdot \delta L^{(\text{sED})} = 0.
\end{equation}
Therefore, the Noether current is simply 
\begin{equation}
    J^{(\text{sED})}_\lambda =  \lambda (\dd \star F - \lambda \star j) - \dd (\lambda \star F),
\end{equation}
from which we read off
\begin{equation}
    C^{(\text{sED})}_\lambda = \lambda (\dd \star F - \star j), \quad Q^{(\text{sED})}_\lambda = - \lambda \star F.
\end{equation}

\subsection{Non-Abelian Chern-Simons theory}
\label{app:CS}

We start with the Lagrangian form for 3d non-Abelian Chern--Simons theory
\begin{equation}\label{eq: CS lagrangian}
    L^{(\text{CS})} = \frac{1}{2}\Tr(A \wedge \dd{A} + \frac{2}{3} A \wedge A \wedge A),
\end{equation}
Under an infinitesimal gauge transformation \eqref{eq: infinitesimal gauge action}, this Lagrangian changes by a boundary term, which we can write as
\begin{equation}\begin{split}
    \hat{\lambda}\fins\fd L^{(\text{CS})} &= \frac{1}{2}\sd\Tr(A\wedge \sd \lambda).\\
\end{split}\end{equation}
We can then take
\begin{equation}\begin{split}
    \alpha^{(\text{CS})}_\lambda&:=\frac{1}{2}\Tr(A\wedge\sd\lambda),
\end{split}\end{equation}
where $\alpha_{X}$ was defined for general gauge transformations in \eqref{eq: covariant L general}. 
The equations of motion and symplectic potential form for this Lagrangian follow from
\begin{equation}\begin{split}
    \delta L^{(\text{CS})} 
    &=\Tr(\fd A\wedge F) +\frac{1}{2}\sd\Tr(\fd A\wedge A)\\
    &\implies\\
    E^{(\text{CS})}:=\Tr(\fd A\wedge F), \qquad \theta^{(\text{CS})}&:=\frac{1}{2}\Tr(\fd A\wedge A), \qquad
    \omega^{(CS)}:=-\frac{1}{2}\Tr(\fd A\wedge \fd A).
\end{split}\end{equation}
Referring to equations \eqref{eq: define J general} and \eqref{eq: decompose J general}, we write the gauge current and its decomposition as 
\begin{equation}\begin{split}
    J^{(\text{CS})}_{\lambda}&:=\hat{\lambda}_{\lambda}\fins\theta^{(\text{CS})} -\alpha^{(\text{CS})}_\lambda\\
    &=-\Tr(\lambda F)+\sd\Tr(\lambda A)\\
    &\implies\\
    C^{(\text{CS})}_\lambda&:=-\Tr(\lambda F), \qquad Q^{(\text{CS})}_\lambda:=\Tr(\lambda A).
\end{split}\end{equation}
We next seek the analogous decomposition of the terms $\fL_{\hat{X}}\theta-\fd\alpha_{\hat{X}}$ (referring to equation \eqref{eq: define tilde C, tilde q general}), which leads to 
\begin{equation}\begin{split}
    \fL_{\hat{\lambda}}\theta^{(\text{CS})}-\delta\alpha^{(\text{CS})}_{\hat{\lambda}}
    &=-\Tr(\fd\lambda~ F)+\sd\Tr(\fd\lambda~ A)\\
    &\implies\\
    \mathcal{C}^{(\text{CS})}_{\lambda}&:=-\Tr(\delta\lambda~ F),
    \qquad
    \mathcal{Q}^{(\text{CS})}_{\lambda}:=\Tr(\delta\lambda~ A).
\end{split}\end{equation}
Again, for generality we have allowed the gauge parameters $\lambda$ to be field dependent.  We can now write the boundary piece \eqref{eq: boundary term general} for non-Abelian Chern-Simons as
\begin{equation}\label{eq: boundary term, CS}
    B^{(\text{CS})}_\lambda:=\mathcal{Q}^{(\text{CS})}_\lambda-\fd Q^{(\text{CS})}_\lambda
    =-\Tr(\lambda~ \fd A).
\end{equation}
One can confirm that the variation of the constraint decomposes according to equation \eqref{eq: delta C general 2} as
\begin{equation}\begin{split}\label{eq: delta C CS}
    \delta C^{(\text{CS})}_{\lambda }
    &=-\hat{\lambda}\cdot \omega^{(\text{CS})} + \mathcal{C}^{(\text{CS})}_{\lambda}+\extd B^{(\text{CS})}_\lambda\\
    &=-\hat{\lambda}\cdot \omega^{(\text{CS})} - \Tr(\delta\lambda~ F)-\sd\Tr(\lambda~ \fd A).
\end{split}\end{equation}

Let us now perform the perturbative analysis similar to the Yang--Mills case. We will start from the form $\eqref{eq: delta C CS}$ with presymplectic current \eqref{eq:CSpresymplcurrent}, hence 
\begin{equation}
    \delta C^{(\text{CS})}_{\lambda} 
    =  \Tr(\DD \lambda \wedge \delta A - \delta\lambda F)-\sd\Tr(\lambda \fd A).
\end{equation}
From this expression follows 
\begin{equation}
    \begin{split}
        C_\lambda^{(\text{CS},1)} &=  \Tr(\DD \lambda \wedge \dv{A}{\eps} - \dv{\lambda}{\eps} F)  \eval_{\eps = 0} -  \dd \Tr(\lambda \dv{A}{\eps}) \eval_{\eps = 0}
        \\
        &=  \Tr(\DD^{(0)} \lambda^{(0)} \wedge A^{(1)} - \lambda^{(1)} F^{(0)}) -  \dd \Tr(\lambda^{(0)} A^{(1)}).
    \end{split}
\end{equation}
With the assumptions of background invariance $\DD^{(0)} \lambda_s^{(0)} = 0$, the background being a solution to $F^{(0)} = 0$, and field-independence of $\lambda_s$, we arrive at the first-order constraint density quoted in the main body \eqref{eq:CSconstraint1}
\begin{equation}
    C_{\lambda_s}^{(\text{CS},1)} = -  \dd \Tr(\lambda_s A^{(1)}).
\end{equation}
Then, for the second-order constraint density we can compute
\begin{equation}
    \begin{split}
        C_\lambda^{(\text{CS},2)} &=  \Tr( \left(\DD \dv{\lambda}{\eps} + \comm{\dv{A}{\eps}}{\lambda} \right) \wedge \dv{A}{\eps} + \DD \lambda \wedge \dv[2]{A}{\eps}) \eval_{\eps = 0}
        \\
        &\hspace{10pt} -  \Tr( \dv[2]{\lambda}{\eps} F +\dv{\lambda}{\eps} \DD \dv{A}{\eps}) \eval_{\eps = 0}
        \\
        &\hspace{10pt} -  \dd \Tr(\dv{\lambda}{\eps} \dv{A}{\eps} + \lambda \dv[2]{A}{\eps}) \eval_{\eps = 0}
        \\&=  \Tr( \left(\DD^{(0)} \lambda^{(1)} + \comm{A^{(1)}}{\lambda^{(0)}} \right) \wedge A^{(1)} + \DD^{(0)} \lambda^{(0)} \wedge A^{(2)})
        \\&\hspace{10pt} -  \Tr( \lambda^{(2)} F^{(0)}  + \lambda^{(1)} \DD^{(0)} A^{(1)})
        \\&\hspace{10pt} -  \dd \Tr(\lambda^{(1)} A^{(1)} + \lambda^{(0)} A^{(2)}).
    \end{split}
\end{equation}
Finally, under same background symmetry assumptions we arrive at \eqref{eq:CSconstraint2}
\begin{equation}
    C_{\lambda_s}^{(\text{CS},2)} =  \Tr( \comm{A^{(1)}}{\lambda_s} \wedge A^{(1)})
    -  \dd \Tr(\lambda_s A^{(2)}).
\end{equation}

\section{Perturbative constraints with explicit parameter dependence}
\label{App: perturbative constraints with explicit dependence}
The formalism outlined in section \ref{subsec: perturbative constraints} for establishing perturbative constraints assumes that the non-perturbative constraints have no \textit{explicit} dependence on the perturbative parameter.
We can also consider the case that the Lagrangian, and therefore constraints, depend explicitly on the perturbative parameter $\varepsilon$ (as in the case of gravity when the matter Lagrangian and gravitational Lagrangian differ by an explicit power of $\kappa^2$). To establish a perturbative expansion of the constraint, we assume the Lagrangian has been rescaled by a multiple of $\varepsilon$ such that the $\varepsilon\rightarrow 0$ limit gives a finite result. 
We can then utilize equation \eqref{eq: delta C general 2} to write the analogues \eqref{eq: first order constraint, general} and \eqref{eq: second order constraint, general}.  
The background ``on-shell" condition is now
$E[\Phi^{(0)},\epsilon=0;\delta\Phi]=0$, which also implies $\mathcal{C}_X[\Phi^{(0)},\varepsilon=0;\delta\Phi]=0$. Note that the $\epsilon=0$ argument is important because the order at which various field equations should be imposed can now differ based on the explicit $\epsilon$ dependence.  For example in the gravity case, the $g^{(0)}$ equation of motion enters at a lower order than the $\phi^{(0)}$ equation of motion for some matter scalar $\Phi$, because the latter only enters in a matter Lagrangian which is offset by $\kappa^2$ from the gravity Lagrangian.

Working on shell of $\Phi^{(0)}$,
\begin{equation}\begin{split}\label{eq: first order constraint with explicit parameter, general}
     &C_{X}^{(1)}:=\left(\left(\fL_{\hat{V}}+\frac{\partial}{\partial\varepsilon}\right)C_{X}[\Phi,\varepsilon]\right)\bigg|_{\varepsilon=0}\\
     &= \left(\mathcal{C}_X\left[\Phi,\varepsilon;\frac{\sd\Phi}{\sd\varepsilon}\right]-\omega\left[\Phi,\varepsilon;\delta_X \Phi,\frac{\extd \Phi}{\extd \varepsilon}\right]+ \sd B_{X}\left[\Phi,\varepsilon; \frac{\extd \Phi}{\extd \varepsilon}\right]
     +\frac{\partial}{\partial\varepsilon}C_X[\Phi,\epsilon]\right)\bigg|_{\varepsilon=0}\\
     &= \mathcal{C}_X\left[\Phi^{(0)},\varepsilon=0;\Phi^{(1)}\right]
     -\omega\left[\Phi^{(0)},\varepsilon=0;\delta_X\Phi^{(0)}, \Phi^{(1)}\right]+ \sd B_{X}\left[\Phi^{(0)},\varepsilon=0; \Phi^{(1)}\right]
     +\left(\frac{\partial}{\partial\varepsilon}C_X[\Phi^{(0)},\epsilon]\right)\bigg|_{\varepsilon=0}\,,
\end{split}\end{equation}
and
\begin{equation}\begin{split}\label{eq: second order constraint with explicit parameter, general}
     C_{X}^{(2)}&:=\left(\left(\fL_{\hat{V}}+\frac{\partial}{\partial\varepsilon}\right)^2 C_{X}[\Phi,\varepsilon]\right)\bigg|_{\varepsilon=0}\\
     &= \left(2\frac{\partial}{\partial\varepsilon}\left(\mathcal{C}_X\left[\Phi,\varepsilon;\frac{\sd\Phi}{\sd\varepsilon}\right]-\omega\left[\Phi,\varepsilon;\delta_X \Phi,\frac{\extd \Phi}{\extd \varepsilon}\right]+ \sd B_{X}\left[\Phi,\varepsilon; \frac{\extd \Phi}{\extd \varepsilon}\right]\right)
     +\frac{\partial^2}{\partial\varepsilon^2}C_X[\Phi,\epsilon]\right)\bigg|_{\varepsilon=0}\\
     &\qquad+\left(\frac{\sd}{\sd\varepsilon}\left(\mathcal{C}_X\left[\Phi,\varepsilon=0;\frac{\sd\Phi}{\sd\varepsilon}\right]-\omega\left[\Phi,\varepsilon=0;\delta_X \Phi,\frac{\extd \Phi}{\extd \varepsilon}\right]+ \sd B_{X}\left[\Phi,\varepsilon=0; \frac{\extd \Phi}{\extd \varepsilon}\right]\right)\right)\bigg|_{\varepsilon=0}
     \\
     &= \left(2\frac{\partial}{\partial\varepsilon}\left(\mathcal{C}_X\left[\Phi^{(0)},\varepsilon;\Phi^{(1)}\right]-\omega\left[\Phi^{(0)},\varepsilon;\delta_X \Phi^{(0)},\Phi^{(1)}\right]+ \sd B_{X}\left[\Phi^{(0)},\varepsilon; \Phi^{(1)}\right]\right)
     +\frac{\partial^2}{\partial\varepsilon^2}C_X[\Phi^{(0)},\epsilon]\right)\bigg|_{\varepsilon=0}\\
     &\qquad+\left(\frac{\sd}{\sd\varepsilon}\left(\mathcal{C}_X\left[\Phi,\varepsilon=0;\Phi^{(1)}\right]-\omega\left[\Phi,\varepsilon=0;\delta_X \Phi^{(0)},\Phi^{(1)}\right]+ \sd B_{X}\left[\Phi,\varepsilon=0; \Phi^{(1)}\right]\right)\right)\bigg|_{\varepsilon=0}
     \\
     &\qquad+\left(
     -\omega\left[\Phi^{(0)},\varepsilon=0;\delta_X \Phi^{(1)},\Phi^{(1)}\right]
     -\omega\left[\Phi^{(0)},\varepsilon=0;\delta_X \Phi^{(0)},\Phi^{(2)}\right]+ \sd B_{X}\left[\Phi^{(0)},\varepsilon=0; \Phi^{(2)}\right]\right)\,.
\end{split}\end{equation}
These expressions are rather complicated, but they simplify substantially if the explicit dependence on $\varepsilon$ is at least quadratic (as in the case of the Einstein-Hilbert action coupled some matter Lagrangian differing by an explicit overall factor of $\kappa^2$):
\begin{equation}\begin{split}\label{eq: first order constraint, quadratic parameter, general}
     C_{X}^{(1)}&= -\omega\left[\Phi^{(0)},\varepsilon=0;\delta_X\Phi^{(0)}, \Phi^{(1)}\right]+ \sd B_{X}\left[\Phi^{(0)},\varepsilon=0; \Phi^{(1)}\right]\,,
\end{split}\end{equation}
and
\begin{equation}\begin{split}\label{eq: second order constraint, quadratic parameter, general}
     &C_{X}^{(2)}=\\
     &\left(\frac{\partial^2}{\partial\varepsilon^2}C_X[\Phi^{(0)},\epsilon]\right)\bigg|_{\varepsilon=0}
     +\left(\frac{\sd}{\sd\varepsilon}\left(\mathcal{C}_X\left[\Phi,\varepsilon=0;\Phi^{(1)}\right]-\omega\left[\Phi,\varepsilon=0;\delta_X \Phi^{(0)},\Phi^{(1)}\right]+ \sd B_{X}\left[\Phi,\varepsilon=0; \Phi^{(1)}\right]\right)\right)\bigg|_{\varepsilon=0}
     \\
     &\qquad+\left(
     -\omega\left[\Phi^{(0)},\varepsilon=0;\delta_X \Phi^{(1)},\Phi^{(1)}\right]
     -\omega\left[\Phi^{(0)},\varepsilon=0;\delta_X \Phi^{(0)},\Phi^{(2)}\right]+ \sd B_{X}\left[\Phi^{(0)},\varepsilon=0; \Phi^{(2)}\right]\right)\,.
\end{split}\end{equation}
We see in that in such cases, when there is a background symmetry $\fd_{X_s}\Phi^{(0)}=0$, the first-order constraint density is again an exact form, and in the second-order constraint, the second order fields $\Phi^{(2)}$ enter in an identical exact form:
\begin{equation}\begin{split}\label{eq: first order constraint, quadratic parameter with sym, general}
     C_{X_s}^{(1)}&= \sd B_{X_s}\left[\Phi^{(0)},\varepsilon=0; \Phi^{(1)}\right]\,,
\end{split}\end{equation}
and
\begin{equation}\begin{split}\label{eq: second order constraint, quadratic parameter with sym, general}
     &C_{X_s}^{(2)}=
     \left(\frac{\partial^2}{\partial\varepsilon^2}C_{X_s}[\Phi^{(0)},\epsilon]\right)\bigg|_{\varepsilon=0}
     +\left(\frac{\sd}{\sd\varepsilon}\left(\mathcal{C}_{X_s}\left[\Phi,\varepsilon=0;\Phi^{(1)}\right]+ \sd B_{X_s}\left[\Phi,\varepsilon=0; \Phi^{(1)}\right]\right)\right)\bigg|_{\varepsilon=0}
     \\
     &\qquad+\left(
     -\omega\left[\Phi^{(0)},\varepsilon=0;\delta_{X_s} \Phi^{(1)},\Phi^{(1)}\right]+ \sd B_{X_s}\left[\Phi^{(0)},\varepsilon=0; \Phi^{(2)}\right]\right)\,.
\end{split}\end{equation}

\subsection{Gravitational case}\label{app_lalala}

Suppose we now want to find perturbative solutions around some matter-free background as in Sec.~\ref{sec:gravitonsandmatter}, and treating $\kappa=\sqrt{32\pi G_N}$ as a formal parameter (taking the place of $\varepsilon$ in section \ref{subsec: perturbative constraints}).
Retaining the notation that the background configuration is $\Phi^{(0)}:=\Phi|_{\kappa=0}$, it is useful to rescale the constraint and all associated quantities in \eqref{eq: Hamilton's equation} by $\kappa^2$, so that the leading order gravitational Lagrangian density is finite in the $\kappa\rightarrow 0$ limit. The matter Lagrangian then comes with an explicit power of $\kappa^2$, so it will not contribute to the constraints until second order. The first-order constraint density (for an on-shell background configuration) is given by
\begin{equation}\begin{split}
    C^{(1)}_\xi:=\left[\left(V\fins\fd+\frac{\partial}{\partial\kappa}\right)C_\xi\right]\bigg|_{\kappa=0}
     = -\omega\left[\Phi^{(0)},\kappa=0;\sL_\xi\Phi^{(0)}, \Phi^{(1)}\right]+ \sd B_{\xi}\left[\Phi^{(0)},\kappa=0; \Phi^{(1)}\right]
\end{split}\end{equation}
while the second-order constraint density (for an on-shell background \textit{and} first order configuration) is
\begin{equation}\begin{split}\label{eq: second order constraint, diffeo app}
     &C_\xi^{(2)}
      =\left[\left(V\fins\fd+\frac{\partial}{\partial\kappa}\right)^2C_\xi\right]\bigg|_{\kappa=0}
      =\left[\hat{V}\fins\fd~ \hat{V}\fins\fd C_\xi+\frac{\partial^2 C_\xi}{\partial\kappa^2}\right]\bigg|_{\kappa=0}\\
      &
      =
      -\omega\left[\Phi^{(0)},\kappa=0;\sL_\xi\Phi^{(1)},\Phi^{(1)}\right]
      +\left(\frac{\extd}{\extd\kappa}\left(
      - \omega\left[\Phi,\kappa=0; \sL_\xi\Phi^{(0)},\Phi^{(1)}\right]
      +\sd B_\xi\left[\Phi,\kappa=0;\Phi^{(1)}\right]
      \right)\right)\bigg|_{\kappa=0}
      \\
      &\quad
      -\omega\left[\Phi^{(0)},\kappa=0;\sL_\xi\Phi^{(0)},\Phi^{(2)}\right]
      +\sd B_\xi\left[\Phi^{(0)},\kappa=0;\Phi^{(2)}\right]+\frac{\partial^2}{\partial\kappa^2}C_\xi[\Phi^{(0)}]\bigg|_{\kappa=0}\,.
\end{split}\end{equation}
In both constraints we have used the fact that the explicit $\kappa$ dependence is quadratic.  For all but the very last term of the second constraint, setting $\kappa=0$ eliminates the contribution from the matter Lagrangian.  This has the implication that these terms depend only on the metric (so that the $\Phi^{(n)}$ in those terms can be replaced with just the metric degrees of freedom $g^{(n)}$). For any background metric with a Killing symmetry $\xi_s$, we then have
\begin{equation}\begin{split}\label{eq: first order constraint, diffeo with sym app}
     C_{\xi_s}^{(1)}
     &= \sd B_{\xi_s}\left[\Phi^{(0)},\kappa=0; \Phi^{(1)}\right],
\end{split}\end{equation}
and
\begin{equation}\begin{split}\label{eq: second order constraint, diffeo with sym app}
     C_{\xi_s}^{(2)}
      =&-\omega\left[\Phi^{(0)},\kappa=0;\sL_{\xi_s}\Phi^{(1)},\Phi^{(1)}\right]
      +\left(\frac{\extd}{\extd\kappa}\sd B_{\xi_s}\left[\Phi,\kappa=0;\Phi^{(1)}\right]
      \right)\bigg|_{\kappa=0}\\
      &~
      +\sd B_{\xi_s}\left[\Phi^{(0)},\kappa=0;\Phi^{(2)}\right]
      +\frac{\partial^2}{\partial\kappa^2}C_\xi[\Phi^{(0)},\kappa]\bigg|_{\kappa=0}\,.
\end{split}\end{equation}

Let us now show how to write the $\omega$ term in~\eqref{eq: second order constraint, diffeo with sym app} in terms of a stress tensor for gravitons and matter, in the case of general relativity. Note that setting $\kappa=0$ in $\omega$ leaves just the Einstein-Hilbert contribution. For the (rescaled by $\kappa^2$) Einstein-Hilbert action we have
\begin{equation}\begin{split}
    \fd L_{\text{EH}} = -2 G^{\mu\nu}\fd g_{\mu\nu}\epsilon+\sd \theta_{\text{EH}}\\
\end{split}\end{equation}
and for the corresponding contribution to the current we have
\begin{equation}\begin{split}
    \hat{\xi_s}\fins\theta_{\text{EH}}-\xi\sins L_{\text{EH}}
    =\sd q_\xi +4\xi_\beta \epsilon_\alpha G^{\alpha\beta}
\end{split}\end{equation}
where $q_\xi$ is the standard gravitational charge aspect (multiplied by $\kappa^2$). This allows us to write
\begin{equation}
    -\hat{\xi}\fins\omega_{\text{EH}}
    =\xi\sins \left(-2G^{\mu\nu}\fd g_{\mu\nu}\epsilon\right)
    -\sd \left(\xi\sins\theta_{\text{EH}}-\fd q_\xi\right)
    +4\fd\left(\xi_\beta\epsilon_\alpha G^{\alpha\beta}\right),
\end{equation}
which then gives
\begin{equation}\begin{split}
    -\hat{V}\fins\hat{\xi}\fins\omega_{\text{EH}}
    &=-\omega_{\text{EH}}\left[g;\sL_\xi g, \frac{\sd g}{\sd\kappa}\right]\\
    &=-2G^{\mu\nu} \frac{\sd g_{\mu\nu}}{\sd\kappa}\xi^\alpha\epsilon_\alpha
    -\sd B_{\xi,\text{EH}}\left[g;\frac{\sd g}{\sd\kappa}\right]
    +4\frac{\sd}{\sd\kappa}\left(\xi_\beta\epsilon_\alpha G^{\alpha\beta}\right),
\end{split}\end{equation}
where $B_{\xi\text{EH}}:=\xi\sins\theta_{\text{EH}}-\fd q_\xi$.  Acting again with $\hat{V}\fins \fd$ gives
\begin{equation}\begin{split}
    -\omega_{\text{EH}}\left[g;\sL_\xi \frac{\sd g}{\sd\kappa}, \frac{\sd g}{\sd\kappa}\right]+\dots
    &=-2\frac{\sd}{\sd\kappa}\left(G^{\mu\nu} \frac{\sd g_{\mu\nu}}{\sd\kappa}\xi^\alpha\epsilon_\alpha\right)
    -\sd \frac{\sd}{\sd\kappa}B_{\xi,\text{EH}}\left[g;\frac{\sd g}{\sd\kappa}\right]
    +4\frac{\sd^2}{\sd\kappa^2}\left(\xi_\beta\epsilon_\alpha G^{\alpha\beta}\right),
\end{split}\end{equation}
where the $\dots$ on the right-hand side indicates pieces that will vanish when $\sL_\xi g=0$. If we evaluate this at $\kappa=0$, on-shell of the background and first order metric (note that although $G^{\mu\nu}$ is not the full metric equations of motion, it \textit{is} sufficient for the zero-th and first order equations of motion since the matter stress tensor only enters at order $\kappa^2$), we get
\begin{equation}\begin{split}\label{eq: intermediate omega EH}
    -\omega_{\text{EH}}\left[g^{(0)};\sL_\xi  g^{(1)}, g^{(1)}\right]+\dots
    &=
    -\sd \left(\frac{\sd}{\sd\kappa}B_{\xi,\text{EH}}\left[g;g^{(1)}\right]\big|_{\kappa=0}+B_{\xi,\text{EH}}\left[g^{(0)};g^{(2)}\right]\right)\\
    &\qquad
    +4 \qty(G^{\alpha(2)}_{~\beta}(g^{(1)},g^{(1)}) + G^{\alpha(1)}_{~\beta}(g^{(2)})) \xi^\beta\epsilon^{(0)}_\alpha.
\end{split}\end{equation}
The terms involving $g^{(2)}$ cancel out (this is just the off-shell first-order constraint) leaving only
\begin{equation}\begin{split}\label{eq: final omega EH}
    -\omega_{\text{EH}}\left[g^{(0)};\sL_\xi g^{(1)}, g^{(1)}\right]+\dots
    &=
    -\sd \frac{\sd}{\sd\kappa}B_{\xi,\text{EH}}\left[g;g^{(1)}\right]\big|_{\kappa=0}
    +4 G^{\alpha(2)}_{~\beta}(g^{(1)},g^{(1)})\xi^\beta\epsilon^{(0)}_\alpha.
\end{split}\end{equation}
Employing this in equation \eqref{eq: second order constraint, diffeo with sym app} lets us re-express the second-order constraint as
\begin{equation}\begin{split}
     C_{\xi_s}^{(2)}
      &=4G^{\alpha(2)}_{~\beta}[g^{(1)},g^{(1)}]\xi^\beta\epsilon^{(0)}_{\alpha}
      +\sd B_{\xi_s}\left[\Phi^{(0)},\kappa=0;\Phi^{(2)}\right]
      +\frac{\partial^2}{\partial\kappa^2}C_\xi[\Phi^{(0)}]\bigg|_{\kappa=0}\\
      &=\left(4G^{\alpha(2)}_{~\beta}[g^{(1)},g^{(1)}]-2 T^{\alpha(0)}_{~\beta}\right)\xi^\beta\epsilon^{(0)}_{\alpha}
      +\sd B_{\xi_s,\text{EH}}\left[g^{(0)},\Phi^{(2)}\right],
\end{split}\end{equation}
where $T^{\alpha(0)}_{~\beta}$ is the background matter stress tensor (which comes from the final term in~\eqref{eq: second order constraint, diffeo with sym app}). In particular, changing the notation $g^{(1)}\to h$, and integrating over a Cauchy surface $\Sigma$, one recovers Eq.~\eqref{2ndorder}.

\printbibliography

@article{Kudler-Flam:2023qfl,
    author = "Kudler-Flam, Jonah and Leutheusser, Samuel and Satishchandran, Gautam",
    title = "{Generalized Black Hole Entropy is von Neumann Entropy}",
    eprint = "2309.15897",
    archivePrefix = "arXiv",
    primaryClass = "hep-th",
    month = "9",
    year = "2023"
}

@article{AliAhmad:2024vdw,
    author = "Ali Ahmad, Shadi and Chemissany, Wissam and Klinger, Marc S. and Leigh, Robert G.",
    title = "{Relational Quantum Geometry}",
    eprint = "2410.11029",
    archivePrefix = "arXiv",
    primaryClass = "hep-th",
    month = "10",
    year = "2024"
}

@article{Witten:2023xze,
    author = "Witten, Edward",
    title = "{A background-independent algebra in quantum gravity}",
    eprint = "2308.03663",
    archivePrefix = "arXiv",
    primaryClass = "hep-th",
    doi = "10.1007/JHEP03(2024)077",
    journal = "JHEP",
    volume = "03",
    pages = "077",
    year = "2024"
}

@article{Saraykar2,
    author = {Saraykar, R. V. and Joshi, N. E.},
    title = {Erratum: Linearization stability of Einstein equations coupled with self‐gravitating scalar fields [J. Math. Phys. 22, 343 (1981)]},
    journal = {Journal of Mathematical Physics},
    volume = {23},
    number = {9},
    pages = {1738-1738},
    year = {1982},
    month = {09},
    issn = {0022-2488},
    doi = {10.1063/1.525544},
    url = {https://doi.org/10.1063/1.525544},
}

@article{Fursaev:1998hr,
    author = "Fursaev, Dmitri V.",
    title = "{Energy, Hamiltonian, Noether charge, and black holes}",
    eprint = "hep-th/9809049",
    archivePrefix = "arXiv",
    doi = "10.1103/PhysRevD.59.064020",
    journal = "Phys. Rev. D",
    volume = "59",
    pages = "064020",
    year = "1999"
}

@article{DVEHK,
    author = {De Vuyst, Julian and Eccles, Stefan and H\"ohn, Philipp A. and Kirklin, Josh},
    title = "{Crossed products and quantum reference frames: on the observer-dependence of gravitational entropy}",
    eprint = "2412.15502",
    archivePrefix = "arXiv",
    primaryClass = "hep-th",
    month = "12",
    year = "2024"
}

@article{Arms:1986vk,
    author = "Arms, J. M. and Anderson, I. M.",
    title = "{Perturbations of conservation laws in field theories}",
    doi = "10.1016/0003-4916(86)90206-X",
    journal = "Annals Phys.",
    volume = "167",
    pages = "354--389",
    year = "1986"
}

@article{Abbott:1981ff,
    author = "Abbott, L. F. and Deser, Stanley",
    title = "{Stability of Gravity with a Cosmological Constant}",
    reportNumber = "CERN-TH-3136",
    doi = "10.1016/0550-3213(82)90049-9",
    journal = "Nucl. Phys. B",
    volume = "195",
    pages = "76--96",
    year = "1982"
}

@article{Alonso-Monsalve:2024oii,
    author = "Alonso-Monsalve, Elba and Harlow, Daniel and Jefferson, Patrick",
    title = "{Phase space of Jackiw-Teitelboim gravity with positive cosmological constant}",
    eprint = "2409.12943",
    archivePrefix = "arXiv",
    primaryClass = "hep-th",
    reportNumber = "MIT-CTP/5769",
    month = "9",
    year = "2024"
}

@article{Kudler-Flam:2023hkl,
    author = "Kudler-Flam, Jonah and Leutheusser, Samuel and Rahman, Adel A. and Satishchandran, Gautam and Speranza, Antony J.",
    title = "{A covariant regulator for entanglement entropy: proofs of the Bekenstein bound and QNEC}",
    eprint = "2312.07646",
    archivePrefix = "arXiv",
    primaryClass = "hep-th",
    month = "12",
    year = "2023"
}

@article{Yngvason_2005,
   title={The role of type III factors in quantum field theory},
   volume={55},
   ISSN={0034-4877},
   url={http://dx.doi.org/10.1016/S0034-4877(05)80009-6},
   DOI={10.1016/s0034-4877(05)80009-6},
   number={1},
   journal={Reports on Mathematical Physics},
   publisher={Elsevier BV},
   author={Yngvason, Jakob},
   year={2005},
   month=feb, pages={135–147} 
}

@article{DeVuyst:2024pop,
    author = {De Vuyst, Julian and Eccles, Stefan and H\"ohn, Philipp A. and Kirklin, Josh},
    title = "{Gravitational entropy is observer-dependent}",
    eprint = "2405.00114",
    archivePrefix = "arXiv",
    primaryClass = "hep-th",
    month = "4",
    year = "2024"
}

@article{Hoehn:2023axh,
    author = {H\"ohn, Philipp A. and Russo, Andrea and Smith, Alexander R. H.},
    title = "{Matter relative to quantum hypersurfaces}",
eprint = "2308.12912",
    archivePrefix = "arXiv",
    primaryClass = "quant-ph",
    doi = "10.1103/PhysRevD.109.105011",
    journal = "Phys. Rev. D",
    volume = "109",
    number = "10",
    pages = "105011",
    year = "2024"
}

@article{Chen:2024rpx,
    author = "Chen, Chang-Han and Penington, Geoff",
    title = "{A clock is just a way to tell the time: gravitational algebras in cosmological spacetimes}",
    eprint = "2406.02116",
    archivePrefix = "arXiv",
    primaryClass = "hep-th",
    month = "6",
    year = "2024"
}

@article{Moncrief:1977hy,
    author = "Moncrief, V.",
    title = "{Gauge Symmetries of Yang-Mills Fields}",
    doi = "10.1016/0003-4916(77)90018-5",
    journal = "Annals Phys.",
    volume = "108",
    pages = "387--400",
    year = "1977"
}

@article{Gotay:1997eg,
    author = "Gotay, Mark J. and Isenberg, James and Marsden, Jerrold E.",
    title = "{Momentum maps and classical relativistic fields. Part 1: Covariant Field Theory}",
    eprint = "physics/9801019",
    archivePrefix = "arXiv",
    month = "11",
    year = "1997"
}

@article{arms1981symmetry,
  title={Symmetry and bifurcations of momentum mappings},
  author={Arms, Judith M and Marsden, Jerrold E and Moncrief, Vincent},
  journal={Communications in Mathematical Physics},
  volume={78},
  pages={455--478},
  year={1981},
  publisher={Springer}
}

@article{Taub1,
    author = {Taub, A. H.},
    title = {Approximate Stress Energy Tensor for Gravitational Fields},
    journal = {Journal of Mathematical Physics},
    volume = {2},
    number = {6},
    pages = {787-793},
    year = {1961},
    month = {11},
    doi = {10.1063/1.1724224},
    url = {https://doi.org/10.1063/1.1724224},
}

@article{Arms_1981, title={The structure of the solution set for the Yang-Mills equations}, volume={90}, DOI={10.1017/S0305004100058813}, number={2}, journal={Mathematical Proceedings of the Cambridge Philosophical Society}, author={Arms, Judith M.}, year={1981}, pages={361–372}}

@article{Taub2,
  title={Stability of general relativistic gaseous masses and variational principles},
  author={Taub, AH},
  journal={Communications in Mathematical Physics},
  volume={15},
  number={3},
  pages={235--254},
  year={1969},
  publisher={Springer}
}

@book{Wald:1984rg,
    author = "Wald, Robert M.",
    title = "{General Relativity}",
    doi = "10.7208/chicago/9780226870373.001.0001",
    publisher = "Chicago Univ. Pr.",
    address = "Chicago, USA",
    year = "1984"
}

@article{Deser:1973zza,
    author = "Brill, D and Deser, Stanley",
    title = "{Instability of Closed Spaces in General Relativity}",
    doi = "10.1007/BF01645610",
    journal = "Commun. Math. Phys.",
    volume = "32",
    pages = "291",
    year = "1973"
}

@article{Moncrief:1976un,
    author = "Moncrief, V.",
    title = "{Space-Time Symmetries and Linearization Stability of the Einstein Equations. 2.}",
    doi = "10.1063/1.522814",
    journal = "J. Math. Phys.",
    volume = "17",
    pages = "1893--1902",
    year = "1976"
}

@article{Moncrief:1978te,
    author = "Moncrief, V.",
    title = "{Invariant States and Quantized Gravitational Perturbations}",
    doi = "10.1103/PhysRevD.18.983",
    journal = "Phys. Rev. D",
    volume = "18",
    pages = "983--989",
    year = "1978"
}

@article{Moncrief:1979bg,
    author = "Moncrief, V.",
    title = "{Quantum linearization instabilities}",
    doi = "10.1007/BF00756792",
    journal = "Gen. Rel. Grav.",
    volume = "10",
    pages = "93--97",
    year = "1979"
}

@article{Arms:1982ea,
    author = "Arms, J. M. and Marsden, J. E. and Moncrief, V.",
    title = "{The structure of the space of solutions of Einstein's equations. II. Several Killing fields and the Einstein-Yang-Mills equations}",
    doi = "10.1016/0003-4916(82)90105-1",
    journal = "Annals Phys.",
    volume = "144",
    pages = "81--106",
    year = "1982"
}

@article{Arms:1979au,
    author = "Arms, J. M.",
    title = "{Linearization stability of gravitational and gauge fields}",
    doi = "10.1063/1.524094",
    journal = "J. Math. Phys.",
    volume = "20",
    pages = "443--453",
    year = "1979"
}

@article{Arms1,
    author = {Arms, Judith M.},
    title = "{Linearization stability of the Einstein–Maxwell system}",
    journal = {Journal of Mathematical Physics},
    volume = {18},
    number = {4},
    pages = {830-833},
    year = {1977},
    month = {04},
    issn = {0022-2488},
    doi = {10.1063/1.523312},
    url = {https://doi.org/10.1063/1.523312},
}

@article{Kudler-Flam:2024psh,
    author = "Kudler-Flam, Jonah and Leutheusser, Samuel and Satishchandran, Gautam",
    title = "{Algebraic Observational Cosmology}",
    eprint = "2406.01669",
    archivePrefix = "arXiv",
    primaryClass = "hep-th",
    month = "6",
    year = "2024"
}

@article{Moncrief1,
    author = {Moncrief, Vincent},
    title = "{Spacetime symmetries and linearization stability of the Einstein equations. I}",
    journal = {Journal of Mathematical Physics},
    volume = {16},
    number = {3},
    pages = {493-498},
    year = {1975},
    month = {03},
    issn = {0022-2488},
    doi = {10.1063/1.522572},
    url = {https://doi.org/10.1063/1.522572},
}

@article{Brill:1968ca,
    author = "Brill, D. R. and Deser, Stanley",
    title = "{Variational methods and positive energy in general relativity}",
    doi = "10.1016/0003-4916(68)90131-0",
    journal = "Annals Phys.",
    volume = "50",
    pages = "548--570",
    year = "1968"
}

@article{Chataignier:2024eil,
    author = {Chataignier, Leonardo and H\"ohn, Philipp A. and Lock, Maximilian P. E. and Mele, Fabio M.},
    title = "{Relational Dynamics with Periodic Clocks}",
    eprint = "2409.06479",
    archivePrefix = "arXiv",
    primaryClass = "quant-ph",
    month = "9",
    year = "2024"
}

@article{Strohmaier:2023hhy,
    author = "Strohmaier, Alexander and Witten, Edward",
    title = "{Analytic States in Quantum Field Theory on Curved Spacetimes}",
    eprint = "2302.02709",
    archivePrefix = "arXiv",
    primaryClass = "math-ph",
    doi = "10.1007/s00023-024-01419-0",
    journal = "Annales Henri Poincare",
    volume = "25",
    number = "10",
    pages = "4543--4590",
    year = "2024"
}

@article{Kolchmeyer:2024fly,
    author = "Kolchmeyer, David K. and Liu, Hong",
    title = "{Chaos and the Emergence of the Cosmological Horizon}",
    eprint = "2411.08090",
    archivePrefix = "arXiv",
    primaryClass = "hep-th",
    reportNumber = "MIT-CTP/5805",
    month = "11",
    year = "2024"
}

@article{Deser:1968zzf,
    author = "Deser, Stanley and Brill, D. R. and Faddeev, L. D.",
    title = "{Sign of Gravitational Energy}",
    doi = "10.1016/0375-9601(68)90533-1",
    journal = "Phys. Lett. A",
    volume = "26",
    pages = "538--539",
    year = "1968"
}

@article{Saraykar:1981qm,
    author = "Saraykar, R. V. and Joshi, N. E.",
    title = "{Linearization stability of Einstein equations coupled with self-gravitating scalar fields}",
    doi = "10.1063/1.524885",
    journal = "J. Math. Phys.",
    volume = "22",
    pages = "343--347",
    year = "1981"
}

@article{Deser:1973zzb,
    author = "Deser, Stanley and Choquet-Bruhat, Y.",
    title = "{On the Stability of Flat Space}",
    doi = "10.1016/0003-4916(73)90484-3",
    journal = "Annals Phys.",
    volume = "81",
    pages = "165--178",
    year = "1973"
}

@incollection{choquet1979maximal,
  title={Maximal hypersurfaces and positivity of mass},
  author={Choquet-Bruhat, Yvonne and Fischer, Arthur E and Marsden, Jerrold E},
  booktitle={Isolated gravitating systems in general relativity},
  year={1979}
}

@article{Khavkine:2013iei,
    author = "Khavkine, Igor",
    title = "{Topology, rigid cosymmetries and linearization instability in higher gauge theories}",
    eprint = "1303.2406",
    archivePrefix = "arXiv",
    primaryClass = "math-ph",
    reportNumber = "ITP-UU-13-08, SPIN-13-06",
    doi = "10.1007/s00023-014-0321-9",
    journal = "Annales Henri Poincare",
    volume = "16",
    number = "1",
    pages = "255--288",
    year = "2015"
}

@inproceedings{fischer1980structure,
  title={The structure of the space of solutions of Einstein's equations. I. One Killing field},
  author={Fischer, Arthur E and Marsden, Jerrold E and Moncrief, Vincent},
  booktitle={Annales de l'institut Henri Poincar{\'e}. Section A, Physique Th{\'e}orique},
  volume={33},
  number={2},
  pages={147--194},
  year={1980}
}

@article{Christensen:1979iy,
    author = "Christensen, S. M. and Duff, M. J.",
    title = "{Quantizing Gravity with a Cosmological Constant}",
    reportNumber = "NSF-ITP-79-01",
    doi = "10.1016/0550-3213(80)90423-X",
    journal = "Nucl. Phys. B",
    volume = "170",
    pages = "480--506",
    year = "1980"
}

@article{Hollands:2012sf,
    author = "Hollands, Stefan and Wald, Robert M.",
    title = "{Stability of Black Holes and Black Branes}",
    eprint = "1201.0463",
    archivePrefix = "arXiv",
    primaryClass = "gr-qc",
    doi = "10.1007/s00220-012-1638-1",
    journal = "Commun. Math. Phys.",
    volume = "321",
    pages = "629--680",
    year = "2013"
}

@article{Donnelly:2016rvo,
    author = "Donnelly, William and Giddings, Steven B.",
    title = "{Observables, gravitational dressing, and obstructions to locality and subsystems}",
    eprint = "1607.01025",
    archivePrefix = "arXiv",
    primaryClass = "hep-th",
    doi = "10.1103/PhysRevD.94.104038",
    journal = "Phys. Rev. D",
    volume = "94",
    number = "10",
    pages = "104038",
    year = "2016"
}

@article{Giddings:2018umg,
    author = "Giddings, Steven B. and Kinsella, Alex",
    title = "{Gauge-invariant observables, gravitational dressings, and holography in AdS}",
    eprint = "1802.01602",
    archivePrefix = "arXiv",
    primaryClass = "hep-th",
    doi = "10.1007/JHEP11(2018)074",
    journal = "JHEP",
    volume = "11",
    pages = "074",
    year = "2018"
}

@article{Donnelly:2015hta,
    author = "Donnelly, William and Giddings, Steven B.",
    title = "{Diffeomorphism-invariant observables and their nonlocal algebra}",
    eprint = "1507.07921",
    archivePrefix = "arXiv",
    primaryClass = "hep-th",
    reportNumber = "NSF-KITP-15-133",
    doi = "10.1103/PhysRevD.93.024030",
    journal = "Phys. Rev. D",
    volume = "93",
    number = "2",
    pages = "024030",
    year = "2016",
    note = "[Erratum: Phys.Rev.D 94, 029903 (2016)]"
}

@article{Altas:2019qcv,
    author = "Altas, Emel and Tekin, Bayram",
    title = "{Second Order Perturbation Theory in General Relativity: Taub Charges as Integral Constraints}",
    eprint = "1903.11982",
    archivePrefix = "arXiv",
    primaryClass = "hep-th",
    doi = "10.1103/PhysRevD.99.104078",
    journal = "Phys. Rev. D",
    volume = "99",
    number = "10",
    pages = "104078",
    year = "2019"
}

@article{Frob:2022ciq,
    author = {Fr\"ob, Markus B. and Much, Albert and Papadopoulos, Kyriakos},
    title = "{Noncommutative geometry from perturbative quantum gravity}",
    eprint = "2207.03345",
    archivePrefix = "arXiv",
    primaryClass = "gr-qc",
    doi = "10.1103/PhysRevD.107.064041",
    journal = "Phys. Rev. D",
    volume = "107",
    number = "6",
    pages = "064041",
    year = "2023"
}

@article{Altas:2017fcp,
    author = "Altas, Emel and Tekin, Bayram",
    title = "{Linearization instability for generic gravity in AdS spacetime}",
    eprint = "1705.10234",
    archivePrefix = "arXiv",
    primaryClass = "hep-th",
    doi = "10.1103/PhysRevD.97.024028",
    journal = "Phys. Rev. D",
    volume = "97",
    number = "2",
    pages = "024028",
    year = "2018"
}

@article{Altas:2018dci,
    author = "Altas, Emel and Tekin, Bayram",
    title = "{Linearization Instability of Chiral Gravity}",
    eprint = "1804.05602",
    archivePrefix = "arXiv",
    primaryClass = "hep-th",
    doi = "10.1103/PhysRevD.97.124068",
    journal = "Phys. Rev. D",
    volume = "97",
    number = "12",
    pages = "124068",
    year = "2018"
}

@book{Marsden_lectures,
author = {Marsden, Jerrold E.},
title = {Lectures on Geometric Methods in Mathematical Physics},
publisher = {Society for Industrial and Applied Mathematics},
year = {1981},
doi = {10.1137/1.9781611970326},
address = {},
edition   = {},
URL = {https://doi.org/10.1137/1.9781611970326},
eprint = {https://doi.org/10.1137/1.9781611970326}
}

@incollection{taub2011variational,
  title={Variational principles in general relativity},
  author={Taub, AH},
  booktitle={Relativistic Fluid Dynamics},
  pages={205--300},
  year={2011},
  publisher={Springer}
}

@article{Altas:2021htf,
    author = "Altas, Emel and Kilicarslan, Ercan and Tekin, Bayram",
    title = "{Einstein\textendash{}Yang\textendash{}Mills theory: gauge invariant charges and linearization instability}",
    eprint = "2105.11744",
    archivePrefix = "arXiv",
    primaryClass = "hep-th",
    doi = "10.1140/epjc/s10052-021-09460-7",
    journal = "Eur. Phys. J. C",
    volume = "81",
    number = "7",
    pages = "648",
    year = "2021"
}

@article{Frob:2023gng,
    author = {Fr\"ob, Markus B. and Much, Albert and Papadopoulos, Kyriakos},
    title = "{Non-commutative coordinates from quantum gravity}",
    eprint = "2303.17238",
    archivePrefix = "arXiv",
    primaryClass = "gr-qc",
    doi = "10.22323/1.436.0307",
    journal = "PoS",
    volume = "CORFU2022",
    pages = "307",
    year = "2023"
}

@article{Faulkner:2024gst,
    author = "Faulkner, Thomas and Speranza, Antony J.",
    title = "{Gravitational algebras and the generalized second law}",
    eprint = "2405.00847",
    archivePrefix = "arXiv",
    primaryClass = "hep-th",
    doi = "10.1007/JHEP11(2024)099",
    journal = "JHEP",
    volume = "11",
    pages = "099",
    year = "2024"
}

@article{Giddings:2022hba,
    author = "Giddings, Steven B. and Perkins, Julie",
    title = "{Perturbative quantum evolution of the gravitational state and dressing in general backgrounds}",
    eprint = "2209.06836",
    archivePrefix = "arXiv",
    primaryClass = "hep-th",
    doi = "10.1103/PhysRevD.110.026012",
    journal = "Phys. Rev. D",
    volume = "110",
    number = "2",
    pages = "026012",
    year = "2024"
}

@article{Sewell:1982zz,
    author = "Sewell, Geoffrey L.",
    title = "{Quantum fields on manifolds: PCT and gravitationally induced thermal states}",
    doi = "10.1016/0003-4916(82)90285-8",
    journal = "Annals Phys.",
    volume = "141",
    pages = "201--224",
    year = "1982"
}

@article{Higuchi:1991tk,
    author = "Higuchi, A.",
    title = "{Quantum linearization instabilities of de Sitter space-time. 1}",
    doi = "10.1088/0264-9381/8/11/009",
    journal = "Class. Quant. Grav.",
    volume = "8",
    pages = "1961--1981",
    year = "1991"
}

@article{Higuchi:1991tm,
    author = "Higuchi, A.",
    title = "{Quantum linearization instabilities of de Sitter space-time. 2}",
    doi = "10.1088/0264-9381/8/11/010",
    journal = "Class. Quant. Grav.",
    volume = "8",
    pages = "1983--2004",
    year = "1991"
}

@article{Losic:2006ht,
    author = "Losic, B. and Unruh, W. G.",
    title = "{On leading order gravitational backreactions in de Sitter spacetime}",
    eprint = "gr-qc/0604122",
    archivePrefix = "arXiv",
    doi = "10.1103/PhysRevD.74.023511",
    journal = "Phys. Rev. D",
    volume = "74",
    pages = "023511",
    year = "2006"
}

@article{Fredenhagen:1984dc,
    author = "Fredenhagen, Klaus",
    title = "{On the Modular Structure of Local Algebras of Observables}",
    reportNumber = "CPT-84/P-1604",
    doi = "10.1007/BF01206179",
    journal = "Commun. Math. Phys.",
    volume = "97",
    pages = "79",
    year = "1985"
}

@article{Buchholz:1986bg,
    author = "Buchholz, D. and Fredenhagen, K. and D'Antoni, C.",
    title = "{The Universal Structure of Local Algebras}",
    reportNumber = "DESY-86-158",
    doi = "10.1007/BF01239019",
    journal = "Commun. Math. Phys.",
    volume = "111",
    pages = "123",
    year = "1987"
}

@article{Buchholz:1995gr,
    author = "Buchholz, Detlev and Verch, Rainer",
    title = "{Scaling algebras and renormalization group in algebraic quantum field theory}",
    eprint = "hep-th/9501063",
    archivePrefix = "arXiv",
    reportNumber = "DESY-95-004",
    doi = "10.1142/S0129055X9500044X",
    journal = "Rev. Math. Phys.",
    volume = "7",
    pages = "1195--1240",
    year = "1995"
}

@article{Carette:2023wpz,
    author = "Carette, Titouan and G\l{}owacki, Jan and Loveridge, Leon",
    title = "{Operational Quantum Reference Frame Transformations}",
    eprint = "2303.14002",
    archivePrefix = "arXiv",
    primaryClass = "quant-ph",
    month = "3",
    year = "2023"
}

@article{Giacomini:2017zju,
    author = "Giacomini, Flaminia and Castro-Ruiz, Esteban and Brukner, \v{C}aslav",
    title = "{Quantum mechanics and the covariance of physical laws in quantum reference frames}",
    eprint = "1712.07207",
    archivePrefix = "arXiv",
    primaryClass = "quant-ph",
    doi = "10.1038/s41467-018-08155-0",
    journal = "Nature Commun.",
    volume = "10",
    number = "1",
    pages = "494",
    year = "2019"
}

@article{Chandrasekaran:2022cip,
    author = "Chandrasekaran, Venkatesa and Longo, Roberto and Penington, Geoff and Witten, Edward",
    title = "{An algebra of observables for de Sitter space}",
    eprint = "2206.10780",
    archivePrefix = "arXiv",
    primaryClass = "hep-th",
    doi = "10.1007/JHEP02(2023)082",
    journal = "JHEP",
    volume = "02",
    pages = "082",
    year = "2023"
}

@article{delaHamette:2021oex,
    author = {de la Hamette, Anne-Catherine and Galley, Thomas D. and H\"ohn, Philipp A. and Loveridge, Leon and M\"uller, Markus P.},
    title = "{Perspective-neutral approach to quantum frame covariance for general symmetry groups}",
    eprint = "2110.13824",
    archivePrefix = "arXiv",
    primaryClass = "quant-ph",
    month = "10",
    year = "2021"
}

@article{Hoehn:2020epv,
    author = {H\"ohn, Philipp A. and Smith, Alexander R. H. and Lock, Maximilian P. E.},
    title = "{Equivalence of Approaches to Relational Quantum Dynamics in Relativistic Settings}",
    eprint = "2007.00580",
    archivePrefix = "arXiv",
    primaryClass = "gr-qc",
    doi = "10.3389/fphy.2021.587083",
    journal = "Front. in Phys.",
    volume = "9",
    pages = "181",
    year = "2021"
}

@article{Hoehn:2019fsy,
    author = {H\"ohn, Philipp A. and Smith, Alexander R. H. and Lock, Maximilian P. E.},
    title = "{Trinity of relational quantum dynamics}",
    eprint = "1912.00033",
    archivePrefix = "arXiv",
    primaryClass = "quant-ph",
    doi = "10.1103/PhysRevD.104.066001",
    journal = "Phys. Rev. D",
    volume = "104",
    number = "6",
    pages = "066001",
    year = "2021"
}

@article{Sorce:2023fdx,
    author = "Sorce, Jonathan",
    title = "{Notes on the type classification of von Neumann algebras}",
    eprint = "2302.01958",
    archivePrefix = "arXiv",
    primaryClass = "hep-th",
    reportNumber = "MIT-CTP/5527",
    doi = "10.1142/S0129055X24300024",
    journal = "Rev. Math. Phys.",
    volume = "36",
    number = "02",
    pages = "2430002",
    year = "2024"
}

@article{Jensen:2023yxy,
    author = "Jensen, Kristan and Sorce, Jonathan and Speranza, Antony J.",
    title = "{Generalized entropy for general subregions in quantum gravity}",
    eprint = "2306.01837",
    archivePrefix = "arXiv",
    primaryClass = "hep-th",
    doi = "10.1007/JHEP12(2023)020",
    journal = "JHEP",
    volume = "12",
    pages = "020",
    year = "2023"
}

@article{Witten:2021unn,
    author = "Witten, Edward",
    title = "{Gravity and the crossed product}",
    eprint = "2112.12828",
    archivePrefix = "arXiv",
    primaryClass = "hep-th",
    doi = "10.1007/JHEP10(2022)008",
    journal = "JHEP",
    volume = "10",
    pages = "008",
    year = "2022"
}

@article{Goeller:2022rsx,
    author = {Goeller, Christophe and H\"ohn, Philipp A. and Kirklin, Josh},
    title = "{Diffeomorphism-invariant observables and dynamical frames in gravity: reconciling bulk locality with general covariance}",
    eprint = "2206.01193",
    archivePrefix = "arXiv",
    primaryClass = "hep-th",
    month = "6",
    year = "2022"
}

@article{AliAhmad:2024wja,
    author = "Ali Ahmad, Shadi and Chemissany, Wissam and Klinger, Marc S. and Leigh, Robert G.",
    title = "{Quantum reference frames from top-down crossed products}",
    eprint = "2405.13884",
    archivePrefix = "arXiv",
    primaryClass = "hep-th",
    doi = "10.1103/PhysRevD.110.065003",
    journal = "Phys. Rev. D",
    volume = "110",
    number = "6",
    pages = "065003",
    year = "2024"
}

@article{AliAhmad:2024saq,
    author = "Ali Ahmad, Shadi and Klinger, Marc S.",
    title = "{Emergent Geometry from Quantum Probability}",
    eprint = "2411.07288",
    archivePrefix = "arXiv",
    primaryClass = "hep-th",
    month = "11",
    year = "2024"
}

@article{Carrozza:2022xut,
    author = {Carrozza, Sylvain and Eccles, Stefan and H\"ohn, Philipp A.},
    title = "{Edge modes as dynamical frames: charges from post-selection in generally covariant theories}",
    eprint = "2205.00913",
    archivePrefix = "arXiv",
    primaryClass = "hep-th",
    doi = "10.21468/SciPostPhys.17.2.048",
    journal = "SciPost Phys.",
    volume = "17",
    number = "2",
    pages = "048",
    year = "2024"
}

@article{Penington:2023dql,
    author = "Penington, Geoff and Witten, Edward",
    title = "{Algebras and States in JT Gravity}",
    eprint = "2301.07257",
    archivePrefix = "arXiv",
    primaryClass = "hep-th",
    month = "1",
    year = "2023"
}

@article{HK,
    author = {H\"ohn, Philipp A. and Kirklin, Josh},
    title = "Fighting nonlocality with nonlocality: observables and microcausality in QED",
    journal = "forthcoming",
    year = "2025"
}

@article{Kolchmeyer:2023gwa,
    author = "Kolchmeyer, David K.",
    title = "{von Neumann algebras in JT gravity}",
    eprint = "2303.04701",
    archivePrefix = "arXiv",
    primaryClass = "hep-th",
    doi = "10.1007/JHEP06(2023)067",
    journal = "JHEP",
    volume = "06",
    pages = "067",
    year = "2023"
}

@book{haag2012local,
  title={Local quantum physics: Fields, particles, algebras},
  author={Haag, Rudolf},
  year={2012},
  publisher={Springer Science \& Business Media}
}

@article{Witten:2023qsv,
    author = "Witten, Edward",
    title = "{Algebras, regions, and observers}",
    eprint = "2303.02837",
    archivePrefix = "arXiv",
    primaryClass = "hep-th",
    journal = "Proc. Symp. Pure Math.",
    volume = "107",
    pages = "247--276",
    year = "2024"
}

@article{Strohmaier:2023opz,
    author = "Strohmaier, Alexander and Witten, Edward",
    title = "{The Timelike Tube Theorem in Curved Spacetime}",
    eprint = "2303.16380",
    archivePrefix = "arXiv",
    primaryClass = "hep-th",
    doi = "10.1007/s00220-024-05009-3",
    journal = "Commun. Math. Phys.",
    volume = "405",
    number = "7",
    pages = "153",
    year = "2024"
}

@article{Aguilar-Gutierrez:2023odp,
    author = "Aguilar-Gutierrez, Sergio E. and Bahiru, Eyoab and Esp\'\i{}ndola, Ricardo",
    title = "{The centaur-algebra of observables}",
    eprint = "2307.04233",
    archivePrefix = "arXiv",
    primaryClass = "hep-th",
    doi = "10.1007/JHEP03(2024)008",
    journal = "JHEP",
    volume = "03",
    pages = "008",
    year = "2024"
}

@article{AliAhmad:2023etg,
    author = "Ali Ahmad, Shadi and Jefferson, Ro",
    title = "{Crossed product algebras and generalized entropy for subregions}",
    eprint = "2306.07323",
    archivePrefix = "arXiv",
    primaryClass = "hep-th",
    doi = "10.21468/SciPostPhysCore.7.2.020",
    journal = "SciPost Phys. Core",
    volume = "7",
    pages = "020",
    year = "2024"
}

@article{Gomez:2023wrq,
    author = "Gomez, C.",
    title = "{Entanglement, Observers and Cosmology: a view from von Neumann Algebras}",
    eprint = "2302.14747",
    archivePrefix = "arXiv",
    primaryClass = "hep-th",
    month = "2",
    year = "2023"
}

@article{Gomez:2023upk,
    author = "Gomez, Cesar",
    title = "{Clocks, Algebras and Cosmology}",
    eprint = "2304.11845",
    archivePrefix = "arXiv",
    primaryClass = "hep-th",
    month = "4",
    year = "2023"
}

@article{Carrozza:2021gju,
    author = {Carrozza, Sylvain and H\"ohn, Philipp A.},
    title = "{Edge modes as reference frames and boundary actions from post-selection}",
    eprint = "2109.06184",
    archivePrefix = "arXiv",
    primaryClass = "hep-th",
    doi = "10.1007/JHEP02(2022)172",
    journal = "JHEP",
    volume = "02",
    pages = "172",
    year = "2022"
}

@article{sorce2024analyticity,
    author = "Sorce, Jonathan",
    title = "{Analyticity and the Unruh effect: a study of local modular flow}",
    eprint = "2403.18937",
    archivePrefix = "arXiv",
    primaryClass = "hep-th",
    reportNumber = "MIT-CTP/5699",
    doi = "10.1007/JHEP09(2024)040",
    journal = "JHEP",
    volume = "24",
    pages = "040",
    year = "2020"
}

@article{Bisognano:1975ih,
    author = "Bisognano, J. J and Wichmann, E. H.",
    title = "{On the Duality Condition for a Hermitian Scalar Field}",
    doi = "10.1063/1.522605",
    journal = "J. Math. Phys.",
    volume = "16",
    pages = "985--1007",
    year = "1975"
}

@article{WittenRevModPhys.90.045003,
  title = {APS Medal for Exceptional Achievement in Research: Invited article on entanglement properties of quantum field theory},
  author = {Witten, Edward},
  journal = {Rev. Mod. Phys.},
  volume = {90},
  issue = {4},
  pages = {045003},
  numpages = {38},
  year = {2018},
  month = {10},
  publisher = {American Physical Society},
  doi = {10.1103/RevModPhys.90.045003},
  url = {https://link.aps.org/doi/10.1103/RevModPhys.90.045003}
}

@article{leutheusser2023causal,
    author = "Leutheusser, Samuel and Liu, Hong",
    title = "{Causal connectability between quantum systems and the black hole interior in holographic duality}",
    eprint = "2110.05497",
    archivePrefix = "arXiv",
    primaryClass = "hep-th",
    reportNumber = "MIT-CTP/5335",
    doi = "10.1103/PhysRevD.108.086019",
    journal = "Phys. Rev. D",
    volume = "108",
    number = "8",
    pages = "086019",
    year = "2023"
}

@article{leutheusser2023emergent,
    author = "Leutheusser, Samuel Aaron Wehlau and Liu, Hong",
    title = "{Emergent Times in Holographic Duality}",
    eprint = "2112.12156",
    archivePrefix = "arXiv",
    primaryClass = "hep-th",
    reportNumber = "MIT-CTP/5382",
    doi = "10.1103/PhysRevD.108.086020",
    journal = "Phys. Rev. D",
    volume = "108",
    number = "8",
    pages = "086020",
    year = "2023"
}

@article{Fewster:2024pur,
    author = "Fewster, Christopher J. and Janssen, Daan W. and Loveridge, Leon Deryck and Rejzner, Kasia and Waldron, James",
    title = "{Quantum reference frames, measurement schemes and the type of local algebras in quantum field theory}",
    eprint = "2403.11973",
    archivePrefix = "arXiv",
    primaryClass = "math-ph",
    month = "3",
    year = "2024"
}

@book{Takesaki1979,
  author = "Takesaki, M.",
  title = {Theory of Operator Algebras I},
  ISBN = {9781461261889},
  url = {http://dx.doi.org/10.1007/978-1-4612-6188-9},
  DOI = {10.1007/978-1-4612-6188-9},
  publisher = {Springer New York},
  year = {1979}
}

@article{Wall_2015,
   title={A Second Law for higher curvature gravity},
   volume={24},
   ISSN={1793-6594},
   url={http://dx.doi.org/10.1142/S0218271815440149},
   DOI={10.1142/s0218271815440149},
   number={12},
   journal={International Journal of Modern Physics D},
   publisher={World Scientific Pub Co Pte Lt},
   author={Wall, Aron C.},
   year={2015},
   month=oct, pages={1544014} }

@article{Araki1963AGO,
title = {A Generalisation Of Borchers' Theorem},
author = {Araki, H},
abstractNote = {Borchers' theorem on causal dependence of rings of operators in quantum field theory was generaiized by a new method of proof, based on the uniqueness theorem for hyperbolic partial differential equations. (auth)},
doi = {},
url = {https://www.osti.gov/biblio/4665531}, 
journal = {Helvetica Physica Acta (Switzerland)},
year = {1963},
month = {1}
}

@Article{Borchers1961,
author="Borchers, H. J.",
title="{\"U}ber die Vollst{\"a}ndigkeit lorentzinvarianter Felder in einer zeitartigen R{\"o}hre",
journal="Il Nuovo Cimento (1955-1965)",
year="1961",
month={2},
day="01",
volume="19",
number="4",
pages="787--793",
issn="1827-6121",
doi="10.1007/BF02733373",
url="https://doi.org/10.1007/BF02733373"
}

@INBOOK{Witten_Crnkovic,
       author = {{Crnkovic}, C. and {Witten}, E.},
        title = "{Covariant description of canonical formalism in geometrical theories.}",
    booktitle = {Three Hundred Years of Gravitation},
         year = 1987,
        pages = {676-684},
       adsurl = {https://ui.adsabs.harvard.edu/abs/1987thyg.book..676C}
}

@article{Lee:1990nz,
    author = "Lee, J. and Wald, Robert M.",
    title = "{Local symmetries and constraints}",
    doi = "10.1063/1.528801",
    journal = "J. Math. Phys.",
    volume = "31",
    pages = "725--743",
    year = "1990"
}

@article{Wald:1993nt,
    author = "Wald, Robert M.",
    title = "{Black hole entropy is the Noether charge}",
    eprint = "gr-qc/9307038",
    archivePrefix = "arXiv",
    reportNumber = "EFI-93-42",
    doi = "10.1103/PhysRevD.48.R3427",
    journal = "Phys. Rev. D",
    volume = "48",
    number = "8",
    pages = "R3427--R3431",
    year = "1993"
}

@article{Jacobson:1993vj,
    author = "Jacobson, Ted and Kang, Gungwon and Myers, Robert C.",
    title = "{On black hole entropy}",
    eprint = "gr-qc/9312023",
    archivePrefix = "arXiv",
    reportNumber = "MCGILL-93-22, NSF-ITP-93-152, UMDGR-94-75",
    doi = "10.1103/PhysRevD.49.6587",
    journal = "Phys. Rev. D",
    volume = "49",
    pages = "6587--6598",
    year = "1994"
}

@article{Wald:1999wa,
    author = "Wald, Robert M. and Zoupas, Andreas",
    title = "{A General definition of 'conserved quantities' in general relativity and other theories of gravity}",
    eprint = "gr-qc/9911095",
    archivePrefix = "arXiv",
    doi = "10.1103/PhysRevD.61.084027",
    journal = "Phys. Rev. D",
    volume = "61",
    pages = "084027",
    year = "2000"
}

@article{Iyer:1994ys,
    author = "Iyer, Vivek and Wald, Robert M.",
    title = "{Some properties of Noether charge and a proposal for dynamical black hole entropy}",
    eprint = "gr-qc/9403028",
    archivePrefix = "arXiv",
    doi = "10.1103/PhysRevD.50.846",
    journal = "Phys. Rev. D",
    volume = "50",
    pages = "846--864",
    year = "1994"
}

@article{Iyer:1995kg,
    author = "Iyer, Vivek and Wald, Robert M.",
    title = "{A Comparison of Noether charge and Euclidean methods for computing the entropy of stationary black holes}",
    eprint = "gr-qc/9503052",
    archivePrefix = "arXiv",
    doi = "10.1103/PhysRevD.52.4430",
    journal = "Phys. Rev. D",
    volume = "52",
    pages = "4430--4439",
    year = "1995"
}

@article{Harlow:2019yfa,
    author = "Harlow, Daniel and Wu, Jie-Qiang",
    title = "{Covariant phase space with boundaries}",
    eprint = "1906.08616",
    archivePrefix = "arXiv",
    primaryClass = "hep-th",
    doi = "10.1007/JHEP10(2020)146",
    journal = "JHEP",
    volume = "10",
    pages = "146",
    year = "2020"
}

@article{anderson1992introduction,
  title={Introduction to the variational bicomplex},
  author={Anderson, Ian M},
  journal={Contemporary Mathematics},
  volume = {132},
  year={1992}
}

@article{Wald:1990mme,
    author = "Wald, Robert M.",
    title = "{On identically closed forms locally constructed from a field}",
    doi = "10.1063/1.528839",
    journal = "J. Math. Phys.",
    volume = "31",
    number = "10",
    pages = "2378",
    year = "1990"
}

@article{Seifert:2006kv,
    author = "Seifert, Michael D. and Wald, Robert M.",
    title = "{A General variational principle for spherically symmetric perturbations in diffeomorphism covariant theories}",
    eprint = "gr-qc/0612121",
    archivePrefix = "arXiv",
    doi = "10.1103/PhysRevD.75.084029",
    journal = "Phys. Rev. D",
    volume = "75",
    pages = "084029",
    year = "2007"
}

@article{KirklinGSL,
author={Kirklin, Josh},
title = "{Generalised second law beyond the semiclassical regime}",
    eprint = "2412.01903",
    archivePrefix = "arXiv",
    primaryClass = "hep-th",
    month = "12",
    year = "2024"
}

@article{AliAhmad:2024eun,
    author = "Ali Ahmad, Shadi and Klinger, Marc S. and Lin, Simon",
    title = "{Semifinite von Neumann algebras in gauge theory and gravity}",
    eprint = "2407.01695",
    archivePrefix = "arXiv",
    primaryClass = "hep-th",
    month = "7",
    year = "2024"
}

@article{Dong_2014,
   title={Holographic entanglement entropy for general higher derivative gravity},
   volume={2014},
   ISSN={1029-8479},
   url={http://dx.doi.org/10.1007/JHEP01(2014)044},
   DOI={10.1007/jhep01(2014)044},
   number={1},
   journal={Journal of High Energy Physics},
   publisher={Springer Science and Business Media LLC},
   author={Dong, Xi},
   year={2014},
   month=jan }

@article{Camps:2013zua,
    author = "Camps, Joan",
    title = "{Generalized entropy and higher derivative Gravity}",
    eprint = "1310.6659",
    archivePrefix = "arXiv",
    primaryClass = "hep-th",
    doi = "10.1007/JHEP03(2014)070",
    journal = "JHEP",
    volume = "03",
    pages = "070",
    year = "2014"
}

@article{Kaplan:2024xyk,
    author = "Kaplan, Molly and Marolf, Donald and Yu, Xuyang and Zhao, Ying",
    title = "{De Sitter quantum gravity and the emergence of local algebras}",
    eprint = "2410.00111",
    archivePrefix = "arXiv",
    primaryClass = "hep-th",
    month = "9",
    year = "2024"
}

@article{Kijowski:1973gi,
    author = "Kijowski, J.",
    title = "{A finite-dimensional canonical formalism in the classical field theory}",
    doi = "10.1007/BF01645975",
    journal = "Commun. Math. Phys.",
    volume = "30",
    pages = "99--128",
    year = "1973"
}

@article{Kijowski:1976ze,
    author = "Kijowski, J. and Szczyrba, W.",
    title = "{A Canonical Structure for Classical Field Theories}",
    doi = "10.1007/BF01608496",
    journal = "Commun. Math. Phys.",
    volume = "46",
    pages = "183--206",
    year = "1976"
}

\end{document}